\documentclass[%
twocolumn,
superscriptaddress,
preprintnumbers,
nofootinbib,
amsmath,amssymb,
aps,
prd,
floatfix,
]{revtex4-1}

\usepackage[utf8]{inputenc}
\usepackage{graphicx}
\usepackage[caption=false]{subfig}
\usepackage{bm}% bold math

\usepackage[pdftex]{color}
\usepackage[dvipsnames]{xcolor}
\usepackage{hyperref}% add hypertext capabilities
\hypersetup{
    colorlinks=true,     % false: boxed links; true: colored links
    linkcolor=blue,      % color of internal links
    citecolor=blue,      % color of links to bibliography
    filecolor=blue,      % color of file links
    urlcolor=blue        % color of external links
}

\usepackage{mathtools} % ensuremath etc
\usepackage{xspace} % So math macros can be used in text as well -- http://tex.stackexchange.com/questions/31091/space-after-latex-commands

\newcommand{\vev}[1]{\ensuremath{\langle #1 \rangle}}

\begin{document}

\preprint{MIT-CTP/5312,FERMILAB-PUB-25-0090-T}

\title{Flow-based sampling for multimodal and extended-mode distributions in lattice field theory}
\author{Daniel C. Hackett}
\affiliation{Center for Theoretical Physics, Massachusetts Institute of Technology, Cambridge, MA 02139, USA}
\affiliation{The NSF AI Institute for Artificial Intelligence and Fundamental Interactions}
\affiliation{Fermi National Accelerator Laboratory, Batavia, IL 60510, USA}
\author{Chung-Chun Hsieh}
\affiliation{Department of Physics and Center for Theoretical Physics, National Taiwan University, Taipei, Taiwan 106}
\affiliation{Department of Physics, University of Maryland, College Park, MD 20742 USA}
\affiliation{Maryland Center for Fundamental Physics, University of Maryland, College Park, MD 20742, USA}
\author{Sahil Pontula}
\affiliation{Center for Theoretical Physics, Massachusetts Institute of Technology, Cambridge, MA 02139, USA}
\affiliation{Department of Electrical Engineering and Computer Science, Massachusetts Institute of Technology, Cambridge, MA 02139, USA}
\author{Michael~S.~Albergo}
\affiliation{Society of Fellows, Harvard University, Cambridge, MA, 02138, USA}
\affiliation{The NSF AI Institute for Artificial Intelligence and Fundamental Interactions}
\author{Denis~Boyda}
\affiliation{Center for Theoretical Physics, Massachusetts Institute of Technology, Cambridge, MA 02139, USA}
\affiliation{The NSF AI Institute for Artificial Intelligence and Fundamental Interactions}
\author{Jiunn-Wei~Chen}
\affiliation{Department of Physics and Center for Theoretical Physics, National Taiwan University, Taipei, Taiwan 106}
\affiliation{Physics Division, National Center for Theoretical Sciences, Taipei 10617, Taiwan}
\affiliation{Leung Center for Cosmology and Particle Astrophysics, National Taiwan University, Taipei, Taiwan 106}
\author{Kai-Feng Chen}
\affiliation{Department of Physics and Center for Theoretical Physics, National Taiwan University, Taipei, Taiwan 106}
\author{Kyle~Cranmer}
\affiliation{Department of Physics and Data Science Institute, University of Wisconsin--Madison, Madison, WI 53706, USA}
\author{Gurtej Kanwar}
\affiliation{Higgs Centre for Theoretical Physics, University of Edinburgh, Edinburgh EH9 3FD, UK}
\author{Phiala E. Shanahan}
\affiliation{Center for Theoretical Physics, Massachusetts Institute of Technology, Cambridge, MA 02139, USA}
\affiliation{The NSF AI Institute for Artificial Intelligence and Fundamental Interactions}

\date{\today}

\begin{abstract}
Recent results have demonstrated that samplers constructed with flow-based generative models are a promising new approach for configuration generation in lattice field theory. In this paper, we present a set of training- and architecture-based methods to construct flow models for targets with multiple separated modes (i.e.~vacua) as well as targets with extended/continuous modes. 
We demonstrate the application of these methods to modeling two-dimensional real and complex scalar field theories in their symmetry-broken phases. In this context we investigate different flow-based sampling algorithms, including a composite sampling algorithm where flow-based proposals are occasionally augmented by applying updates using traditional algorithms like HMC.
\end{abstract}

\maketitle

\renewcommand{\baselinestretch}{0.97}\normalsize % HACK: get VI.D in left column
\tableofcontents
\renewcommand{\baselinestretch}{1.0}\normalsize

% ToC finishes as block above text
\onecolumngrid
~\\
~\\
\twocolumngrid

% TOC in left col, intro starts in right
% \newpage

\section{Introduction}

Quantum field theories (QFTs) are the mathematical and conceptual frameworks through which we describe particle-based theories of Nature. For example, the Standard Model of particle physics encapsulates our current best understanding of fundamental particles and their interactions in terms of a QFT. 
In many cases, deriving physics results from QFTs
requires calculations in strong-coupling regimes
where perturbative methods are inapplicable.
Lattice quantum field theory (LQFT) provides a non-perturbative approach 
based on numerically evaluating path integrals, rendered finite and regularized by discretizing the quantum fields onto a Euclidean spacetime lattice~\cite{morningstar2007}. This method is commonly applied to 
Quantum Chromodynamics (QCD)~\cite{Lehner:2019wvv,Kronfeld:2019nfb,Cirigliano:2019jig,Detmold:2019ghl,Bazavov:2019lgz,Joo:2019byq}, 
as well as QCD-like components of speculative beyond--Standard-Model theories~\cite{Brower:2019oor,DeGrand:2015zxa,Svetitsky:2017xqk,Kribs:2016cew} and 
many-body systems in condensed matter physics~\cite{Mathur2016}.

There are growing efforts to apply machine learning techniques to accelerate sampling for lattice field theories with the goal of enabling calculations which are currently computationally intractable~\cite{Wang2017,Huang:2017,song2017nice,levy2018generalizing,LiWang2018NNRG,Pawlowski:2018qxs,Wu:2019,Albergo:2019eim,Kanwar:2020xzo,Boyda:2020hsi,Pawlowski_2020,Nicoli:2020evf,Nicoli:2020njz,Nagai:2020jar,Lawrence:2021izu,Albergo:2021bna,Albergo:2021vyo,DelDebbio:2021qwf,Tomiya:2021ywc,Foreman:2021ixr,Gabrie:2021tlu,Foreman:2021rhs,Wu:2021tfb,deHaan:2021erb,Foreman:2021ljl,Abbott:2022hkm,Albergo:2022qfi,Caselle:2022acb,
% Caselle:2022esc,
Matthews:2022sds,Gerdes:2022eve,Finkenrath:2022ogg,Singha:2022lpi,Chen:2022ytr,Bialas:2022iro,Albandea:2023wgd,Abbott:2022zhs,Tomiya:2022meu,Jin:2022bgq,Abbott:2022zsh,Bacchio:2022vje,Boyle:2022xor,Albandea:2023ais,Nicoli:2023qsl,Bacchio:2023all,Nagai:2023fxt,Caselle:2023mvh,Cranmer:2023xbe,Abbott:2023thq,R:2023dcr,Komijani:2023fzy,Singha:2023xxq,Bialas:2023fyj}.
% \cite{Wang2017,Huang:2017,song2017nice,levy2018generalizing,LiWang2018NNRG,Wu:2019,Albergo:2019eim,Kanwar:2020xzo,Boyda:2020hsi,Pawlowski_2020,Nicoli:2020evf,Nicoli:2020njz,Albergo:2021bna}.
One such approach uses flow models to independently sample field configurations. In this framework, the flow model defines a variational proposal distribution that can be optimized to approximate the physical distribution of interest. Flow-based MCMC~\cite{Albergo:2019eim} then uses proposals generated by flow models with a corrective Metropolis-Hastings accept/reject step, which grants theoretical guarantees of asymptotic exactness in the same way that more traditional sampling methods such as hybrid/Hamiltonian Monte Carlo (HMC)~\cite{Duane:1987de} do. 

For both traditional methods and flow-based MCMC these guarantees of exactness are asymptotic, and in practice a finite ensemble of samples may not be sufficient to achieve a representative sampling and unbiased results~\cite{Albergo:2022qfi}.
However, as we discuss and demonstrate in this work, traditional methods and flow-based MCMC can exhibit poor sampling behavior in different ways, and can be combined to produce efficient samplers where either method would fail individually.
Distributions with nontrivial mode structure provide a clear example where this is the case, and are the focus of this work.

The remainder of the introduction comprises a detailed overview of this extended exploration.

First, Sec.~\ref{sec:challenges} sets up the problem to be solved. This section reviews the algorithms of interest---traditional MCMC methods and emerging flow-based approaches---and 
the separate problems faced by each in sampling distributions with nontrivial mode structure.
In particular, finite-sample pathologies have been demonstrated for traditional methods in cases where the sampling algorithm remains trapped in one or several of many modes of a multimodal distribution, such as when ``topological freezing'' occurs~\cite{DelDebbio:2004xh,Schaefer:2009xx,Schaefer:2010hu,Hasenbusch:2017fsd,Albergo:2022qfi}. 
Meanwhile, the efficiency of flow-based MCMC depends on how well the model approximates the target distribution: models that undersample particular regions of the target result in inefficient samplers due to long runs of rejected proposals, requiring many proposal steps before asymptotic convergence.
This poses a particular problem in the context of multimodal distributions: self-training procedures are prone to producing ``mode-collapsed'' models that sample only single modes of the target.
A similar problem arises in modeling extended-mode distributions.
Avoiding these pathologies is critical to accurately modeling these distributions.

Section~\ref{sec:augmented-MCMC} reviews a general and widely employed algorithmic framework, which we refer to in generality as ``composite'' or ``augmented MCMC'', where multiple interleaved update types are used to advance the Markov chain. This scheme provides a simple way to combine flow-based MCMC with traditional sampling algorithms. When used in combination, the two approaches may correct each other's respective deficiencies, with flow proposals providing rapid mixing between modes while updates with traditional methods populate regions undersampled by the flow model. This further motivates the need to construct models which capture the full mode structure of the theory of interest.

To this end, without specializing to a particular example, Sections \ref{sec:arch_methods} and \ref{sec:multimodal-approaches} present a broad (but not exhaustive) zoo of methods to construct models for distributions with nontrivial mode structure.
Section~\ref{sec:arch_methods} presents architecture-based approaches, which specialize the construction of the flow model to the mode structure of interest.
The methods discussed are statistical mixtures of multiple flow models, symmetry-equivariant flows, and topology matching.
Section~\ref{sec:multimodal-approaches} presents several training-based approaches, which specialize the training scheme to prevent mode collapse.
These include a novel self-training scheme which admits symmetrization of the training data, as well as two different schemes---applicable to arbitrary mode structures---which regulate training by slowly varying the loss to anneal into the desired mode structure.
Different approaches require different degrees of a priori knowledge and engineering to employ, as explored in the discussion.

Sections~\ref{sec:multimodal-comparison} and \ref{sec:extended} present numerical demonstrations of these methods.
As testbeds, we study the particular examples of two-dimensional real (Sec.~\ref{sec:multimodal-comparison}) and complex (Sec.~\ref{sec:extended}) scalar field theories. 
With quadratic and quartic interactions, the real theory has a $Z_2$ global symmetry (negation of the field) and a corresponding broken-symmetry phase for which the distribution of fields is bimodal. 
Similarly, the complex theory possesses a global U(1) symmetry (multiplication by a phase), which produces an extended ``ring-shaped'' mode when spontaneously broken.
In the bimodal real case, we further explore modeling the theory with a symmetry-breaking linear interaction, providing a model for the more general case of multiple modes with different shapes and relative weights.
In each case, we demonstrate the efficacy of all approaches discussed in Secs.~\ref{sec:arch_methods} and \ref{sec:multimodal-approaches} in constructing models with the desired mode structure.

Considering the example of the bimodal real theory specifically, Sec.~\ref{sec:sampling} explores how the models constructed in Sec.~\ref{sec:multimodal-comparison} can be used for sampling. First, the models are investigated for use with flow-based MCMC as the exclusive driver of the Markov process. Through a careful analysis of the asymptotic behavior of one such Markov process, we identify subtle issues with this approach that arise at high statistics due to undersampling in the ``inner tails'' (regions of low density between modes). We conclude that---despite their quality---these multimodal flow models are not useful for sampling by themselves. We go on to apply the methods of Sec.~\ref{sec:augmented-MCMC} to construct a composite sampling algorithm which interleaves steps of flow-based MCMC and HMC. Analyzing the asymptotic properties of the resulting Markov process, we find that incorporating HMC steps resolves the problems exhibited in pure flow-based MCMC. The result is a sampler with the advantages of flow-based MCMC---i.e.~fast mixing, especially between modes---without its asymptotic problems.

Finally, Sec.~\ref{sec:conclusion} closes with a discussion of the results and provides an outlook.

\section{Challenges in sampling multimodal and extended-mode distributions}
\label{sec:challenges}

In field theory, expectation values of physical observables are obtained by evaluating path integrals over (field) configuration space.
By discretizing the QFT onto a lattice in Euclidean spacetime, one can evaluate path integrals nonperturbatively. The expectation value of an observable $\mathcal{O}$ can be expressed as:
\begin{equation}
\langle\mathcal{O}\rangle=\frac{1}{Z}\int{d\phi ~ \mathcal{O}(\phi) ~ e^{-S_E(\phi)}}, \;\,\, Z=\int{\mathcal{D}\phi ~ e^{-S_E(\phi)}},
\label{eqn:path-integral-vev}
\end{equation}
where $\int d\phi$ denotes the path integral over the space of field configurations $\phi$ and $S_E$ is the Euclidean action. With the field degrees of freedom discretized on a lattice, this integral is finite but very high-dimensional. However, it can be recast as an expectation value $\vev{\mathcal{O}}_p$ with respect to the distribution $p(\phi) = e^{-S_E(\phi)}/Z$, as long as $S_E$ is real.
The path integral of Eq.~\eqref{eqn:path-integral-vev} can then be approximated in a systematically improvable way by computing stochastic estimators of observable expectation values
\begin{equation}
\vev{\mathcal{O}}_p \approx \frac{1}{N} \sum_{i=1}^N{\mathcal{O}(\phi_i)}
\end{equation}
over $N$ samples of field configurations $\phi_i \sim p$ (i.e. drawn from the distribution $p$).

In the following, we distinguish between three general approaches to sampling high-dimensional distributions relevant for lattice field theories:
\begin{enumerate}
    \item \emph{Update-based sampling} --- sampling using MCMC, producing each new field configuration by perturbing the previous configuration in the Markov chain. This class of samplers includes the Metropolis-Hastings accept/reject step applied with a proposal based on local updates~\cite{Metropolis:1953am,Hastings:1970aa}, heat bath updates to subsets of variables~\cite{Creutz:1979dw,Cabibbo:1982zn,Kennedy:1985nu}, and HMC~\cite{Duane:1987de}.
    \item \emph{Flow-based sampling} --- using flow models to generate independent configurations distributed according to an approximation of the desired probability density. Asymptotically unbiased estimates of expectation values can be computed using e.g.~reweighting\footnote{Also known as ``importance sampling'' in the statistics literature.} or independence Metropolis~\cite{tierney1994markov}. Though the latter is an MCMC sampler, we distinguish this approach from updated-based sampling based on the independence of the proposed updates.
    \item \emph{Combined approaches} --- combination of different sampling approaches, such as update-based and flow-based sampling methods. For example, one can perform a composition of flow-based MCMC and HMC.
\end{enumerate}

High-dimensional distributions with nontrivial mode structure present challenges to both update-based samplers and flow-based samplers for different reasons. We explore these difficulties in the following sections, and find that the distinct challenges faced in flow-based sampling for these distributions suggest new approaches.

\subsection{Update-based methods: MCMC and the Metropolis algorithm}

For many high-dimensional distributions, exact sampling schemes are not known. In these cases, MCMC provides a general approach to drawing samples with asymptotically correct statistics~\cite{brooks2011handbook}.
MCMC methods implement importance sampling by constructing a Markov chain of consecutive configurations $\phi_0, \phi_1, \ldots, \phi_N$ generated by a Markov process which satisfies balance with respect to the target distribution.
The balance condition guarantees that the target distribution is fixed under the stochastic updates that define the Markov process and, assuming ergodicity, results in samples distributed according to the target distribution in the limit of a large number of updates~\cite{meyn2012markovErgodicity}. 

Appropriate Markov processes can be conveniently constructed using the Metropolis algorithm~\cite{Metropolis:1953am,Hastings:1970aa}: given the current configuration $\phi_i$, one determines the next configuration $\phi_{i+1}$ by proposing an update $\phi'$ with (conditional) probability $T(\phi_i \rightarrow \phi')$ and then accepting or rejecting the proposal with probability
\begin{equation}
p_{\text{acc.}}(\phi_i \rightarrow \phi') = \min\left[ 1, ~
\frac{T(\phi' \rightarrow \phi_i)}{T(\phi_i \rightarrow \phi')} \frac{p(\phi')}{p(\phi_i)}
\right].
\label{eqn:mcmc-accept-rate}
\end{equation}
If the proposal is accepted, the next configuration is $\phi_{i+1} = \phi'$; if rejected, then the previous configuration is repeated, $\phi_{i+1} = \phi_i$.
The resulting Markov process satisfies a stronger form of balance, known as ``detailed balance'', guaranteeing the correct stationary distribution.
Any method of generating proposals is acceptable as long as it is ergodic and $T(\phi_i \rightarrow \phi')/T(\phi' \rightarrow \phi_i)$ is computable.

As familiar in the lattice field theory context, MCMC algorithms require an initial thermalization or ``burn-in'' period before they begin generating target-distributed samples.
Separately, a Markov chain must be sufficiently long to reach the regime where expectation values evaluated over the chain converge smoothly in $1/\sqrt{N}$ to their asymptotic values, i.e.~that the true sampling distributions of means over chains of length $N$ are approximately Gaussian.
Away from this regime, finite-sample error estimates can severely underestimate the true sampling variance, leading to unreliable results.
As discussed below, achieving a slow approach to this smoothly converging regime can be a practical issue for either update-based or flow-based methods, especially when sampling multimodal distributions.

\emph{Sampling challenges for update-based methods:}
In update-based sampling methods, each Markov chain step produces a new configuration that is a perturbation of the previous configuration. 
To satisfy detailed balance, these steps preferentially move towards directions of higher probability density. When the target distribution is multimodal with widely separated modes, this approach encounters severe difficulties: regions between modes must be traversed by the sampler to successfully sample from all modes, but by detailed balance this can happen only rarely when updates are local perturbations.
As modes become increasingly separated, the characteristic ``tunneling time'' for the sampler to move from one mode to another grows rapidly, sometimes referred to as ``freezing''.
The Markov chain generated by the sampler will not yield smoothly converging results until many such tunneling events have occurred.
This effect thus rapidly increases the sample sizes required to obtain unbiased results, presenting an obstacle for traditional MCMC methods.

\subsection{Flow-based methods and reverse KL self-training}
\label{sec:flow-review}

Flow models are variational ansatz\"e for probability distributions, constructed using normalizing flows~\cite{rezende2016variational,dinh2016density,papamakarios2019normalizing}.
Each model is a generative parametrization (i.e.~exact/direct sampler) for a ``model distribution'' $\tilde{p}$ and comprises a simple, tractably sampleable prior distribution $r(z)$ and an invertible function or ``flow'', $f(z)$.
The flow maps from the latent-space variables $z$ of the prior distribution to the target-space variables $\phi$ of the model distribution.
Sampling proceeds simply by drawing independent samples from the prior distribution $z \sim r$ and applying the flow to obtain independent samples from the model, $f(z) = \phi \sim \tilde{p}$.
Conservation of probability gives the model density as
\begin{equation}
\tilde{p}(\phi)=\tilde{p}(f(z))=\left\lvert \det \frac{\partial f(z)}{\partial z} \right\rvert^{-1} r(z)
\label{eqn:model-prob-def}
\end{equation}
so $\tilde{p}(\phi)$ can be evaluated if $r(z)$ is known and the Jacobian determinant is practically computable. 
When applying $f$ to flow ``forwards''---from the prior space to the target space---this expression can be used to compute the model density for each sample as it is generated.
When applying $f^{-1}$ to flow ``backwards'', it allows evaluation of the model density for any given configuration, including those generated using other samplers.
Flows are often built from a sequence of simple ``coupling layers'', parametrized by neural networks, which each partition the variables and transform one subset conditioned on the other; this guarantees invertibility and yields a triangular Jacobian whose determinant can be computed efficiently \cite{dinh2016density}.

Given a flow model sampler for a density $\tilde{p}(\phi)$, we can obtain asymptotically correct results for observables under the target density $p(\phi)$ in several ways. 
One approach to statistical correction is to apply reweighting with weights
\begin{equation}
    w(\phi) \equiv \frac{p(\phi)}{\tilde{p}(\phi)}
\end{equation}
such that expectation values are computed as
\begin{equation}
    \vev{\mathcal{O}}_p = \vev{w \mathcal{O}}_{\tilde{p}} \approx \frac{1}{N} \sum_i w(\phi_i) ~ \mathcal{O}(\phi_i)
\end{equation}
using model samples $\phi_i \sim \tilde{p}$.
Alternately, one may use the flow to generate proposals for the Metropolis algorithm (flow-based MCMC). Because the samples drawn from the flow are independent, the conditional proposal density simplifies as ${T(\phi_i \rightarrow \phi') = \tilde{p}(\phi')}$ and the acceptance rate is
\begin{equation}\begin{aligned}
p_{\text{acc.}}(\phi_i \rightarrow \phi') &= \min\left(1, \frac{\tilde{p}(\phi_i)}{\tilde{p}(\phi')} \frac{p(\phi')}{p(\phi_i)} \right)
\\ &= \min\left(1, \frac{w(\phi')}{w(\phi_i)}\right).
\label{eqn:im-accept-rate}
\end{aligned}\end{equation}
This special case is known as ``independence Metropolis''~\cite{tierney1994markov}.
Unlike update-based methods where proposals are perturbations of previous samples, each accepted sample is a global update of the field, such that the only autocorrelations in the Markov chain arise from rejections.
Thus, if flow models can be constructed with sufficiently well-behaved distributions of reweighting factors, they can naturally circumvent problems in sampling like critical slowing down~\cite{Albergo:2019eim} and topological freezing~\cite{Kanwar:2020xzo}.

Many previous works using flow models~\cite{Albergo:2019eim,Kanwar:2020xzo,Boyda:2020hsi,Nicoli:2020njz,Gabrie:2021tlu,DelDebbio:2021qwf,Albergo:2021bna} have optimized models by minimizing a stochastic estimate of a modified reverse Kullback-Leibler (KL) divergence,
\begin{equation}
\begin{split}
L(\tilde{p}, p) & = D_{\mathrm{KL}}(\tilde{p}||p)-\log Z \\
 & = \int{\mathcal{D}\phi ~ \tilde{p}(\phi)(\log \tilde{p}(\phi) - \log p(\phi) - \log Z)} \\
 & = \int{\mathcal{D}\phi ~ \tilde{p}(\phi)(\log \tilde{p}(\phi) + S(\phi))} \\
 & \approx \frac{1}{N}\sum_{i=1}^N {\log \tilde{p}(\phi_i) + S(\phi_i)}, \quad (\phi_i \sim \tilde{p}) .
\end{split}
\label{eqn:reverse-KL-loss}
\end{equation}
In each step of ``reverse KL self-training'' the loss $L$ is estimated by drawing a batch of configurations $\phi_i$ with accompanying densities $\tilde{p}(\phi_i)$ from the model, computing the action of each resulting configuration to obtain $\exp[-S(\phi_i)] \propto p(\phi_i)$, then varying the model parameters to minimize the reverse KL divergence using standard stochastic gradient descent methods.
This self-training procedure does not require pre-generation of training data with another sampling method, and because a new batch is drawn for each training step, the training dataset is effectively infinite in size.
However, as discussed in the next section, self-training procedures---and especially those using the reverse KL loss---have inherent difficulties in the context of distributions with nontrivial mode structure.

\emph{Sampling challenges for flow-based MCMC:}
\label{sec:flow-based-mcmc-challenges}
If the prior distribution of a flow model has support over all field space and its flow is invertible, then the model distribution also has support over all field space. 
Flow-based MCMC therefore provides an ergodic Markov chain step, 
guaranteeing asymptotically unbiased expectations under the target.
However, while nonzero, the support may be arbitrarily small in some region so that the model yields configurations from that part of field space only rarely.
This admits a pathology where, when these regions are also severely underweighted in the model relative to the target, estimates computed using the flow-based sampler will converge extremely slowly: the sampler must eventually propose some configuration $\phi$ in this region for which $p(\phi) / \tilde{p}(\phi) \gg 1$, which is always accepted because this factor saturates the acceptance probability Eq.~\eqref{eqn:mcmc-accept-rate}; subsequent proposals to move from $\phi \rightarrow \phi'$ are almost certainly rejected because this factor appears in inverse in the acceptance probability, causing a long chain of rejections.
Asymptotically, this yields long autocorrelation times and poor statistical efficiency. 
At finite sample sizes, as with update-based approaches failing to tunnel between modes, such rejection streaks can individually change observable and error estimates significantly; if such an event can occur, the chain has not entered the smoothly converging regime and any estimate from it is untrustworthy.
Although this problem can occur generically, it can be especially pronounced with flow models that severely underweight entire modes of multimodal distributions or entire regions of extended modes.
When using such ``mode-collapsed'' models with flow-based MCMC, samples from missed regions will be proposed only rarely, with each proposal triggering a long rejection streak. 

This is not an abstract worry: training with reverse KL is known to experience ``mode seeking" behavior~\cite{minka2005divergence}.
The reverse KL loss consists of an entropy term $\tilde{p} \log{\tilde{p}}$ and cross-entropy $\tilde{p} \log{p}$. The dynamics of reverse KL training cause initial overconcentration on peaks from the cross-entropy term, until later in training when the entropy term begins to dominate.
When training a model onto multimodal distributions, this often manifests as a tendency for the sampler to initially collapse onto a subset of modes; the later dominance of the entropy term corresponds to the increasing relative cost of mismodeling the inner tails between modes, which can eventually lead to the discovery of the missed modes.
The multimodal distribution is the true minimum of the loss, but training the system into this minimum can be difficult (requiring intractably long training times) or impossible (if the training dynamics are incapable of finding the global minimum from a local one).
In the case of theories with extended modes, this effect manifests as collapse onto a subspace of the extended mode.
Like multimodal distributions, the extended mode is the true minimum, but may be very difficult to find without inductive-bias--based assumptions incorporated into the model.

Collapse is potentially an issue for any self-training procedure: if the model never samples from some region, it is not penalized for badly mismodeling that region.
Further, because $\log Z$ and thus the exact minimum of Eq.~\eqref{eqn:reverse-KL-loss} is unknown, we cannot detect whether a model has missed modes using model samples alone.

This tendency to overconcentrate can be counteracted by training using data with the mode structure incorporated by hand (Sec.~\ref{sec:forwards-kl}), by varying the target from unimodal to multimodal over the course of training (Sec.~\ref{sec:adiabatic-retraining}), or by using additional regularization terms in the loss (Sec.~\ref{sec:flow-distance-reg}), as we explore below. To complement these training augmentations, we also explore architectures which explicitly incorporate the mode structure, as described in Sec.~\ref{sec:arch_methods}.

\subsection{Combined methods: composite \& augmented MCMC}
\label{sec:augmented-MCMC}

Any Markov process that satisfies balance with respect to a target distribution is an exact sampler for that distribution, and provides asymptotically correct results when used to produce a Markov chain.
The same holds for a Markov chain generated using multiple different types of update intermixed.
This presents an opportunity: composing Markov chain updates with complementary properties avoids pathologies of single update types.
This general principle is exploited ubiquitously to produce more efficient MCMC samplers, most frequently when one update type (an ``augmentation'') is mixed in specifically to improve the performance of the primary update type.
Throughout this work we refer to this general class of algorithms as ``augmented MCMC''.
In practice, these schemes typically take advantage of another benefit afforded by composition, that not all types of updates in the composition need to be ergodic individually; for example, this is the case for methods like overrelaxation~\cite{Adler:1987ce,Creutz:1987xi}, ``overrelaxed HMC'' as in Ref.~\cite{Nicoli:2020njz}, post-hoc symmetrization of target samples, and even HMC itself, all of which may be thought of as different special cases of augmented MCMC.

For example, considering HMC as a sampler for the joint distribution of fields and their conjugate momenta, we can think of the momentum refresh between HMC trajectories as an augmentation step which improves the properties of the update defined by the integration and accept/reject step.\footnote{The accept/reject steps guarantee exactness if ergodicity is satisfied, but without momentum refreshing this sampler will follow the same trajectory as unrefreshed molecular dynamics, which is not always ergodic.}
Without momentum refreshing, HMC is deterministic except for the accept/reject step, so identical configurations are proposed repeatedly until accepted; this can trigger extremely long sequences of rejections if the acceptance probability is small.
Mixing in updates of the momenta results in vastly improved performance, truncating these long rejection runs and moreover often allowing HMC to move quickly out of regions of phase space where the symplectic integrator used suffers from significant integration error.

Augmented MCMC is particularly useful in the context of multimodal distributions, where update-based MCMC algorithms face difficulties. For example, for some theories we have sufficient knowledge of the mode structure of the theory to implement ``mode-hopping'' transformations that move the field between modes. Such knowledge could arise from a mean-field description or a spontaneously broken global symmetry.
In these cases, we can augment the usual updates with occasional applications of a randomly chosen mode-hopping transformation followed by an accept-reject step.

To define a general augmented MCMC approach, suppose we have some set of volume-preserving transformations $\{t_a\}$ and a corresponding set of probabilities of applying each transformation $\rho(t_a; \phi)$ with $\sum_a \rho(t_a;\phi) = 1$.
Every $n$th configuration, we propose an update to the current configuration $\phi$ by randomly selecting a transformation $t$ from the set $\{t_a\}$ with probability $\rho(t; \phi)$, then accepting the proposed $\phi' = t \circ \phi$ with probability
\begin{equation}\begin{aligned}
p_{\text{acc.}} &= \min \left( 1, 
    \frac{T(\phi' \rightarrow \phi)}{T(\phi \rightarrow \phi')} \frac{p(\phi')}{p(\phi)}
\right)
\\&= \min \left( 1, \frac{\rho(t^{-1}; \phi')}{\rho(t; \phi)} \frac{p(\phi')}{p(\phi)}
\right).
\end{aligned}\end{equation}
The set $\{t_a\}$ need not include the identity and may represent a continuous set of transformations. However, for the acceptance probability to be well-defined, the set must include the inverse of every transformation present and $\rho(t^{-1}; t\circ\phi)$ must be nonzero whenever $\rho(t; \phi)$ is. 

The augmentation approach can be applied straightforwardly to systems with spontaneous symmetry breaking by selecting $\{t_a\}$ to be the relevant group of symmetry transformations, in which case $p(\phi) = p(t \circ \phi)$ and $p_{\text{acc.}}$ simplifies. If, in addition, $\rho(t_a;\phi)=\rho(t_a^{-1};\phi)$, e.g.~when all transformations are applied with equal probability, then the proposed update is always accepted. 
Overrelaxation is a specific case of this kind of update~\cite{Gattringer}, and this limiting-case algorithm used with mode-hopping transformations was referred to as overrelaxed HMC in Ref.~\cite{Nicoli:2020njz}.
In some cases\footnote{The transition function used to construct the original Markov chain must be equivariant (as satisfied by e.g.~HMC) or invariant (e.g.~flow proposals) with respect to the symmetries applied.} this allows augmentation (symmetrization) to be applied as a postprocessing step, in which cases the algorithm is functionally equivalent to simple post-hoc symmetrization.
If the symmetry is also broken explicitly, these simplifications do not apply, but the more general algorithm is still valid.

This algorithm can also be useful in cases where we lack a symmetry description of the mode structure, but have some other means of deriving a set of transformations relating the modes. For example, we may expect a mean-field description, where the distribution concentrates about local minima of the action, to provide an approximation of the dynamics of the theory. In this case, the relevant transformations would be (schematically) a set of offsets $\Delta$ relating the mean-field minima, $\phi \rightarrow \phi + \Delta$. 
The field dependence of the probabilities $\rho(\Delta; \phi)$ can be set to ensure that only transformations that move the system between modes are selected.
Ref.~\cite{sminchisescu2003mode} proposes a version of this approach.

While most of the discussion of this section has focused on the case where one update type is supplemented by another, non-ergodic one, there are also advantages to composing updaters which are individually ergodic.
In Sec.~\ref{sec:sampling} we discuss and investigate an algorithm defined by the composition of HMC and flow-based MCMC. 
In different limits, this may be considered as HMC augmented by flow-based MCMC to speed mixing between modes, or as flow-based MCMC augmented by HMC to regulate slow convergence issues.
Ref.~\cite{Gabrie:2021tlu} presented a related approach, ``adaptive MCMC'' (involving Metropolis-adjusted Langevin dynamics instead of HMC), which also samples and trains simultaneously.

As discussed in Sec.~\ref{sec:multimodal-comparison}, results computed using HMC augmented with mode-hopping transformations serve as a benchmark for the other methods presented in the rest of this paper. Augmented HMC is also used to generate training data for the HMC training procedure described in Sec.~\ref{sec:hmc-training}.

\section{Architectural approaches for multimodal and extended-mode distributions}
\label{sec:arch_methods}

This section presents different approaches to architecture design which can aid construction of models for theories with nontrivial mode structure.
We first discuss how mixtures can be built out of multiple mode-collapsed flow models with sample-independent mixture weights to obtain coverage over all modes, which requires only that a flow model can be trained to sample each mode.
As discussed in the previous section, sometimes sufficient knowledge of the theory is available to implement mode-hopping transformations.
Explicitly incorporating this knowledge, we present two ``single-model mixture'' extensions of this standard construction where the component models only differ by composition with different mode-hopping transformations, including one which uses sample-dependent mixture weights to set the relative mode weights adaptively.
We also describe how this prior knowledge of the mode structure can be incorporated into other aspects of the flow architecture, including ``equivariant flows'' that explicitly respect the symmetries of the theory and ``topology-matched priors'' that share the mode structure of the target.

\subsection{Equivariant flows}

In many physical cases of interest, a spontaneously broken global symmetry gives rise to the mode structure.
For these, models invariant under the same symmetry must identically model all modes, and thus are immune to mode collapse during training.
Previous work has successfully applied this idea to model the bimodal phase of real scalar field theory by enforcing $Z_2$ invariance~\cite{Nicoli:2020njz,DelDebbio:2021qwf,Nicoli:2023qsl}.
In these studies, equivariant flows were constructed by restricting the structure of the neural networks parametrizing the flows.
However, it is possible to implement equivariance for arbitrary global symmetries without such restrictions. This section presents two methods for doing so.

\subsubsection{Symmetrization}

Symmetrizing a map to enforce invariance or equivariance is a standard technique in machine learning.
In an LQFT context, this has previously been investigated to enforce permutation equivariance in SU(N) eigenvalue flows~\cite{Boyda:2020hsi}, as well as to implement boundary conditions for pseudofermion distributions~\cite{Albergo:2021bna}. 
Here, we consider its application to encoding global symmetries associated with mode structure.
Consider a general transformation 
$\phi' = h(\phi)$
and a finite group of global symmetries $G$ that acts multiplicatively like $\phi \rightarrow g \phi$.
We can symmetrize this transformation as
\begin{equation}
    h_\mathrm{sym}(\phi) = \frac{1}{|G|}\sum_{g\in G} g^{-1} h(g \phi),
    \label{eq:sym-transf}
\end{equation}
for $g \in G$ and $|G|$ the order of $G$. For discrete symmetries, Eq.~\eqref{eq:sym-transf} exactly encodes the symmetry (e.g., for real $\phi^4$ theories). However, applying this method to a continuous symmetry requires approximation with a discrete subgroup, as investigated in Sec.~\ref{sec:extended}.

\subsubsection{Canonicalization}
\label{sec:canonical}

``Canonicalization'' refers to the general class of methods that treat a symmetry by collapsing its orbit to a canonical point singled out by some condition. One familiar example is gauge fixing, as applied in both perturbative calculations and numerically in lattice QCD. For a multimodal distribution arising from spontaneous breaking of a global symmetry, canonicalization projects the distribution to a single mode. For an extended mode corresponding to a broken continuous symmetry, canonicalization projects the model to a point along its orbit. While most often thought of as breaking the symmetry, canonicalization can equivalently be considered as constructing and operating on a new set of degrees of freedom which transform trivially under the symmetry of interest. As with symmetrization, this was previously investigated to enforce permutation equivariance in SU(N) eigenvalue flows~\cite{Boyda:2020hsi}.

We can build a canonicalized transformation by first constructing a pair of objects, $c(\phi)$ and $\overline{c}(\phi)$ such that $c(g \phi) = c(\phi) g^{-1}$ and $\overline{c}(g \phi) = g \overline{c}(\phi)$. Applying these as described above, we obtain
\begin{equation}
	h_\text{canon}(\phi) = \overline{c}(\phi) h(c(\phi) \phi)
\end{equation}
which is equivariant because the argument of $h$ is invariant while $\overline{c}$ is equivariant. 
The choice of $c$ and $\overline{c}$ is not unique. Although unnecessary, one can choose $c(\phi) \overline{c}(\phi) = 1$. Applied to a variable-partitioned transformation, if $c$ and $\overline{c}$ are functions only of the frozen degrees of freedom, then the canonicalized transformation is also invertible with a triangular Jacobian.

\subsection{Topology matching}
\label{sec:topo-matching}

Flow models are most often constructed with diffeomorphic flows. Such flows are unable to change the topology of a distribution, as defined by the structure of its regions of zero density.\footnote{The notion of the topology of a distribution is unrelated to gauge field topology.} In principle, this does not pose an obstacle for typical LQFT distributions, which formally have support everywhere and thus trivial topology in the exact sense. In practice, however, LQFT distributions with nontrivial mode structure---i.e., with regions of finite density separated by regions with nearly zero density---resemble topologically nontrivial distributions sufficiently to cause difficulties in modeling. This difficulty can be quantified in terms of the Lipschitz constants of different flow architectures \cite{Cornish2019RelaxingBC}. However, problems for modeling arise only due to \emph{mismatches} in topology between the prior and target distributions. This suggests a straightforward remedy: use instead a prior distribution with the same (approximate) topology as the target.

Exploitation of this abstract idea in modeling requires a priori knowledge of the target mode structure. This is not available in the general case. However, when the mode structure arises from spontaneous breaking of a known global symmetry, the mode topology can be determined, for example using a mean-field analysis. Application of this idea then requires only devising an easy-to-sample distribution with the same topology for use as the prior. We present example applications of this idea in Secs.~\ref{sec:real-topo-matching} and \ref{sec:cx-topo-matching} to both the discrete- and continuous-symmetry cases.

\subsection{Mixture models}
\label{sec:mixture-models}

One approach to modeling multimodal/extended-mode targets is to build mixture models using other (potentially mode-collapsed) flow models as components.
Generically, (additive) mixture models are defined by a sampling procedure which yields a sample drawn from one randomly selected component model, with different probabilities of selecting each component which may be sample-dependent.
The density of the mixture is a weighted average over the densities of the component models, where the mixture weights of each component are implicitly defined by the sampling procedure.

\subsubsection{Mixtures of multiple models}

Suppose that we have a set of samplers for several different distributions $\{\tilde{p}_a\}_{a \in \{ 1 \ldots M \}}$; we may use these to construct and sample from a mixture model.
To sample from the mixture, we choose one of the models from the set at random according to a corresponding set of probabilities $\rho_a$ normalized as $\sum_a \rho_a = 1$ and use it to generate a sample $\phi$.
The density of $\phi$ under the mixture is then
\begin{equation}
\tilde{p}_{\text{mix}}(\phi) = \sum_{a=1}^M \rho_a \tilde{p}_a(\phi) \, .
\label{eqn:mix-model-logq}
\end{equation}
Given a sample $\phi$, computing the density $\tilde{p}_{\text{mix}}(\phi)$ requires evaluating the density of $\phi$ under all $M$ components.
For flow models, this is of comparable expense per component to generating a sample, resulting in an $M$-fold increase in cost relative to sampling from just one component.
Samples drawn from a mixture of models may be reweighted or used for independence Metropolis proposals just as with samples drawn from a single flow model.
The mixture weights $\rho_a$ are free parameters that can be chosen to optimize statistical efficiency.

For effective sampling, the models in the mixture must together provide coverage of all modes of the target distribution.
This approach can be more economical than training a single flow model of all modes of the target.
In cases where the modes of the target distribution are widely separated, with little density between them, a mixture of unimodal models can provide a good approximation of the highest-density regions but may underestimate the density between modes, thus compromising the quality of the model.
In theories with many modes, this problem can be alleviated by building the mixture out of models with overlap on subsets of the modes, with appropriate support between those captured.
However, obtaining a full set of models may require specialized training procedures that target low-weight modes which are unlikely to be found by unguided training.

\subsubsection{Single-model symmetrized mixtures}
\label{sec:symm-mixtures}

Ref.~\cite{Boyda:2020hsi} introduced a procedure to explicitly restore a discrete symmetry that is not respected by a given flow model by constructing a mixture.
In this construction, each mixture component is the same flow model $\tilde{p}$, composed with an application of a symmetry transformation $t$ which acts as $\phi \rightarrow \phi' = t \circ \phi$. If $\{t_a\}$ is the group of (volume-preserving) transformations associated with the symmetry (including the identity), the density of some sample $\phi$ under the mixture is
\begin{equation}
\tilde{p}_{\text{mix}}(\phi) = \frac{1}{|\{t_a\}|} \sum_a \tilde{p}(t_a^{-1} \circ \phi)
\label{eqn:symm-mix-model}
\end{equation}
where the components are combined with equal weights $\rho_a = 1/|\{t_a\}|$ to explicitly encode the symmetry.
By leaving these weights free to be tuned instead, this procedure can naturally be extended to distributions described by weakly broken symmetries, although this approach will eventually break down when the symmetry is strongly broken and the shapes of the modes become increasingly dissimilar.

This construction may be applied as a postprocessing step to flow model samples, requiring only that when we generate each sample $\phi$ we also compute and record $\tilde{p}(t \circ \phi)$ and $p(t \circ \phi)$ for all $t$ as needed to compute $\tilde{p}_{\text{mix}}(\phi)$, as well as all $\mathcal{O}(t \circ \phi)$ for any observable of interest that does not transform trivially.
Note that this postprocessing scheme is inequivalent to the ex post facto symmetrization discussed in Sec.~\ref{sec:augmented-MCMC}.

\subsubsection{Single-model adaptive mixtures}
\label{sec:adaptive-mixtures}

Applying the construction of the previous section to cases where the mode weights are not guaranteed to be equal by symmetry requires tuning the mixture weights to match the target.
We can instead define parameter-free constructions which adaptively set the mixture weights based on the target density.

In general, suppose we have a flow model and some finite set of invertible transformations $\{t_a\}$ which act on the fields as $\phi'_a = t_a \circ \phi$.
The adaptive mixture sampler is defined by drawing a sample $\varphi$ from the flow model, computing $\varphi'_a = t_a \circ \varphi$ and the corresponding $p(\varphi'_a)$ for each $t_a$, then selecting the final sample $\phi$ to be one of the $\varphi'_a$ with probability $p(\varphi'_a) / \sum_b p(\varphi'_b)$, i.e.~preferentially choosing transitions to higher-density regions of configuration space.
This procedure defines a mixture with density
\begin{equation}
    \tilde{p}_{\text{mix}}(\phi) = p(\phi) \sum_b 
    \frac{ \tilde{p}(t_b^{-1} \circ \phi) }{ \sum_a p(t_a \circ t_b^{-1} \circ \phi) }
    \left|\det \frac{ \partial (t_b^{-1} \circ \phi) }{ \partial \phi }\right|.
\label{eqn:Q-general-adaptive-mix-model}
\end{equation}
As in the previous section, this construction can be applied as a postprocessing step.
This sample-dependent choice yields a tractable mixture density only because the output sample is selected from a deterministically related set of options, from which it follows that for each $\phi$ we can reconstruct all possible untransformed samples that might have led to $\phi$ as $(\varphi)_b = t_b^{-1} \circ \phi$ and all options in the ensuing selections as $(\varphi'_a)_b = t_a \circ t_b^{-1} \circ \phi$.
In contrast, making a sample-dependent choice among samples drawn independently from each component defines a mixture whose density involves an intractable marginalization over all possible draws and choices.

Specializing to the case where $\{t_a\}$ are a finite group of volume-preserving\footnote{This trivializes the Jacobian determinant factor in Eq.~\eqref{eqn:Q-general-adaptive-mix-model}.} transformations, Eq.~\eqref{eqn:Q-general-adaptive-mix-model} simplifies to
\begin{equation}
    \tilde{p}_{\text{mix}}(\phi) = \frac{p(\phi)}{\sum_c p(t_c \circ \phi)} \sum_b \tilde{p}(t_b^{-1} \circ \phi)
\label{eqn:adaptive-mix-model}
\end{equation}
after using group closure to relabel ${t_a \circ t_b^{-1} = t_c}$, allowing the sum in the denominator to be factorized out.
The resulting mixture model explicitly encodes the correct relative weights between states in an orbit of $\{t_a\}$ of the target distribution, i.e.
\begin{equation}
    \frac{\tilde{p}_{\text{mix}}(t \circ \phi)}{\tilde{p}_{\text{mix}}(\phi)} 
    = \frac{p(t \circ \phi)}{p(\phi)}
\end{equation}
for all $t \in \{t_a\}$.
In the limiting case where $\{t_a\}$ are a symmetry group of the action, $p(t_c \circ \phi)$ is equal for all $t_c$ and we recover the symmetrized mixture model of Eq.~\eqref{eqn:symm-mix-model}.
Far from this limit, using this construction with a multimodal target where the modes are too dissimilarly shaped can distort the shapes of the component distributions in the mixture in a way that reduces overlap with the target.
However, for a weakly broken symmetry where the modes of the target are similarly shaped, adaptive mixtures can be applied successfully; we explore this usage in Sec.~\ref{sec:mixture-results}.

Generally, $\{t_a\}$ can be any set of invertible transformations including additional flows, allowing for adaptively weighted selection of flow models in a mixture if each flow is applied to the same draw from a shared prior distribution. This connection can be made more clear by rewriting the Jacobian factor in Eq.~\eqref{eqn:Q-general-adaptive-mix-model} as
\begin{equation}
\left|\det \frac{ \partial (t_b^{-1} \circ \phi) }{ \partial \phi }\right|
= \left|\det \frac{ \partial \phi }{ \partial (t_b^{-1} \circ \phi) }\right|^{-1}
= \left|\det \frac{ \partial (t_b \circ z_b) }{ \partial z_b }\right|^{-1}
\end{equation}
defining $z_b = t_b^{-1} \circ \phi$, the draw from the shared prior that yields $\phi$ after applying $t_b$.
This construction resembles the many-to-one flows of Ref.~\cite{papamakarios2019normalizing}, except each flow is applied to the full latent space rather than non-overlapping subsets.
Note also that although we describe $p$ as the target density, the derivation does not depend on this identification and more generally $p$ can be any non-negative function.

\section{Training approaches for multimodal and extended-mode distributions}
\label{sec:multimodal-approaches}

As discussed in Sec.~\ref{sec:flow-review}, flow-based MCMC produces asymptotically unbiased results.
This guarantee is a consequence of the invertibility of the flows, which holds by construction regardless of the values of the model parameters.
Thus, there is no incorrect way to train a flow model and any optimization procedure is acceptable.
This section describes a set of approaches which exploit this freedom to train flow models which capture the full mode structure of a target distribution.

\subsection{Forwards KL training}
\label{sec:forwards-kl}

This section describes a set of methods which use the forwards KL divergence for the loss, rather than the reverse KL described in Sec.~\ref{sec:flow-review}.
The forwards KL divergence is defined as an expectation value under the target distribution, rather than the model distribution as in the case of the reverse KL. This makes it more natural to optimize the forwards KL divergence using target-distributed samples generated using some other sampling method, whereas reverse KL is most naturally suited for self-training with model-distributed samples.
However, as we introduce below, the forwards KL divergence can also be employed in a self-training scheme using reweighted model samples.
Unlike reverse KL self-training, forwards KL self-training permits data augmentation, allowing the mode structure to be incorporated into the training data by hand.
This can be done in a statistically exact way using the mixture model constructions discussed in Sec.~\ref{sec:mixture-models}.

\subsubsection{Training with target data}
\label{sec:hmc-training}

Using a target-distributed dataset generated using another sampling scheme, we can train the model by minimizing a stochastic estimate of the forwards KL divergence:
\begin{equation}\begin{aligned}
D_{\mathrm{KL}}(p||\tilde{p}) &= \int d\phi ~ p(\phi) \log \frac{p(\phi)}{\tilde{p}(\phi)} \\
&\approx \frac{1}{N} \sum_{i=1}^N \log \frac{p(\phi_i)}{\tilde{p}(\phi_i)}, \quad (\phi_i \sim p)
\label{eqn:fwds-kl-psamp}
\end{aligned}\end{equation}
which is equivalent to maximizing the model likelihood.
A perfect forwards KL training dataset consists of independent and identically distributed samples drawn from the target distribution, with no reuse of configurations.
In lattice field theory, large datasets of independent samples from the target distribution may not be available for training. However, for the simple theories considered later in this paper, it is numerically tractable to generate comprehensive datasets using HMC augmented with mode-hopping transformations as detailed in Sec.~\ref{sec:augmented-MCMC}.
Thus, we are able to use this training scheme as a benchmark for the others described below.

Although an ideal forwards KL training dataset is unattainable on large lattice volumes, it may be useful to train with forwards KL on smaller volumes and then polish with reverse KL on larger ones. Even on larger volumes, it may be useful to use a small dataset drawn from the target distribution for training, if this dataset has good coverage of the various modes of the distribution. 
Although we do not explore this possibility here, even with a small dataset it may be beneficial to combine steps of forwards KL training (recycling the training data) with steps of reverse KL self-training. During training, the forwards KL steps provide a loss penalty that encourages the model to keep support on all modes. Such a dataset can be generated in practice using schemes such as HMC with multiple Markov chains initialized from starting configurations near different modes.

\subsubsection{Forwards KL self-training}
\label{sec:forwards-kl-training}

An alternative to forwards KL training with a dataset sampled from the target distribution $p$ is to use samples from some other distribution $p' \ne p$, reweighting by the factor $p(\phi)/p'(\phi)$ to estimate
\begin{equation}\begin{aligned}
    D_{\mathrm{KL}}^{\text{fwd,rw}}(p||\tilde{p})
    &= \int d\phi ~ p'(\phi) \frac{p(\phi)}{p'(\phi)} \log \frac{p(\phi)}{\tilde{p}(\phi)}
    \\& \approx \frac{1}{N} \sum_i \frac{p(\phi_i)}{p'(\phi_i)} \log \frac{p(\phi_i)}{\tilde{p}(\phi_i)}, \quad (\phi_i \sim p').
    \label{eq:fwd_dkl}
\end{aligned}\end{equation}
Because there is no $p'$ dependence on the LHS, this estimator asymptotically computes the same divergence as Eq.~\eqref{eqn:fwds-kl-psamp}.
Note that the reweighting factors $p(\phi_i)/p'(\phi_i)$ are not functions of the model parameters.
We can use the reweighted forwards KL divergence to define a self-training scheme by drawing the training dataset from the model, i.e.~$\phi \sim p' = \tilde{p}$.
In this scheme, as far as training is concerned it is a coincidence that $p'(\phi) = \tilde{p}(\phi)$, and the reweighting factors are taken to be independent of the model parameters in gradient computations.

Training using reweighted model samples to compute the forwards KL divergence as described above still encounters the difficulties associated with self-training detailed in Sec.~\ref{sec:flow-based-mcmc-challenges}.
However, because $p'$ is arbitrary and the reweighting factors are not functions of the model parameters, we can modify this procedure to train with an augmented dataset with all modes represented.
We can accomplish this simply by taking $p'$ to be a single-model mixture of the model being trained, constructed to sample from all modes as described in Sec.~\ref{sec:mixture-models}.
Using the mixture model construction to generate the training dataset requires additional applications of the flow to compute $\tilde{p}_{\text{mix}}$.
While forwards-pass evaluation is typically a relatively small cost during training, it can become significant if the number of mixture components grows large.
For continuous symmetries, exact mixtures are intractable and must be approximated.

In practice, one can instead try a naive data augmentation scheme, i.e.\ simply applying a random transformation to each configuration, while using the naive model density in the reweighting factors rather than the true mixture density.
This can be thought of as an approximation of sampling from a mixture where the mixture density is (falsely) reported as $\tilde{p}_{\text{mix}}(\phi) \approx \tilde{p}(\phi)$ (which is cheaper to compute).
This approximation holds if either: (1) $\tilde{p}(1 \circ \phi) \gg \tilde{p}(t_a \circ \phi)$ for all $t_a$ that aren't the identity, i.e.~the model is very asymmetric with respect to the mode-hopping transformations and unlikely to draw the transformed samples; or, (2) $\tilde{p}(t_a \circ \phi) \approx \tilde{p}(t_b \circ \phi)$ for all $a \ne b$, i.e.~the model is approximately symmetric.
Early in training (1) is likely to hold for any model, and if the target distribution is approximately symmetric, (2) is likely to hold later in training as the model distribution begins to resemble the target.
We compare the performance of this naive augmentation scheme versus training with mixture samples in Sec.~\ref{sec:forwards-kl-results} below.
This scheme also allows treatment of continuous symmetries, as explored in Sec.~\ref{sec:extended-training-results}.

\subsection{Adiabatic retraining}
\label{sec:adiabatic-retraining}

Distributions over field configurations in lattice QFT generically have tunable parameters that can be used to adjust their complexity, such as the lattice volume (i.e.~the number of degrees of freedom in the discretized field) and the lattice action parameters.
A standard technique exploits
this dependence to more easily and efficiently train models using transfer learning or ``retraining'', wherein a model trained targeting some easier-to-approximate set of parameters is used to initialize the training for harder ones, rather than training a model for the more difficult target from a random initialization.
The obvious generalization is to vary the parameters epoch-by-epoch over some trajectory through parameter space over the course of training.

In an adiabatic retraining scheme, we attempt to vary the target action parameters sufficiently slowly that the model being trained remains a good approximation of the target distribution throughout training, even as the target distribution changes.
This is possible as long as the target distribution changes smoothly with the parameters, the model is sufficiently expressive for all parameters along the trajectory, and the model parameters can change smoothly to track the changing target (or at least that any abrupt reconfiguration is not too violent).
Applied to distributions with nontrivial mode structure, this method requires choosing a trajectory through action parameter space which changes the model sufficiently close to adiabatically to train from a simple unimodal phase into a multimodal/extended-mode one.

\subsection{Flow-distance regularization}
\label{sec:flow-distance-reg}

In this section we discuss ``flow-distance regularization'', a procedure that can train models to multimodal distributions with minimal information about the mode structure.
This is in contrast to forwards KL self-training and adiabatic retraining, which both rely on a priori knowledge of the mode structure: forwards KL self-training requires a known set of transformations that move the system between modes, and adiabatic retraining requires tunable action parameters that smoothly change the mode structure.
These methods are inapplicable absent this information.

The flow-distance regularizer is a term that can be added to the loss to penalize flow functions $f$ that transform prior samples $z$ to significantly different output samples $\phi$. This penalty favors a flow function closer to the identity and thus a model probability density $\tilde{p}$ that is close to the prior density $r$.
The explicit form we consider in this work takes the form $\lVert \phi-z \rVert_2$ so that a configuration $\phi$ contributes to the loss as
\begin{equation}
L(\phi, p) = \log \tilde{p}(\phi) + S(\phi) + \kappa g(t) \lVert \phi-z \rVert_2
\end{equation}
where the L2 norm implies a sum over all degrees of freedom on the lattice, $\kappa$ is the coupling strength of this ``locality constraint", and the schedule $g(t)$ is a function of the time in training $t$.
The schedule may be used to slowly remove the regulator over the course of training.
If the prior distribution is sufficiently broad, i.e.~has nontrivial support where the modes of the target do, this counters mode collapse by additionally penalizing training for moving density off a mode.
Thus if the training schedule is sufficiently slow, the model will smoothly change from a distribution similar to the prior to an approximation of the target.

Similar expressions have been defined in other studies of normalizing flow models, e.g.~regularized neural ODEs~\cite{finlay2020train}, based on the idea of optimal transport as defined using the Kantorovich/Wasserstein distance~\cite{benamou2000computational}.
Note also that flow-distance regularization resembles adiabatic retraining in a way that can be made precise: the regulator may be thought of as amounting to additional terms in the action, induced by the (very complicated) function of the fields defined by the identification $z = f^{-1}(\phi)$.
These additional terms involve the parameters of the model, and the regulator follows the schedule $g(t)$, so the regulated action changes from epoch to epoch over training.
In a loose sense, this may be thought of as an all-purpose mode-regulating operator, applicable to any theory without a priori knowledge of the mode structure and whose efficacy relies only on the broadness of the prior.

\section{Multimodal models for real scalar field theory}
\label{sec:multimodal-comparison}

This section demonstrates numerical applications of the methods described in Secs.~\ref{sec:arch_methods} and \ref{sec:multimodal-approaches} to model the bimodal phase of real scalar field theory in two dimensions.
We first discuss the lattice theory and how we apply augmented HMC to obtain ground-truth samples. 
Next, we present a baseline methodology for constructing and training models for this theory, and demonstrate that it is prone to mode collapse.
We then perturb this baseline to show how both the architecture- and training-based methods can be used to alleviate mode collapse.

\subsection{Real lattice scalar field theory}

We consider the lattice discretization of real scalar $\phi^4$ theory in $d=2$ (Euclidean spacetime) dimensions on a lattice consisting of sites $x_\mu = an_\mu$, where $n_{\mu} \in \mathbb{Z}^d$ and $a$ is the lattice spacing. Working in lattice units where $a=1$, one particular choice of discretization gives the Euclidean lattice action
\begin{equation}
\begin{split}
S_E(\phi) = & \sum_{x} \left( \sum_{\mu=1}^{d}  \frac{1}{2}(\phi(x+\hat{\mu})-\phi(x))^2 \right.\\
 & \left. + \frac{1}{2}  m^2\phi(x)^2+\lambda\phi(x)^4 + \alpha\phi(x)\right)
\end{split}
\label{eqn:phi4-action}
\end{equation}
where periodic boundary conditions are implied (i.e.~$\phi(x+L)=\phi(x)$), and $m^2$ can be negative.
When $\alpha = 0$, the action is invariant under a global $Z_2$ symmetry corresponding to the transformation $\phi \rightarrow -\phi$. The $\alpha=0$ theory has two phases: one symmetric, corresponding to a unimodal distribution, and one where the global $Z_2$ symmetry is spontaneously broken, corresponding to a bimodal distribution. The symmetry guarantees that the two modes have equal weights and are the same shape. The two phases are differentiated by an order parameter, the average magnetization
\begin{equation}
\overline{\phi} \equiv \frac{1}{L^2} \sum_x \phi(x) \, .
\end{equation}
When $\alpha \ne 0$, the global $Z_2$ is explicitly broken. For values of $\alpha$ where the explicit breaking is weak, both phases are still present, however in the bimodal phase the modes have uneven relative weights and are no longer identically shaped.

\begin{figure}
    \centering
    \subfloat[\centering  ]{{
        \includegraphics[width=\linewidth]{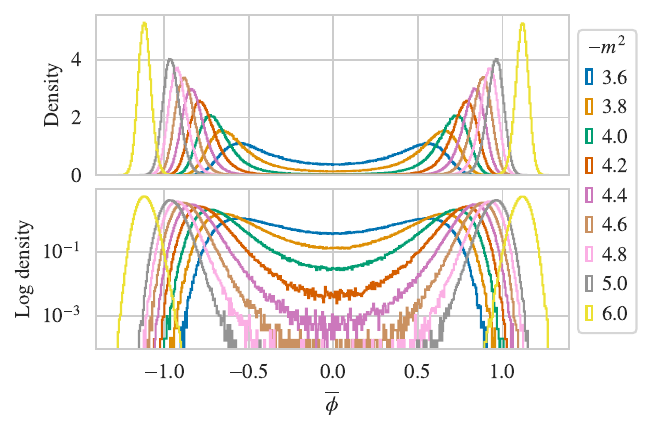}
        \label{fig:hmc-hists-M2}
    }}
    \!
    \subfloat[\centering  ]{{
        \includegraphics[width=\linewidth]{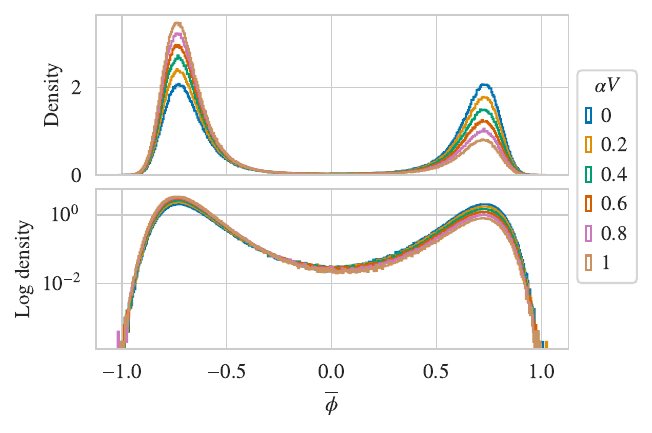}
        \label{fig:hmc-hists-alpha}
    }}
    \caption{
        Distribution of the average magnetization $\overline{\phi}$ in real scalar field theory for \protect\subref{fig:hmc-hists-M2} fixed $\alpha=0$ varying $m^2$ and \protect\subref{fig:hmc-hists-alpha} fixed $m^2=-4$ varying $\alpha$, at $\lambda=1$ on a $10 \times 10$ lattice geometry.
        Each is histogram estimated using $1.28 \times 10^6$ samples generated with AHMC.
    }
    \label{fig:hmc-hists}
\end{figure}

For this exploratory study, we fix the lattice geometry to $10 \times 10$ and the coupling $\lambda$ to 1, and vary the mode structure by adjusting $m^2$ and $\alpha$.
Figure~\ref{fig:hmc-hists} shows how the distribution of the order parameter $\overline{\phi}$ varies for the set of action parameters we consider in this work (estimated using augmented HMC, as discussed below).

\label{sec:aug-hmc-results}

The two modes of the broken phase of $\phi^4$ theory are related by the transformation $\phi \rightarrow -\phi$. Even when $\alpha \ne 0$ and the global $Z_2$ symmetry is explicitly broken, this transformation suffices to move the system between modes as long as the reflected modes overlap with each other.
In this case, the augmented MCMC algorithm discussed in Sec.~\ref{sec:augmented-MCMC} consists of standard HMC transitions mixed with proposed sign flips from the current state $\phi_i$ to $\phi'=-\phi_i$, which is accepted or rejected with probability $\min(1, p(-\phi_i) / p(\phi_i))$. 
When the global $Z_2$ symmetry is unbroken, this flip is always accepted; when explicitly broken, rejections will occur as necessary to give the correct relative mode weights and shapes.

We use AHMC to compute ground truth results for comparison with results derived from flow-based samplers.
We fix the HMC trajectory length to 1 and the number of leapfrog steps to 10; the acceptance rate is $90-100\%$ at all parameters we consider.
The saved configurations are separated by 10 trajectories and a single proposed sign flip, with the frequency chosen to avoid deterministically applying an even number of signs between each retained sample in the symmetric $\alpha=0$ case.
All chains are thermalized for 1000 trajectories from a hot start.
For convenience and efficiency on GPUs, we generate independent chains of 10000 (saved) configurations in parallel rather than running one long chain.
We find the integrated autocorrelation time in the action is $0.5 \lesssim \tau \lesssim 1$ (depending on the parameters) in units of saved configurations, indicating that residual autocorrelations in each chain are small.

\subsection{Baseline model \& training}
\label{sec:real-baseline}

Demonstrating the efficacy of different methods to treat mode collapse requires a collapse-prone baseline to improve upon.
For this baseline, we use real NVP flows~\cite{Albergo:2019eim,dinh2016density} constructed from affine coupling layers.
Each coupling layer partitions the variables by checkerboarding the sites, and updates one parity of site conditioned on the frozen values of the other parity; successive coupling layers update opposite-parity sites such that all variables are updated once after two layers.
As described in Refs.~\cite{Albergo:2019eim,dinh2016density}, this partitioned updating scheme is invertible by construction and gives a triangular Jacobian, allowing efficient computation of the Jacobian determinant.
In each coupling layer, we use a convolutional neural network to parametrize the affine transformation; compared with fully connected networks, this reduces model size and encodes translational symmetry (up to breaking by the checkerboarding).
The implementation follows that of Ref.~\cite{Albergo:2021vyo}.

In particular, for the baseline architecture, the prior distribution is an uncorrelated unit-width
Gaussian distribution for each component of the scalar field (i.e.~on each site).
Each model is constructed from 24 affine coupling layers. The networks parametrizing each affine transformation are made of 4 convolutions with kernel size 5, with leaky ReLU activations (with negative slope = 0.01) after each except the last one, which is followed by a $\tanh$. 
Each network takes 1 channel as input, gives 2 channels as output, and works with 12 channels at all intermediate stages.

The baseline training procedure is reverse KL self-training, as described in Sef.~\ref{sec:flow-review}, applying gradient-based updates with the Adam optimizer~\cite{kingma2014adam}.
Models are trained starting from randomly initialized weights (per PyTorch defaults).
Throughout, by one epoch we mean one gradient computation and optimizer step.
We train in 32-bit precision, but sample in 64-bit precision to avoid introducing bias due to round-off error compromising invertibility.

In some cases, we apply standard ML techniques to improve training performance.
Specifically, we sometimes use gradient norm clipping to stabilize training: after backpropagation, we measure the norm of the gradients before passing them to the optimizer and, if the norm exceeds some threshold value, scale all gradients down by a constant to clip the norm to the threshold.
We also employ a step-scheduler in some cases which drops the learning rate by a factor $\gamma=0.5$ every $n$ epochs (typically $n = 20$k).
We note below where these are applied.

\subsection{Assessing model quality}

In this study, qualitative success is determined simply by whether mode collapse is avoided.
This is straightforwardly assessed from histograms of the average magnetization $\overline{\phi}$.
During training, the average magnetization under the model $\vev{\overline{\phi}}_{\tilde{p}}$ serves as an easily monitored scalar diagnostic of mode collapse.
It indicates collapse when nonzero (or when different from its value under the target, $\vev{\overline{\phi}}_p$, in asymmetric cases).
When $\vev{\overline{\phi}}_{\tilde{p}}$ and the loss are shown in training histories in the following subsections, they are smoothed by averaging blocks of 25 epochs.

We also consider several more quantitative metrics of model quality.
One standard option is the
effective sample size (ESS) per configuration~\cite{Kish:1965,Martino_2017},
\begin{equation} \begin{aligned}
    \text{ESS}/N 
    & \equiv \frac{1}{ \int d\phi ~ \tilde{p}(\phi) w(\phi)^2 }
    \\&\approx \frac{1}{N} \frac{ \left( \sum_i w_i \right)^2 }{ \sum_i w_i^2 }, \quad (\phi_i \sim \tilde{p})
\label{eqn:ess-estimator}
\end{aligned} \end{equation}
where $w_i \equiv w(\phi_i) \equiv p(\phi_i)/\tilde{p}(\phi_i)$.
The second expression is a stochastic estimator evaluated over a sample of $N$ model-distributed configurations. 
This estimator is constructed such that the normalization of $w_i$ cancels in the ratio, allowing evaluation using $\exp[-S(\phi)]$ in place of $p(\phi)$.
The ESS per configuration is directly motivated by a particular use-case for flow models, i.e.~computing reweighted expectation values.
It quantifies the reduction in effective sample size (i.e.~increase in variance) when reweighting from the model to the target distribution.
Its value falls between 0 and 1, achieving the maximum value of 1 for a perfect model $\tilde{p} = p$ where all $w_i = 1$.
In the discussion below, we occasionally use this finite-sample metric as a rough measure of the quality of overlap of a model on to a target.
However, this measure should be interpreted cautiously given the discussion in Sec.~\ref{sec:sampling}, where we find the asymptotic ESS is near zero for many of these models.
We defer further discussion of how this and other metrics relate to 
the performance of different sampling schemes until Sec.~\ref{sec:sampling}.

Given these complications, it is also useful to consider a metric that is agnostic to how the model will be employed to generate target-distributed samples or otherwise compute expectation values under the target.
For this purpose, we also consider the forwards KL divergence.
Using a batch of target-distributed samples generated with augmented HMC, it can be computed as
\begin{equation}\begin{aligned}
    D_{\mathrm{KL}}(p||\tilde{p}) 
    &= \int d\phi ~ p(\phi) \log \frac{p(\phi)}{\tilde{p}(\phi)}
    \\ &\approx \frac{1}{N} \sum_i \log \frac{p(\phi_i)}{\tilde{p}(\phi_i)}, \quad (\phi_i \sim p)
\end{aligned}\end{equation}
using $\log p = \exp[-S]/Z$ normalized using the stochastic estimator~\cite{Nicoli:2020njz}
\begin{equation}
    Z^{-1} \approx \frac{1}{N} \sum_i \frac{\tilde{p}(\phi_i)}{e^{-S(\phi_i)}}, \quad (\phi_i \sim p)
    \label{eqn:stochastic-Z}
\end{equation}
so that $D_{\mathrm{KL}} = 0$ when $p = \tilde{p}$.
For each set of action parameters (i.e.~each target distribution), we evaluate $D_{\mathrm{KL}}$ using a single shared ensemble of AHMC validation data consisting of $N = 10^6$ samples, enabling direct comparisons with perfectly correlated sampling noise between different models of the same target.
We use a shared value of $Z$ to normalize $p$ for each set of parameters, computed using models trained with AHMC data and not accounting for the error in this estimate.

\subsection{Mode collapse}
\label{sec:rev-results}

We first examine the behavior of our baseline methodology, an affine coupling flow trained with reverse KL self-training.
The results provide a concrete demonstration of mode collapse, and establish a sense of the difficulty of training a model which correctly captures the bimodal distribution of interest.

\begin{figure}
    \centering
    \includegraphics[width=\linewidth]{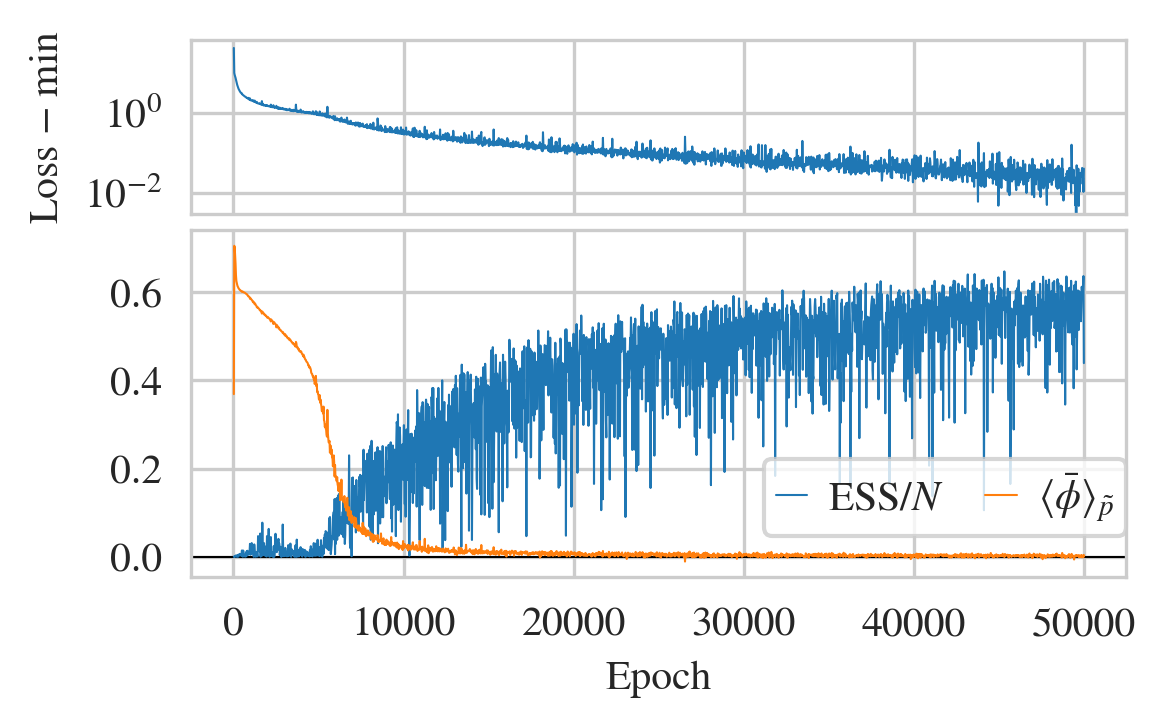}
    \caption{
        Reverse KL training history exhibiting smooth approach from unimodal to bimodal, for a flow model targeting real scalar field theory with $m^2=-3.6$ and $\alpha=0$, trained with batch size 16000.
        The ESS is computed every 25 epochs.
    }
    \label{fig:rev_train_smooth}
\end{figure}

Figure~\ref{fig:rev_train_smooth} shows a training history for a model targeting $m^2 = -3.6$ and $\alpha=0$. 
As seen in Figure~\ref{fig:hmc-hists-M2}, the target distribution for these parameters is bimodal but the modes still overlap substantially.
As displayed by the running estimate of $\vev{\overline{\phi}}_{\tilde{p}}$, the model is initially heavily biased towards one mode, but slowly and gradually learns to evenly weight the two modes.
In this case, collapse is a transient feature of the early dynamics and smoothly removed by further training.

\begin{figure}
    \centering
    \includegraphics[width=\linewidth]{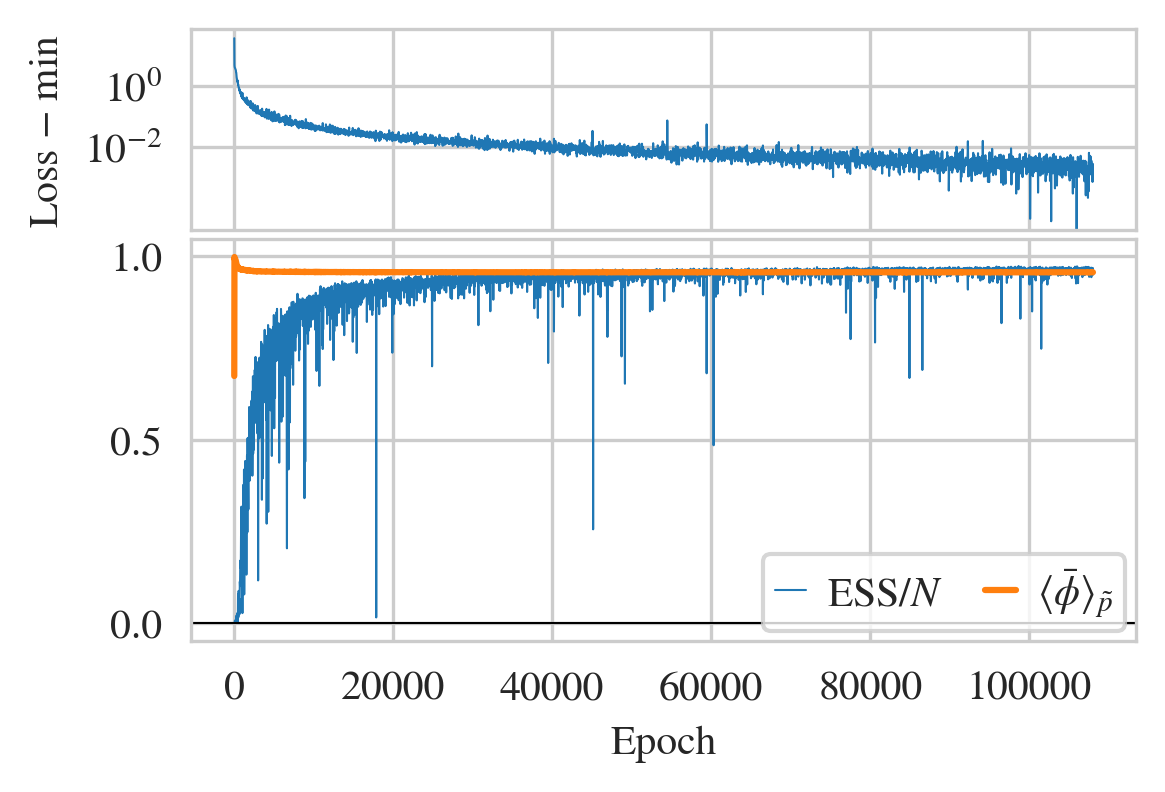}
    \caption{
        Reverse KL training history exhibiting mode collapse, for a flow model targeting real scalar field theory with $m^2=-5$ and $\alpha=0$, trained with batch size 16000.
        The ESS is computed every 25 epochs.
        The similar values of $\text{ESS}/N$ and $\vev{\overline{\phi}}_{\tilde{p}}$ are coincidental.
    }
    \label{fig:rev_train_collapse}
\end{figure}

In the opposing regime, Fig.~\ref{fig:rev_train_collapse} shows a training history for the case where $m^2 = -5$, where the two modes are well-separated.
The model quickly trains into a unimodal distribution, as diagnosed by the nonzero value of $\vev{\overline{\phi}}_{\tilde{p}}$.
The apparently high finite-sample ESS for the mode-collapsed model suggests that it is a good fit to the collapsed-upon mode.
Given sufficient training, the model may have eventually become bimodal, but it was not found to do so in the long (though finite) training time used here.

\begin{figure}
    \centering
    \includegraphics[width=\linewidth]{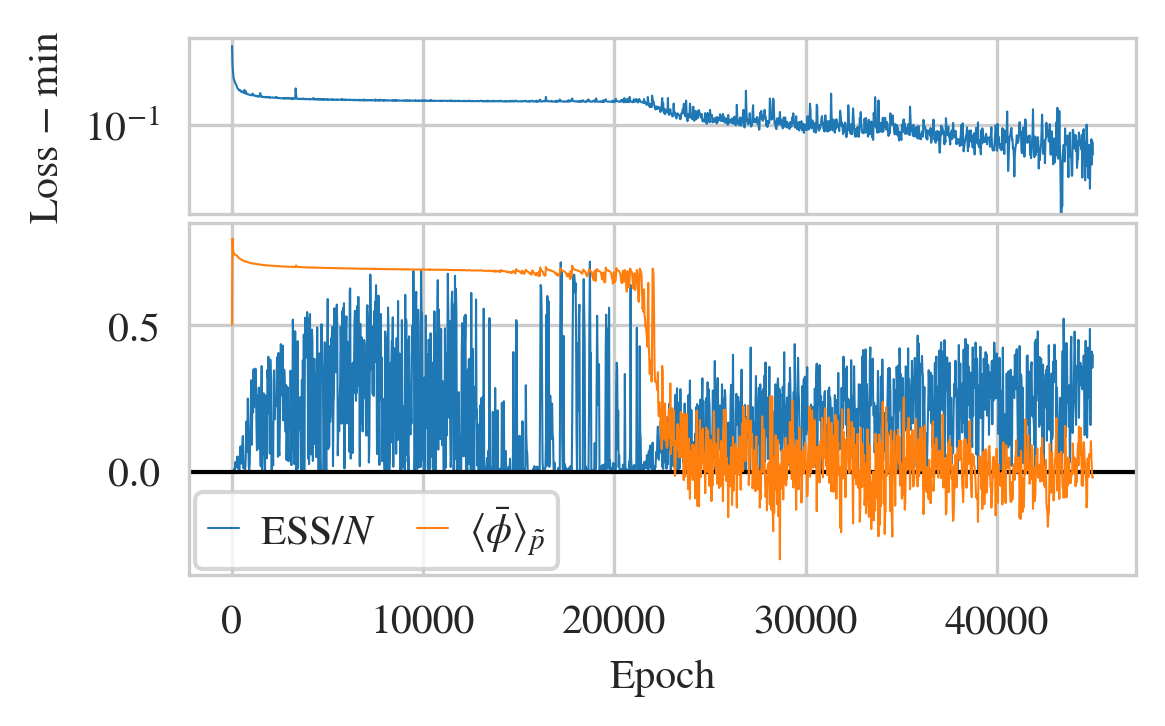}
    \caption{
        Reverse KL training history exhibiting tunneling from unimodal to bimodal, for a flow model targeting  real scalar field theory with 
 ${m^2=-4}$ and $\alpha=0$, training with batch size 16000 and using gradient norm clipping.
        The ESS is computed every 25 epochs.
    }
    \label{fig:rev_train_tunnels}
\end{figure}

Figure~\ref{fig:rev_train_tunnels} presents a training history for a model targeting $m^2=-4$ and $\alpha=0$, demonstrating interesting behavior in the borderline regime between the two extremes discussed above.
The model initially learns to sample from a unimodal distribution.
However, around epoch 15k the model apparently begins to become sensitive to the other mode. At epoch $\approx 22000$, the model tunnels suddenly to a bimodal distribution with $\vev{\overline{\phi}}_{\tilde{p}} \approx 0$. The loss drops as this occurs, consistent with the bimodal distribution being the true minimum of the loss.
Note that, unlike the previous cases, this training used (aggressive) gradient norm clipping; without it, the model stochastically tunnels between unimodal and bimodal before eventually settling into the bimodal distribution (after $\approx (100-150) \times 10^3$ epochs, in one example).

For these simple systems, we find that for $\alpha=0$ and $-m^2 \le 4$, reverse KL self-training produces a bimodal distribution in a tractable amount of time with behavior like that of either Fig.~\ref{fig:rev_train_smooth} or \ref{fig:rev_train_tunnels}.
Similarly, reverse KL training produces a bimodal distribution for all of the values of $\alpha$ we consider at fixed $m^2=-4$, likely because models could be trained for $m^2=-4$ and $\alpha=0$, and the other choices of $\alpha$ used here do not significantly perturb the modal structure.
We have seen that we can (slightly) extend the reach of reverse KL self-training using tools like gradient norm clipping and decaying learning rates, so it may be possible to push further into the broken regime with improved training. 

It is interesting to note that for $\alpha \ne 0$ we observe that the models collapse onto the lower-weight mode of the target distribution at positive $\overline{\phi}$ in early training in ${\gtrsim 50\%}$ of training runs.
This suggests that the baseline architecture  is biased towards the positive-$\overline{\phi}$ peak.

\subsection{Architectural approaches}
\label{sec:real-arch-results}

Here, we test the architectural approaches presented in Sec.~\ref{sec:arch_methods}: equivariant flows, topology matching, and mixture models.
For equivariant flows and topology matching, we aim to demonstrate only qualitative success. Thus, for these methods we consider only a single target distribution, $m^2=-4$, $\lambda=1$, and $\alpha=0$ on a $10 \times 10$ lattice geometry.
For mixture models, we perform a more in-depth exploration to demonstrate different variations on the approach.

\subsubsection{Equivariant flows}

As discussed in Sec.~\ref{sec:arch_methods}, when the mode structure of the target is associated with a spontaneously broken symmetry, mode collapse can be avoided by constructing model distributions which are invariant under the same symmetry. In the case of real scalar field theory, the relevant symmetry is the global $Z_2$ under $\phi \rightarrow -\phi$. Invariant models can thus be constructed from a $Z_2$-invariant prior distribution and a $Z_2$-equivariant flow transformation. 
Various previous efforts \cite{Nicoli:2020njz,DelDebbio:2021qwf,Nicoli:2023qsl} have demonstrated the efficacy of equivariant flows constructed by restricting the structure of the transformations and/or the neural network architecture. Here, we demonstrate that symmetrized and canonicalized transformations can be applied to similar effect without such restrictions.

Given some base transformation kernel, symmetrization provides a unique prescription to construct an equivariant transformation.
For the global $Z_2$ symmetry of interest, applying Eq.~\eqref{eq:sym-transf} to an arbitrary transformation kernel $h(\phi_A|\phi_F)$ yields
\begin{equation}
    \phi'_A = \frac{1}{2} \left[
        h(\phi_A|\phi_F) - h(-\phi_A|-\phi_F)
    \right].
\end{equation}
Specializing further to affine transformations gives
\begin{equation}
    \phi_A' = \left[e^{s(\phi_F)} + e^{s(-\phi_F)} \right] \phi_A + \left[ t(\phi_F) - t(-\phi_F) \right]
\end{equation}
for which equivariance under $(\phi_A, \phi_F) \rightarrow (-\phi_A, -\phi_F)$ is apparent.

In contrast, canonicalization is not unique, and requires a choice of scheme.
For this application, we canonicalize the global $Z_2$ using the sign of the magnetization computed only on the frozen degrees of freedom.
In the notation of Sec.~\ref{sec:canonical}, this is
\begin{equation}
    c(\phi_F) = \overline{c}(\phi_F) = \mathrm{sign} [\sum_x \phi_F(x) ] \equiv \mathrm{sign}[\overline{\phi}_F].
\end{equation}
Applied to affine transformations, this results in the equivariant transformation
\begin{equation}
    \phi'_A = \mathrm{sign}[\overline{\phi}_F] \left[ e^{s(\mathrm{sign}[\overline{\phi}_F] \phi_F)} + t(\mathrm{sign}[\overline{\phi}_F] \phi_F) \right].
\end{equation}
Note that for this choice, the canonicalized transformation is discontinuous at $\overline{\phi}_F = 0$.
In principle, this is a violation of the statistical formalism, as the Jacobian is not well-defined on this surface.
In practice, however, this is not an issue, as the problematic configurations are a set of measure zero.
While this transformation is invertible when $\overline{\phi}_F \neq 0$, it is not a diffeomorphism, and so (favorably) the topological constraints discussed in Sec.~\ref{sec:topo-matching} do not in principle apply.
However, in exchange, discontinuities at $\overline{\phi}_F = 0$ cause practical difficulties with training stability and expressivity near $\overline{\phi}=0$.

\begin{figure}
    \centering
    \includegraphics[width=\linewidth]{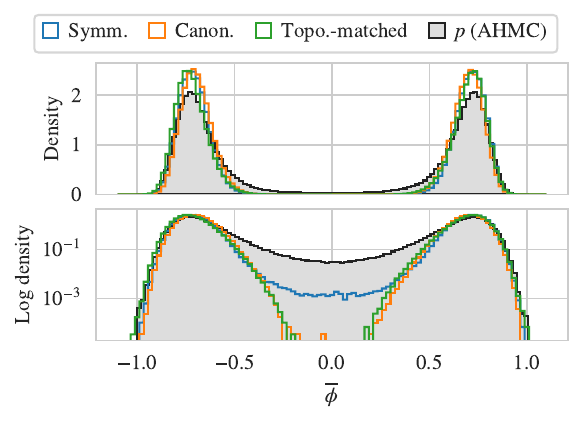}
    \caption{
        For real scalar field theory with $m^2=-4$ and $\alpha=0$, distribution of $\overline{\phi}$ for example models constructed using the different architectural approaches described in the text: equivariance with symmetrization (Symm.) or canonicalization (Canon.), and topology matching using a bimodal prior distribution (Topo.-matched).
        All models were trained for 10000 epochs.
        \textbf{Note:} these histograms are computed using raw model samples, and no statistical corrections have been applied.
        The target distribution ($p$), as measured using AHMC, is shown for comparison.
        All histograms are computed with $1.28 \times 10^6$ samples.
    }
    \label{fig:arch_hists}
\end{figure}

Figure~\ref{fig:arch_hists} demonstrates the qualitative success of these methods over the baseline.
For the same parameters and training time where the baseline methodology collapses (cf.~Fig.~\ref{fig:rev_train_tunnels}), both equivariant approaches yield bimodal models resembling the target.
This result is unsurprising, as exact $Z_2$ invariance forbids mode collapse.
However, in both cases, the models severely underweight the inner tail between the two modes.

\subsubsection{Topology matching}
\label{sec:real-topo-matching}

As discussed in Sec.~\ref{sec:topo-matching}, topological constraints on models constructed using diffeomorphic flows suggest using prior distributions with similar mode structure to the target.
For this application, this requires a bimodal prior distribution.
A simple, tractably sampleable choice is a bimodal mixture of spherical Gaussians with width $\sigma$, offset by $\pm R$ from the origin.
Sampling is accomplished straightforwardly by drawing independent Gaussian noise on each site, then globally adding or subtracting $R$ with equal probability.
The density of the resulting mixture is
\begin{equation}
    r_\text{mix}(z) = \frac{1}{2} \left[ r_\sigma(z-R) + r_\sigma(z+R) \right]
\end{equation}
where $z=z(x)$ is the lattice of scalar degrees of freedom and $r_\sigma(z)$ is the density of a width-$\sigma$ spherical Gaussian centered at zero.

The affine coupling flows employed in this study are able to express exact overall rescalings of the prior samples. Thus, the overall width of the prior can be fixed to 1 without loss of expressivity. However, 
the peak sharpness $\sigma/R$ remains a nonredundant parameter of the model. Here, we treat it as an architecture hyperparameter.

\begin{figure}
    \centering
    \includegraphics[width=\linewidth]{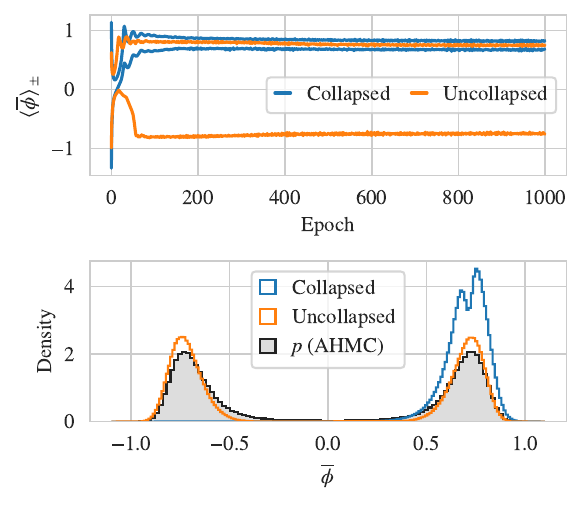}
    \caption{
        Demonstrations of successful (Uncollapsed) and failed (Collapsed) topology matching of real scalar field theory with $m^2=-4$ and $\alpha=0$.
        Models are constructed using the bimodal mixture prior described in the text with $\sigma/R = 1$ and differ only by pseudorandom seed for initialization and training.
        In the top panel, $\vev{\overline{\phi}}_\pm$ denotes the magnetization averaged over samples mapped from the positive- and negative-magnetization modes of the prior distribution.
        Each pair of same-colored lines shows the evolution of these quantities under reverse KL self-training for one of the cases.
        In the collapsed case, both prior modes are mapped to a single target mode.
        The bottom panel shows the distribution of $\overline{\phi}$ for models obtained after 10000 epochs of training with batch size 1024 in each case.
        \textbf{Note:} these histograms are computed using raw model samples, and no statistical corrections have been applied.
        The target distribution ($p$), as measured using AHMC, is shown for comparison.
        All histograms are computed with $1.28 \times 10^6$ samples.
    }
    \label{fig:mix_prior_collapse}
\end{figure}

Figure~\ref{fig:mix_prior_collapse} illustrates typical results of using this bimodal prior with the baseline non-equivariant flow and reverse KL training procedure, applied again to $m^2=-4$, ${\lambda=1}$, and $\alpha=0$ on a $10 \times 10$ lattice.
We observe this modified architecture may still undergo mode collapse, with the flow learning to map both prior modes onto a single mode of the target.
This collapse appears to be similarly robust against further training as in the baseline approach, but occurs early in training.
This admits the practical remedy of repeatedly restarting training until an uncollapsed start is found.
We also find that the prior peak sharpness $\sigma/R$ has a large effect on the frequency of collapse.
With $\sigma/R = 1$, collapse occurs in $\approx 75\%$ of runs, while for $\sigma/R = 0.1$, collapse occurs rarely if ever.
Figure~\ref{fig:arch_hists} compares the distribution of $\overline{\phi}$ obtained in this experiment with the invariant models and the target distribution.
As with the equivariant architectures, the resulting topology-matched model severely undersamples in the inner tails.
 
\subsubsection{Mixture models}
\label{sec:mixture-results}

Constructing a mixture model which is a good approximation to some target distribution requires a set of component models which together provide good coverage of the target: everywhere the target has significant support, at least one component model should have nontrivial density.
The baseline reverse KL self-training is not effective at training samplers for bimodal distributions when the modes of the target are well-separated.
However, the apparently high finite-sample ESS for mode-collapsed models indicate that they are good fits to the collapsed-upon modes.
They are thus good candidates for mixture components in single-model mixtures as described in Sec.~\ref{sec:mixture-models}.

\begin{figure}
    \centering
    \includegraphics[width=\linewidth]{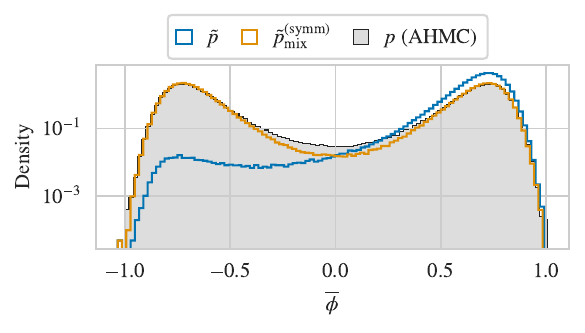}
    \caption{
        Distribution of $\overline{\phi}$ for models of  real scalar field theory, for samples drawn from a unimodal model ($\tilde{p}$) and from a symmetrized mixture model constructed from it ($\tilde{p}_{\text{mix}}^{(\text{symm})}$), both for $m^2=-4$ and $\alpha=0$.
        The mixture samples are generated by postprocessing the samples shown for the unimodal model.
        \textbf{Note:} these histograms are computed using raw model samples, and no statistical corrections have been applied.
        The target distribution ($p$), as measured using AHMC, is shown for comparison.
    }
    \label{fig:mix_hist_a0}
\end{figure}

In the symmetric $\alpha=0$ case, the two modes of the target distribution are identically shaped, so a model for one mode can serve equally well as a model for the other.
Thus, rather than training separate models for each mode, we use a single model to construct a symmetrized mixture as discussed in Sec.~\ref{sec:symm-mixtures}.
The relevant set of transformations are the global $Z_2$ under which $\phi \rightarrow \phi' = t \phi$ with $t \in \{1, -1\}$.
Specializing Eq.~\eqref{eqn:symm-mix-model} to this case, the density of the mixture is
\begin{equation}
    \tilde{p}^{(\text{symm})}_{\text{mix}}(\phi) = \frac{1}{2} \left[ \tilde{p}(\phi) + \tilde{p}(-\phi) \right].
    \label{eqn:phi4-symm-mix-logq}
\end{equation}
Figure~\ref{fig:mix_hist_a0} compares the distribution of $\overline{\phi}$ from a mode-collapsed model to the distribution for a symmetrized mixture constructed from it; as expected, the mixture distribution fits the peaks very well, but is undersupported in the region between the modes.

For the explicitly broken $\alpha \ne 0$ case, we may use a symmetrized mixture model without any formal issues; this will simply result in poor overlap with the target when the relative weights of the modes of the target are too dissimilar.
For the range of $\alpha$ we consider here, we observe (cf.~Fig.~\ref{fig:hmc-hists-alpha}) that while the relative weights of the modes become increasingly tilted, the shapes of the modes remain similar.
This can be treated with a straightforward generalization of the symmetrized mixtures, wherein we randomly apply a sign with probability $p_{\text{flip}} \ne 0.5$.
The mixture is thus over the original flow model with weight $1 - p_{\text{flip}}$, and the flow model composed with a sign flip with weight $p_{\text{flip}}$; the mixture density is
\begin{equation}
    \tilde{p}^{(\text{tune})}_{\text{mix}}(\phi) = (1-p_{\text{flip}}) \tilde{p}(\phi) + p_{\text{flip}} \tilde{p}(-\phi).
\end{equation}
We emphasize that $\phi$ in the above equation should be read as the $\phi$ obtained after applying the random sign.
Here, $p_{\text{flip}}$ is a parameter which can be tuned to maximize some measure of quality of the model.
We optimize $p_{\text{flip}}$ by minimizing the reverse KL divergence on a fixed sample of $10^6$ configurations drawn from the component model, fixing the draw used to determine whether a sign flip is applied to each configuration.

In Sec.~\ref{sec:adaptive-mixtures} we present a construction where the mixture weights are adaptively set using information from the target distribution rather than left as tunable parameters.
Applied to the $Z_2$ transformations relevant here, we can sample from an adaptive mixture by first drawing a sample $\varphi$ from our flow model, then randomly applying a minus sign with probability $p(-\varphi) / [p(\varphi) + p(-\varphi)]$ to produce $\phi = \pm \varphi$.
In the symmetric $\alpha=0$ case where $p(\varphi) = p(-\varphi)$ this reduces to a symmetrized mixture.
Away from this limit, we can specialize Eq.~\eqref{eqn:adaptive-mix-model} to obtain the density of the final sample,
\begin{equation}
    \tilde{p}^{(\text{adapt})}_{\text{mix}}(\phi) = \frac{p(\phi)}{p(\phi) + p(-\phi)} \left[ \tilde{p}(\phi) + \tilde{p}(-\phi) \right].
\end{equation}

\begin{figure}
    \centering
    \includegraphics[width=\linewidth]{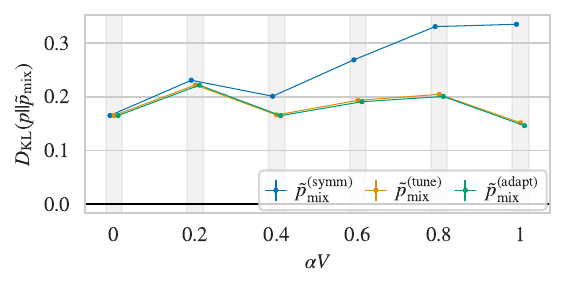}
    \caption{
        Forwards KL divergence (lower indicates better agreement) for the three different mixture model constructions described in the text, estimated using shared ensembles of $10^6$ target samples, for real scalar field theory with fixed $m^2 = -4$ and varying $\alpha$.
        At each $\alpha$, all three mixtures are constructed from the same unimodal component model, obtained using reverse KL self-training.
        Errors are computed by bootstrapping, and are smaller than the markers for all points.
        Additional variance due to seed dependence is not quantified.
    }
    \label{fig:fkl_mix_models_alpha}
\end{figure}

\begin{figure}
    \centering
    \includegraphics[width=\linewidth]{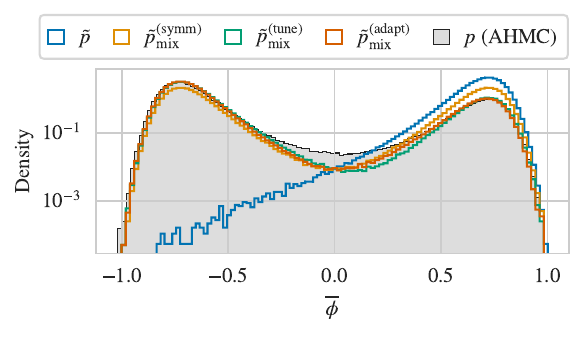}
    \caption{
        Distribution of $\overline{\phi}$ for samples drawn from a unimodal model ($\tilde{p}$), as well as various different mixtures constructed from it ($\tilde{p}_{\text{mix}}^{(\ldots)})$ as described in the text, all for real scalar field theory with $m^2=-4$ and $\alpha V = 0.8$.
        The mixture samples are generated by postprocessing the samples shown for the unimodal model.
        \textbf{Note:} these histograms are computed using raw model samples, and no statistical corrections have been applied.
        The target distribution ($p$), as measured using AHMC, is shown for comparison.
    }
    \label{fig:mix_hist_a8}
\end{figure}

Figure~\ref{fig:fkl_mix_models_alpha} compares the forwards KL divergence of these different constructions at fixed $m^2 = -4$ varying $\alpha$ (all three mixture constructions coincide when $\alpha = 0$).
At lower values of $\alpha$ where the distribution is more symmetric, all three methods perform roughly equivalently.
As $\alpha$ grows and the distribution becomes more asymmetric, the adaptive mixtures and mixtures with tuned $p_{\text{flip}}$ remain approximately the same, while the symmetrized mixture becomes an increasingly poor approximation of the target.
Adaptive mixtures perform nearly identically with mixtures with tuned $p_{\text{flip}}$, without the need to optimize any mixture parameters.
Figure~\ref{fig:mix_hist_a8} compares the distributions of $\overline{\phi}$ resulting from applying the three different methods at $\alpha = 0.8/V$. The symmetric mixture overweights the lower peak, while adaptive mixtures and mixtures with tuned $p_{\text{flip}}$ produce similar distributions with the correct relative mode weights.

Note that in Fig.~\ref{fig:fkl_mix_models_alpha}, the slight peak at $\alpha V = 0.2$ is most likely a training effect. As mentioned above, for all $\alpha$ we consider, reverse training eventually finds the bimodal distribution. In fact, the models begin to move towards bimodal early on, compromising the available quality of unimodal models (cf.~the drop in ESS in Fig.~\ref{fig:rev_train_tunnels} around epoch 10k).
Further study at parameters deeper in the bimodal phase could provide cleaner results, but the relative performance of the different approaches is clear.

\subsection{Training approaches}
\label{sec:real-training-results}

Here, we test the various different training methods presented in Sec.~\ref{sec:multimodal-approaches}: different variations of forwards KL training, adiabatic retraining, and flow-distance regularization.
We specialize each method to real scalar field theory and apply it to the baseline non-equivariant architecture described in Sec.~\ref{sec:real-baseline}.
Where meaningful, we compare performance between different variations and examine dependence on the parameters of the target.

\subsubsection{Forwards KL training}
\label{sec:forwards-kl-results}

As discussed in Sec.~\ref{sec:forwards-kl}, we can implement a self-training scheme using the forwards KL divergence by generating a training dataset using the model itself, then reweighting to correct for the mismodeling of the target distribution.
This approach allows us to augment the training data in a way is not structurally possible when using reverse KL self-training.

\begin{figure}
    \centering
    \includegraphics[width=\linewidth]{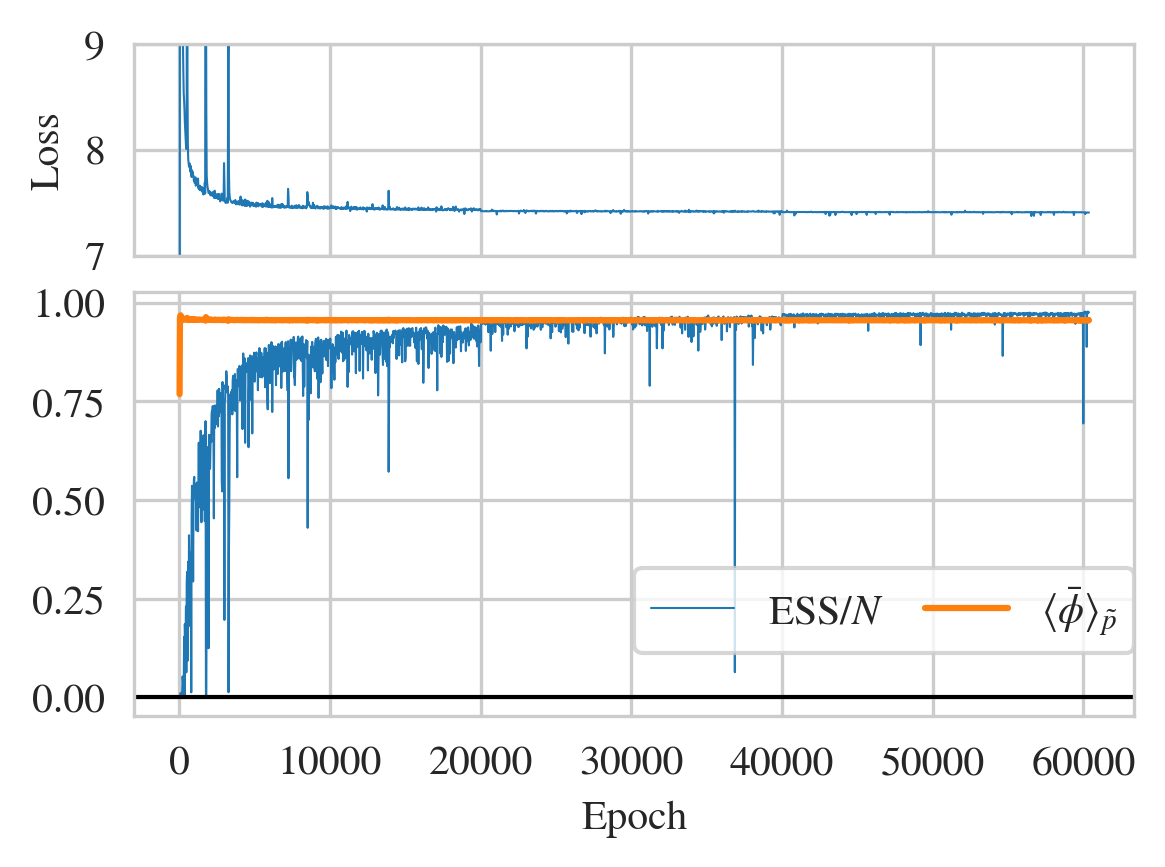}
    \caption{
        Reweighted forwards KL self-training history exhibiting mode collapse, for a flow model targeting real scalar field theory with $m^2=-5$ and $\alpha=0$, training with batch size 16000, using gradient norm clipping, and with the learning rate dropping by a factor of $\gamma=0.5$ every 20000 epochs.
        The ESS is computed every 25 epochs.
        The similar values of $\text{ESS}/N$ and $\vev{\overline{\phi}}_{\tilde{p}}$ are coincidental.
    }
    \label{fig:fwd_train_m5}
\end{figure}

We first examine the behavior of reweighted forwards KL self-training without any augmentation of the training data.
Figure~\ref{fig:fwd_train_m5} shows a typical training history for this approach, applied to the parameters $m^2=-5$ and $\alpha=0$ where the modes are well-separated.
The model collapses onto one of the modes, indicating that mode collapse is a problem of self-training schemes in general and not a specific feature of reverse KL self-training.

\begin{figure}
    \centering
    \includegraphics[width=\linewidth]{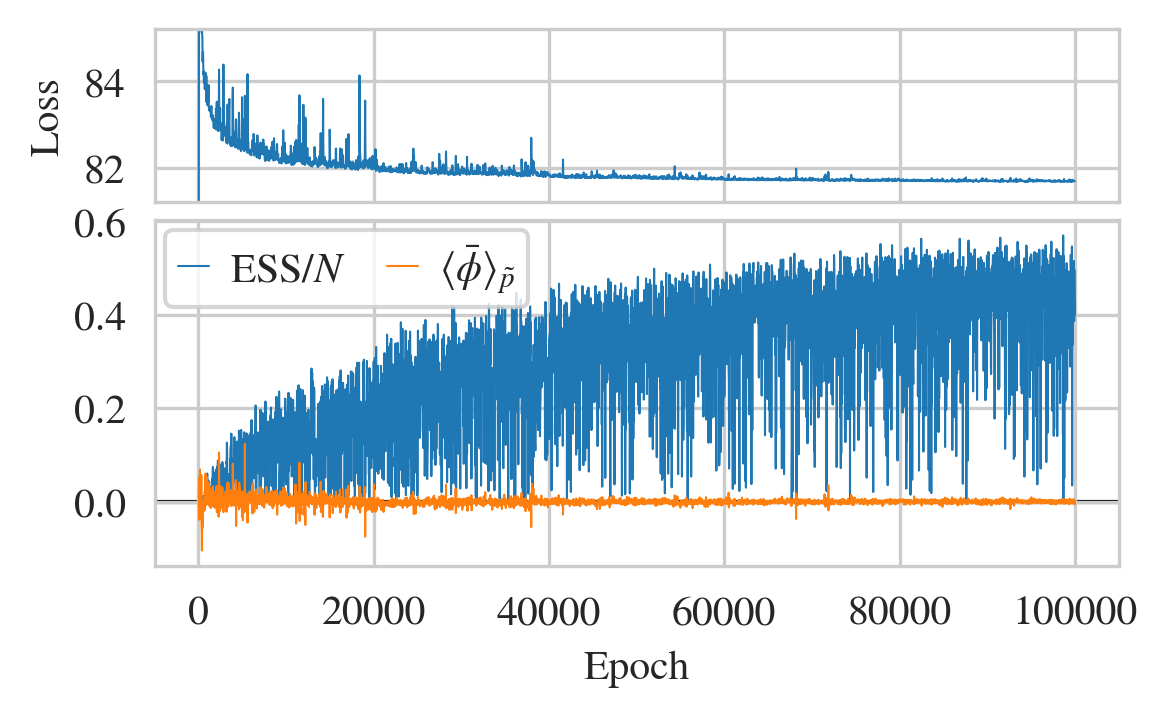}
    \caption{
        Reweighted forwards KL self-training history (with training data sampled from a mixture) exhibiting direct training to a bimodal distribution, for a flow model targeting real scalar field theory with $m^2=-4$ and $\alpha=0$, training with batch size 16000, using gradient norm clipping, and with the learning rate dropping by a factor of $\gamma=0.5$ every 20000 epochs.
        The ESS is computed every 25 epochs.
    }
    \label{fig:fwd_train_m4}
\end{figure}

To improve over reverse KL self-training we must take advantage of the ability to augment the training dataset.
To do so, we generate each batch of training data by sampling from a single-model mixture constructed from the flow model being trained.
We first examine the symmetric $\alpha = 0$ case where all of the mixture constructions are equivalent.
Figure~\ref{fig:fwd_train_m4} shows an example training history for this scheme, targeting $m^2=-4$ and $\alpha=0$.
Over the course of training, $\vev{\overline{\phi}}_{\tilde{p}} \approx 0$ up to per-epoch fluctuations, indicating that the model trains directly into a bimodal distribution.
This is dissimilar to the training dynamics observed in e.g.~Fig.~\ref{fig:rev_train_tunnels} with reverse KL self-training at these parameters, where the model initially collapses onto a single mode before eventually finding the bimodal distribution.

We find this scheme has instabilities in training that must be regulated with gradient norm clipping.
The large gradients that necessitate clipping likely arise in part from batches where a single high-weight configuration (like those discussed in Sec.~\ref{sec:sampling}) dominates the reweighted loss estimate.
We also observe that the value of $\vev{\overline{\phi}}_{\tilde{p}}$ can fluctuate at late training times, but also that step-scheduling the learning rate (i.e.~reducing the learning rate by a factor $\gamma=0.5$ every 20000 epochs) regulates this effect and allows $\vev{\overline{\phi}}_{\tilde{p}}$ to settle to zero (or the appropriate nonzero value, in the asymmetric $\alpha \ne 0$ case).

We also test the naive data augmentation scheme described in Sec.~\ref{sec:forwards-kl} where, rather than properly sampling training data from a mixture, we simply apply a random sign flip to each configuration.
In this case, the reweighting factors use $p' = \tilde{p}$ (in the notation of Eq.~\eqref{eq:fwd_dkl}) for the configurations before negation, while $\log \tilde{p}$ in the loss is computed after negation.
We observe similar success and training dynamics for this scheme as with proper mixture sampling.
We compare the resulting models with those obtained using other forwards KL approaches below.

To set a baseline for these self-training approaches, we train models using target-distributed data generated on-the-fly using AHMC, providing a theoretically infinite training dataset as in self-training schemes.
This training dataset amounts to a limiting-case perfect dataset for reweighted forwards KL self-training: all reweighting factors are equal so no statistical power is lost, and due to the augmentation the training data is guaranteed to encode the appropriate relative mode weights.
Specifically, for a given batch size $B$, we generate our training dataset by running $B$ completely independent AHMC chains, initially equilibrating each for 1000 trajectories. For every epoch of training, we advance all $B$ chains by 10 trajectories to produce a new training dataset for each epoch. We use the AHMC settings provided in Sec.~\ref{sec:aug-hmc-results}. Autocorrelations within each chain are minimal for these settings and targets, so each batch of training data is approximately independent from the last.

\begin{figure}
    \centering
    \includegraphics[width=\linewidth]{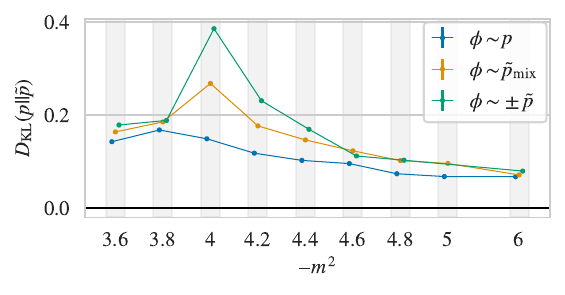}
    \caption{
        Forwards KL divergence (lower indicates better agreement) for models trained using different versions of forwards KL training, estimated using shared ensembles of $10^6$ target samples, for real scalar field theory with fixed $\alpha=0$ and varying $m^2$.
        The label $\phi \sim p$ denotes models trained using data generated by AHMC, $\phi \sim \tilde{p}_{\mathrm{mix}}^{(\text{symm})}$ denotes models self-trained using samples from a symmetrized mixture, and $\phi \sim \pm \tilde{p}$ denotes models self-trained using samples naively augmented by applying a random sign.
        All models are trained with batch size 16000 and using gradient norm clipping.
        Errors are computed by bootstrapping, and are smaller than the markers for all points.
        Additional variance due to seed dependence is not quantified.
    }
    \label{fig:fkl_fwds_vs_M2}
\end{figure}

Figure~\ref{fig:fkl_fwds_vs_M2} compares the forwards KL divergences for models obtained using the self-training schemes and training with AHMC data in the $\alpha = 0$ case as a function of $m^2$.
For most parameters, including those deep in the bimodal phase, all methods perform comparably, with self-training producing only marginally poorer models than training with AHMC data.
The approximate self-training scheme with naive data augmentation performs comparably to drawing training data from a mixture.

\begin{figure}
    \centering
    \includegraphics[width=\linewidth]{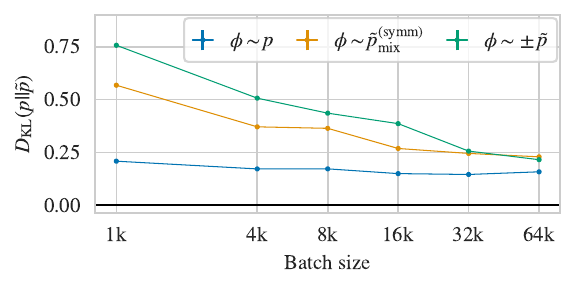}
    \caption{
        Forwards KL divergence as a function of batch size for models trained using different versions of forwards KL training for real scalar field theory with $\alpha=0$ and $m^2=-4$, estimated using shared ensembles of $10^6$ target samples.
        Errors are computed by bootstrapping, and are smaller than the markers for all points.
        Additional variance due to seed dependence is not quantified.
    }
    \label{fig:fkl_fwds_vs_bs}
\end{figure}

As shown in Fig.~\ref{fig:fkl_fwds_vs_bs}, we observe that the final quality of self-trained models depends significantly on the batch size used for training, even when training to apparent convergence.
The dependence is weak, if present at all, for models trained with AHMC data.
It is unclear whether self-trained model quality will converge to AHMC-trained model quality in the limit of infinite training batch sizes.
For sufficiently large batch sizes, the naive data augmentation scheme appears to produce equivalently good models as the mixture scheme.
This batch size dependence is stronger than observed using the baseline reverse KL self-training on unimodal distributions.

As discussed above, when $\alpha \ne 0$ we can consider several in-principle different mixture model constructions.
To augment the data for forwards KL self-training, we only investigate the performance of symmetrized mixtures (where ${p_{\text{flip}}=0.5}$, regardless of the relative weights of the modes of the target) and adaptive mixtures.
For simplicitly, we do not consider the mixture construction with tunable $p_{\text{flip}}$, which is difficult to apply in this context: during training the flow models will in general be bimodal, so $p_{\text{flip}}$ is not directly related to the relative mode weights as it is with mixtures of unimodal models, and the relative mode weights change quickly, meaning the optimal choice of $p_{\text{flip}}$ depends on the current state of the model.
We also test the naive data augmentation scheme for $\alpha \neq 0$, fixing the probability of negating each configuration to $p_{\text{sign}} = 0.5$ to avoid the same complications.

\begin{figure}
    \centering
    \includegraphics[width=\linewidth]{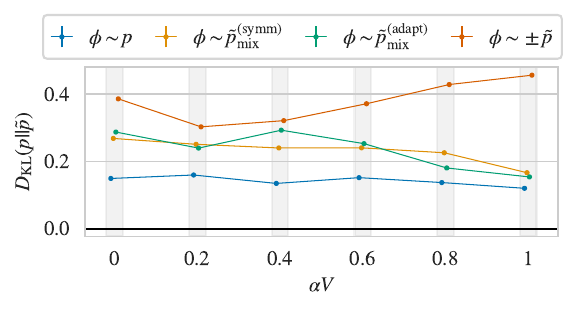}
    \caption{
        Similar to Fig.~\ref{fig:fkl_fwds_vs_M2}, but for real scalar field theory with fixed $m^2=-4$ and varying $\alpha V$.
        Additional variance due to seed dependence is not quantified,
        but training using data from symmetrized ($\phi \sim \tilde{p}^{(\text{symm})}_{\text{mix}}$) and adaptive ($\phi \sim \tilde{p}^{(\text{adapt})}_{\text{mix}}$) mixtures are identical procedures at $\alpha = 0$, so the discrepancy reflects seed dependence in that case.
    }
    \label{fig:fkl_fwds_vs_alpha}
\end{figure}

Figure~\ref{fig:fkl_fwds_vs_alpha} compares the results of these tests.
The discrepancy between AHMC-trained results and self-trained results extends to non-zero $\alpha$.
The two different mixture approaches perform comparably, although the results using adaptive mixtures have more variance, suggesting less stable training or stronger seed dependence.
The success of the symmetrized mixture scheme implies that the increased variance from reweighting to correct the mode weights is not a large effect.
The naive data augmentation scheme performs generically worse than the mixtures (although comparing with Fig.~\ref{fig:fkl_fwds_vs_alpha}, $m^2=-4$ is a particularly challenging target for this scheme), and predictably produces models of increasingly poor quality as $\alpha$ increases and $p_{\text{sign}}=0.5$ becomes a worse approximation.

\subsubsection{Adiabatic retraining \& flow-distance regularization}
\label{sec:adiabatic-results}

As introduced in Secs.~\ref{sec:adiabatic-retraining} and \ref{sec:flow-distance-reg}, both adiabatic retraining and flow-distance regularization can be thought of as reverse KL self-training with an adiabatically changing the loss function, so we discuss them together.

For adiabatic retraining, we tune the action parameter $m^2$ over the course of training via a schedule function $g(t) \in [0,1]$ so that the effective instantaneous $\phi^4$ action can be explicitly written as
\begin{equation}
S_{\text{inst}}(\phi;t) = S_E(\phi) + g(t)\frac{1}{2}(m_0^2-m^2) \sum_{x} \phi(x)^2
\label{eqn:adiabatic-retraining-inst-action}
\end{equation}
where $S_E$ is given by Eq.~\eqref{eqn:phi4-action}. 
When targeting the symmetric potentials where $\alpha = 0$, the initial parameter $m_0^2 = -3$ is chosen to correspond to a unimodal distribution.
For flow-distance regularization, each configuration $\phi$ contributes to the loss as
\begin{equation}
L(\phi) = \log \tilde{p} + S_E(\phi) + \kappa \tilde{g}(t) \sum_{x} |\phi(x)-z(x)|^2
\label{eqn:flow-distance-loss}
\end{equation}
where $\tilde{g}(t) \in [0,1]$ is a schedule function. The hyperparameter $\kappa$ sets the initial relative contributions of the regularizer term and the action.

In practice, we find that both training processes are more stable if we take the schedule functions $g(t), \tilde{g}(t)$ to be concave functions of $t$ rather than linear interpolations.
We use
\begin{equation}
g(t) = \tilde{g}(t) = \frac{e^{-k\frac{t-t_1}{t_2-t_1}}-e^{-k}}{1-e^{-k}}, \quad t_1 < t \leq t_2
\end{equation}
choosing $t_1, t_2, k$ appropriate to each problem.
The schedule is defined to be flat before $t_1$ so that initially, adiabatic retraining trains a model appropriate to the initial parameters before beginning to interpolate, and flow-distance regularization trains a model close to the identity before the regulator begins to be removed.

In the symmetric case where $\alpha=0$, for $m^2 = -4$ both approaches successfully train bimodal flow models using the schedule parameters in Table~\ref{tab:sched_params}.
The models tend to collapse when the schedule moves too quickly.
Increasingly slow tuning is required as $m^2$ becomes more negative.
We were able to use more aggressive schedules by ``rewinding'' when mode collapse occurs.
This amounts to reloading a recent checkpoint with a new random seed and, if necessary, lowering the learning rate.
We have had mixed success either reloading the optimizer state or resetting the optimizer parameters when rewinding.
Gradient norm clipping also helps make training more robust against collapse.

Both schemes are applicable to explicitly broken potentials with $\alpha \ne 0$.
Flow-distance regularization can be used to train directly into the bimodal target just as in the $\alpha=0$ case, although we find that for greater $\alpha$, the model is more likely to collapse to unimodal and usually needs a slower schedule.
For adiabatic retraining, we observe that the particular trajectory through action parameter space is important, specifically that avoiding mode collapse requires training into a bimodal distribution before training towards a target with explicitly broken symmetry.

\begin{table}[]
\centering
\begin{ruledtabular}
\begin{tabular}{ccccc}
Method & Coupling strength & $t_1$ & $t_2$ & $k$ \\ \hline
Adiabatic retraining & $m_0^2=-3.0$ & 40000 & 120000 & 1 \\
Flow-distance & $\kappa\approx0.1$ & 20000 & 120000 & 1 \\
\end{tabular}
\end{ruledtabular}
\caption{Example schedule parameters for adiabatic retraining and flow-distance regularization, for parameters $m^2=-4$ and $\alpha=0$.}
\label{tab:sched_params}
\end{table}

With the caveat that any interpretation based on the finite-sample ESS estimator may be unreliable given the slow convergence problems discussed in Sec.~\ref{sec:sampling} below, we proceed with an analysis of the training dynamics of these methods.

\begin{figure}
    \centering
    \includegraphics[width=\linewidth]{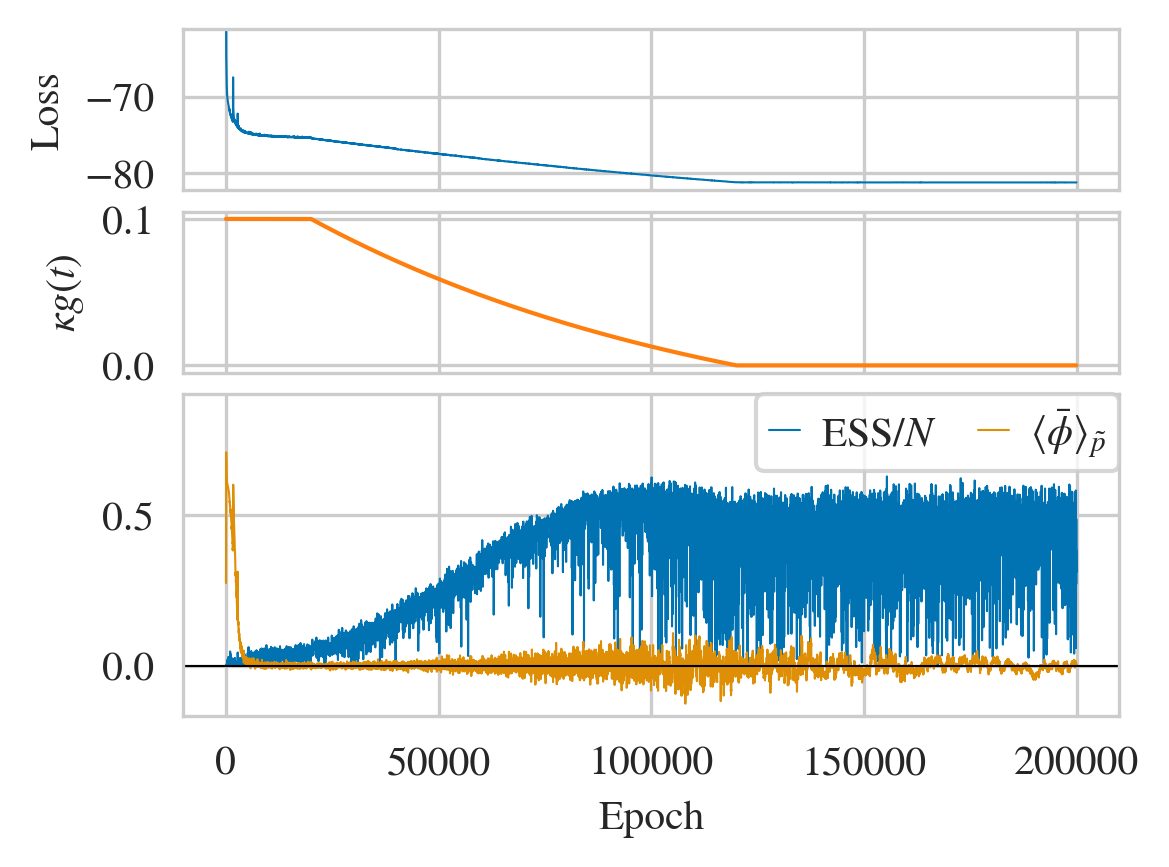}
    \caption{
        Training history for flow-distance regularization, for a flow model targeting real scalar field theory with $m^2 = -4$ and $\alpha=0$, training with batch size 1000.
        The learning rate starts from $10^{-3}$ and drops by a factor of $\gamma = 0.5$ every 20000 epochs.
        The ESS is computed every 25 epochs.
    }
    \label{fig:train_distance}
\end{figure}

Figure~\ref{fig:train_distance} shows the training history of a model for $m^2=-4$ trained using flow distance regularization with the parameters in Table~\ref{tab:sched_params} and a decaying learning rate.
During the initial regulator-dominated part of training when $t < t_1$, the loss approximately plateaus and the magnetization $\vev{\overline{\phi}}_{\tilde{p}}$ goes to zero.
The finite-sample ESS is near zero, indicating poor overlap with the target, as expected for a model which has converged to a prior-like distribution.
As the regulator is slowly removed, the loss changes smoothly, the ESS increases slowly, and the magnetization remains near zero, indicating that the model has not collapsed.
After the regulator is removed, the envelope of the ESS is nearly flat (possibly increasing slowly), indicating that the training schedule finds a nearly converged model without additional unregulated training.
Fluctuations in $\vev{\overline{\phi}}_{\tilde{p}}$ under additional training are damped by the decaying learning rate.

\begin{figure}
    \centering
    \includegraphics[width=\linewidth]{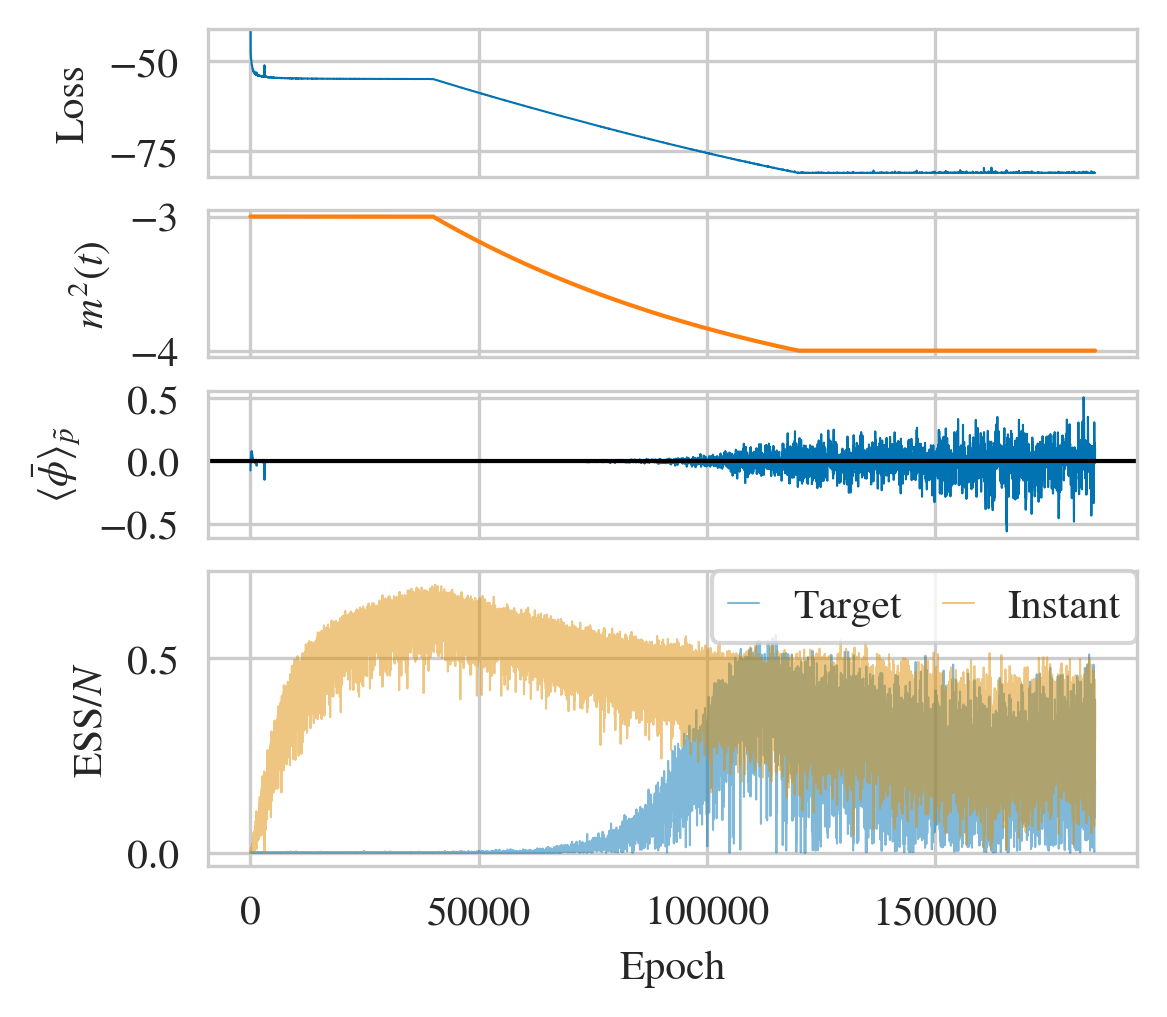}
    \caption{
        Training history for adiabatic retraining, for real scalar field theory with $\alpha=0$ and interpolating from the initial $m_0^2 = -3$ to the final target $m^2 = -4$ as described in the text, training with batch size 4000.
        In the bottom panel, ``Target'' indicates ESS$/N$ for reweighting to the final action with $m^2=-4$, while ``Instant'' is for reweighting to the current instantaneous action with $m^2(t)$ as shown in the second panel.
        The target ESS is evaluated on 16000 configurations every 25 epochs.
        The instantaneous ESS is evaluated every epoch, and smoothed by averaging over blocks of 4 epochs so that it incorporates the same number of samples as the target ESS; the resulting metrics are nevertheless inequivalent and the instantaneous ESS is expected to be more positively biased due to the smaller sample size.
    }
    \label{fig:train_adiabatic}
\end{figure}

Figure~\ref{fig:train_adiabatic} shows the training history of a model for $m^2 = -4$ trained using adiabatic retraining with the schedule parameters in Table~\ref{tab:sched_params}.
During the initial phase of training before $t_1$, the model learns a nontrivial approximation of the instantaneous target with $m^2=-3$ as indicated by the loss and by the ESS onto the instantaneous target (orange), although the ESS is still increasing when the schedule begins to be tuned.
The ESS onto the final target $m^2=-4$ (blue) is near zero, as expected given the different initial and final target distributions.
As the action is tuned towards $m^2=-4$, the loss changes smoothly and $\vev{\overline{\phi}}_{\tilde{p}}$ remains near zero up to per-epoch fluctuations, indicating that the model does not collapse onto a single mode.
Similar to Fig.~\ref{fig:train_distance}, the ESS onto the final target $m^2=-4$ increases slowly as the model becomes an increasingly good approximation of the target.
The increasing variance in the magnetization $\vev{\overline{\phi}}_{\tilde{p}}$ and the change in ESS after the schedule reaches $m^2=-4$ reflect that the interpolation is not fully adiabatic, although this interpretation is confounded by the interpolation beginning with a model which is not yet stable under further training.
The apparent drop in the ESS under additional training after the schedule reaches $m^2=-4$ may suggest that the scheme produces transiently better models than can be achieved by direct training.

On the other hand, the finite-sample ESS onto the instantaneous target in Fig.~\ref{fig:train_adiabatic} changes smoothly over the course of adiabatic retraining; the decrease reflects the fixed architecture's decreasing ability to represent the increasingly complicated target.
The nontrivial overlap with all instantaneous targets throughout training suggests that 
the initial expense of training a model from a random initialization could be amortized over different action parameters, which could help circumvent possible issues with the scaling of training costs~\cite{DelDebbio:2021qwf}.

\subsection{Comparison}
\label{sec:results_comparison}

\begin{figure}
    \centering
    \includegraphics[width=\linewidth]{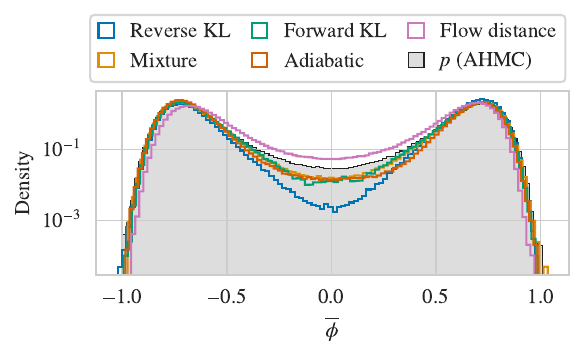}
    \caption{
        Distribution of $\overline{\phi}$ for models constructed using the different approaches, for real scalar field theory with $m^2=-4$ and $\alpha=0$.
        \textbf{Note:} these histograms are computed using raw samples from the model distribution, and no statistical corrections have been applied.
        The distribution of $\overline{\phi}$ for the target distribution $p$, estimated using AHMC, is underlaid in gray for ease of comparison.
    }
    \label{fig:cf_hist_a0}
\end{figure}

\begin{figure}
    \centering
    \includegraphics[width=\linewidth]{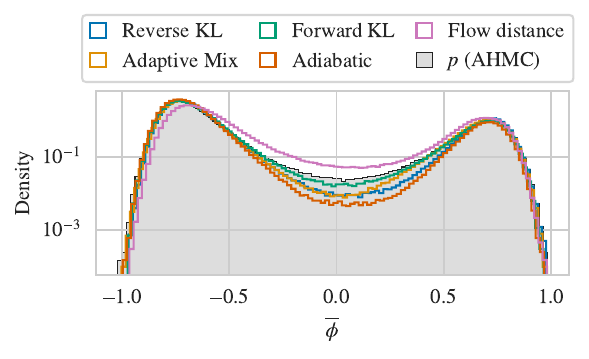}
    \caption{
        As for Fig.~\ref{fig:cf_hist_a0}, but for $m^2=-4$ and $\alpha = 0.8/V$.
    }
    \label{fig:cf_hist_a8}
\end{figure}

As illustrated in Figs.~\ref{fig:arch_hists}, \ref{fig:cf_hist_a0}, and \ref{fig:cf_hist_a8}, all of the methods demonstrated in this section succeed in alleviating the baseline's issues with mode collapse.
While the baseline can produce bimodal models for some parameters, the improved methods can do so more reliably and efficiently.
For other parameters, the improved methods gave bimodal models where the baseline was unable to do so for any reasonable training time.
This is a qualitative improvement in capability.

Comparing Figs.~\ref{fig:arch_hists}, \ref{fig:cf_hist_a0}, and \ref{fig:cf_hist_a8}, underweighting of the inner tail region between the peaks appears to be a nearly generic problem.
As explored in Sec.~\ref{sec:sampling} below, mismodeling of the inner tails can lead to severe issues when using these models with flow-based MCMC.
Learning to model such regions is inherently challenging, as only a small fraction of training data is sampled from them.
Nevertheless, there is no reason to expect improved training methods cannot resolve this issue.

\begin{figure}
    \centering
    \subfloat[\centering  ]{{
        \includegraphics[width=\linewidth]{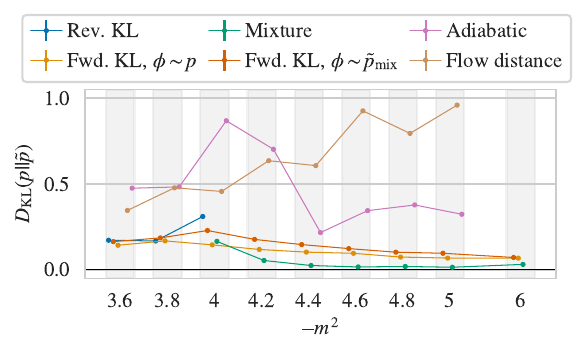}
        \label{fig:cf_fkl_M2}
    }}
    \!
    \subfloat[\centering  ]{{
        \includegraphics[width=\linewidth]{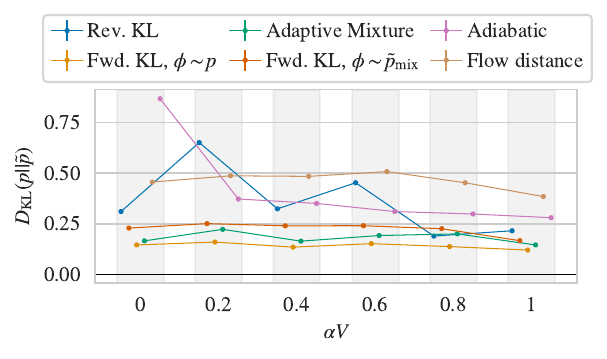}
        \label{fig:cf_fkl_alpha}
    }}
    \caption{
        Forwards KL divergence (lower indicates better agreement) for models constructed using the different approaches described in the text, for real scalar field theory with \protect\subref{fig:cf_fkl_M2} fixed $\alpha=0$ and varying $m^2$ and \protect\subref{fig:cf_fkl_alpha} fixed $m^2=-4$ and varying $\alpha$, estimated using shared ensembles of $10^6$ target-distributed samples.
        Errors are computed by bootstrapping, and are smaller than the markers for all points.
        Additional variance due to seed dependence is not quantified.
        Missing points indicate that we were unable to obtain models for those parameters using the corresponding method.
        These data should not be interpreted as ranking the methods; see the discussion in the text.
    }
    \label{fig:cf_fkl}
\end{figure}

Figure~\ref{fig:cf_fkl} compares forwards KL divergences over a range of different target parameters for the models of Sec.~\ref{sec:real-training-results}.
All models have generally good overlap with their targets (the metric is unbounded from above).
While some hierarchies are apparent in the plot, they should not be interpreted as a ranking of the methods themselves.
There is no reason to expect a similar ranking if the same methods are applied to different target parameters or especially different target theories.
Even small changes in training or architecture hyperparameters could change the picture substantively.
This caveat applies especially to adiabatic retraining and flow-distance regularization, where results are highly sensitive to the choice of schedule.
This sensitivity is apparent in the lack of structure in the data for those methods in Fig.~\ref{fig:cf_fkl}.

Figure~\ref{fig:cf_fkl} also suggests that the degree of bimodality is not uniquely related to the difficulty of the modeling problem.
As the peaks become increasingly well-separated, reverse KL self-training produces worse-quality models, then fails.
In contrast, symmetrized mixtures and forwards KL training methods produce better models for more bimodal targets.
This may be related to the apparent difficulties with modeling inner tails, as more bimodal targets have less mass in these regions to be mismodeled.
For adiabatic retraining and flow-distance regularization, the relationship with bimodality is difficult to disambiguate from training effects.
In the asymmetric case, all methods seem to perform consistently well across the range of $\alpha$ investigated (keeping in mind unquantified variance from training effects).

\section{Extended-mode models for complex scalar field theory}
\label{sec:extended}

In this section, we extend the exploration of the previous section to an example of a system with an extended mode: two-dimensional complex scalar field theory in its symmetry-broken phase.
Similar to the bimodal example of the previous section, self-training procedures are prone to a form of mode collapse in this case as well (``subspace collapse'', as discussed below).
This example provides a qualitatively distinct testbed for the architecture- and training-based approaches to modeling presented in Secs.~\ref{sec:arch_methods} and \ref{sec:multimodal-approaches}.
We demonstrate their application, and show how they can be used to alleviate this problem.

\subsection{Complex lattice scalar field theory}

We consider a discretization of two-dimensional complex $\phi^4$ theory, defined by the Euclidean lattice action
\begin{equation}\begin{aligned}
    S_E(\phi) = 
    & \sum_x \left( 
        \sum_{\mu=1}^d \left| \phi(x+\hat{\mu} - \phi(x)) \right|^2
    \right. \\ & \left. 
        + m^2 |\phi(x)|^2 + \lambda |\phi(x)|^4
    \right)
    \label{eq:lag}
\end{aligned}\end{equation}
where $\phi = (\phi_R + i \phi_I)/\sqrt{2}$.
This action possesses a global U(1) symmetry under phase rotations, $\phi(x) \rightarrow e^{i\theta} \phi(x)$.
Spontaneous breaking of this symmetry gives rise to an extended ``ring-shaped'' mode.
For this demonstration, we consider a single target theory with $m^2=-4$ and $\lambda=3$ on an $8 \times 8$ lattice geometry, which lies in the symmetry-broken phase near the phase transition boundary.

To establish ground-truth, we sample this theory using HMC. A trajectory length of 1 integrated in 20 leapfrog steps gives an acceptance rate $\approx$ 96\%. We save out every 10th configuration. All chains are thermalized for 1000 trajectories from a hot start. The ensemble of $1.28 \times 10^5$ configurations for validation data is generated in 128 parallel streams. For this target, HMC mixes quickly around the extended mode, so we do not apply any additional augmentation.
Fig.~\ref{fig:priors} shows the distribution of the complex magnetization $\overline{\phi}$ for this target theory computed from HMC samples (as well as for examples of the two classes of prior distributions employed in the experiments of this section, described below).

\subsection{Baseline model \& training}

\begin{figure}
    \centering
    \includegraphics[width=\linewidth]{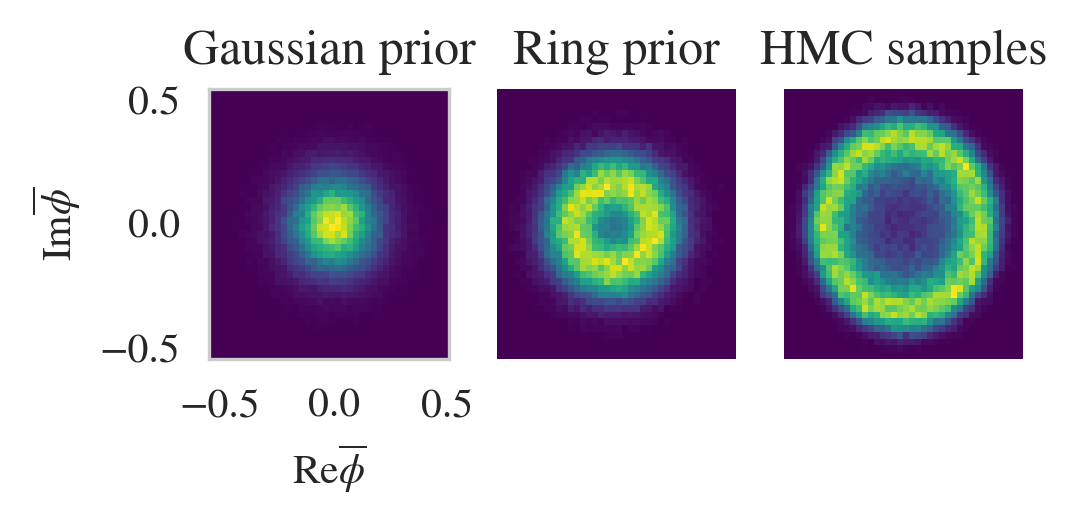}
    \caption{Magnetization distributions for the target complex scalar field theory with $m^2=-4$ and $\lambda=3$ on an $8 \times 8$ lattice geometry (HMC samples), as well as the same for a complex Gaussian prior and an example of the topology-matched "ring prior" (Eq.~\ref{eq:ring_prior}) used to construct models for complex $\phi^4$ theories.}
    \label{fig:priors}
\end{figure}

All models are constructed with flows based on affine coupling transformations, similarly to the real case.
Specifically, the forward transformation for each coupling layer takes the form $\phi_A' = h(\phi_A|\phi_F)$ with components
\begin{equation}
\begin{split}
    h_R(\phi_A|\phi_F)&=e^{s_R(\phi_F)}\phi_{AR}+t_R(\phi_F) \\
    h_I(\phi_A|\phi_F)&=e^{s_I(\phi_F)}\phi_{AI}+t_I(\phi_F)
    \label{eq:complex-affine}
\end{split}
\end{equation}
where $\phi_A,\phi_F$ respectively denote the active and frozen sites in a given coupling layer of the flow, and subscripts $R$ and $I$ respectively denote real and imaginary components. For each such coupling layer, the scaling $(s_R,s_I)$ and offset $(t_R,t_I)$ parameters are obtained from a convolutional neural network (CNN) with two hidden layers of width 8 and kernel size $3\times 3$. 
We partition the variables by checkerboarding over sites, such that $\phi_A$ and $\phi_F$ are sites of opposite parity to explicitly encode nearest neighbor interactions.
We construct the flow from a stack of 32 coupling layers, alternating checkerboard parity in each.
For the baseline architecture, we use this non-equivariant flow with an independent unit complex Gaussian as the prior distribution.
Fig.~\ref{fig:priors} shows the complex magnetization distribution of this prior.
Baseline training is performed using gradient norm clipping, a batch size of 1024, and initial learning rate $10^{-2}$ that is dropped by a factor $\gamma=0.5$ every 20,000 epochs.
In the experiments described below, this baseline is modified in key ways to implement and test the architecture- and training-based methods of Secs.~\ref{sec:arch_methods} and \ref{sec:multimodal-approaches}.

\subsection{Assessing model quality}

As in the real case, collapse may be diagnosed straightforwardly by examination of two-dimensional histograms of the complex magnetization $\overline{\phi}$.
For a diagnostic suitable for monitoring during training, one can compare the average absolute magnetization $\vev{|\overline{\phi}|}_q$ and the absolute average magnetization $|\vev{\overline{\phi}}|_q$, computed with model samples.
For a U(1)-symmetric distribution, $|\vev{\overline{\phi}}| \approx 0$, but $\vev{|\overline{\phi}|} \neq 0$ in general and will be large in the broken phase.
$|\vev{\overline{\phi}}| \neq 0$ suggests U(1) is broken, with $|\vev{\overline{\phi}}| \approx \vev{|\overline{\phi}|}$ indicating almost complete breaking.

For complex $\phi^4$ theories, metrics such as the forwards KL divergence and ESS can be conditioned on the phase of the magnetization to quantify the equivariance of a learned distribution.
These may be computed by evaluating the model density on configurations rotated to a specific magnetization angle.
For instance, a phase-conditioned forwards $D_{KL}$ can be defined as
\begin{align}
\begin{split}
        \widehat{D}_{KL}(p || \tilde{p}, \theta) &= \frac{1}{N} \sum_{n=1}^N  \log{p}(e^{i(\theta - \arg\overline{\phi}_n)} \phi_n) \\&- \log{\tilde{p}}(e^{i(\theta - \arg\overline{\phi}_n)} \phi_n) ,
\end{split}
    \label{eq:dklphase}
\end{align}
evaluated on an ensemble of $N$ target-distributed samples. The phase $e^{i(\theta - \arg\overline{\phi}_n)}$ rotates each field configuration $\phi_n$ so its (complex) magnetization $\overline{\phi}_n$ lies along an angle $\theta$.
Since $p(e^{i\alpha}\phi)=p(\phi)$ by the U(1) global invariance of the theory, Eq.~\eqref{eq:dklphase} gives a direct measure of the equivariance of the model and thus a diagnostic of mode collapse.

Separately and complementarily, radius-conditioned versions of these metrics can be produced by binning configurations by the radius of their complex magnetization, $|\overline{\phi}|$, and evaluating the forwards KL divergence estimator on the subset for each radius. Such metrics quantify the local accuracy of the model in the inner and outer tails of the circularly symmetric mode structure.

\subsection{Subspace collapse}

\begin{figure}
    \centering
    \subfloat[\centering  ]{{
        \includegraphics[width=\linewidth]{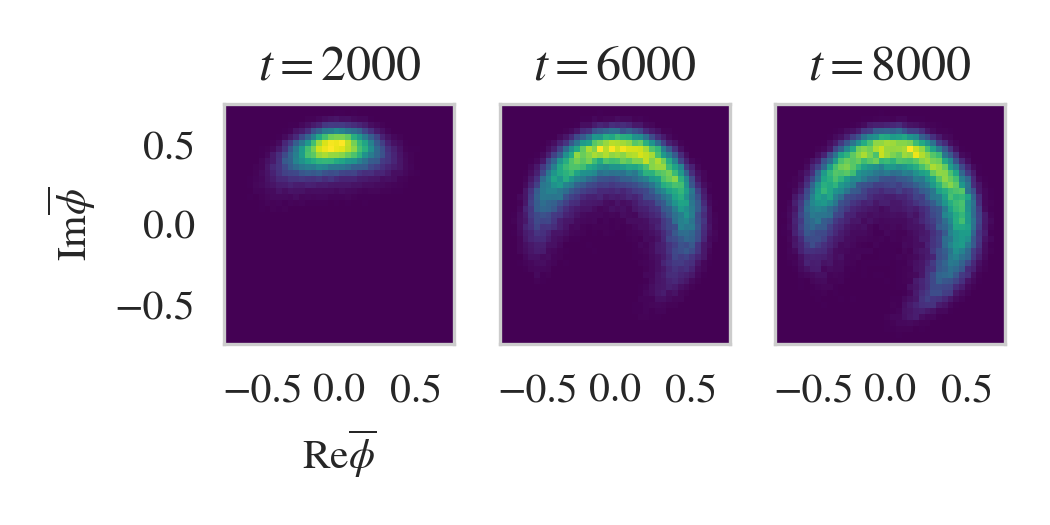}
        \label{fig:gne_heatmap}
    }}
    \!
    \subfloat[\centering  ]{{
        \includegraphics[width=\linewidth]{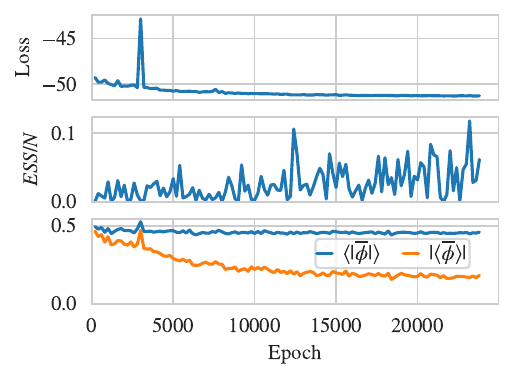}
        \label{fig:gne_dkl}
    }}
    \caption{
        For the target complex scalar field theory, magnetization distribution (a) and training dynamics (b) 
        demonstrating subspace collapse under reverse KL self-training for a model constructed with a Gaussian prior and non-equivariant flow.
        $\langle | \overline{\phi} | \rangle$ remains constant at the radius of the ring, while $| \langle \overline{\phi} \rangle |$ decreases due to cancellation as support spreads around the ring.
        Though the range of phases over which the model has support increases during training, a finite range of phases is left with near-zero support/sampling.
    }
    \label{fig:gne}
\end{figure}

Figure~\ref{fig:gne} illustrates the typical result of the baseline architecture and training applied to a target in the U(1)-broken phase.
The model initially collapses onto a small subspace of the theory's extended mode.
Note that, in contrast to the multimodal case, the collapsed-upon subspace has boundaries within high-density regions of the target.
Under additional training, the model spreads to cover more of the mode; however, unlike in the multimodal case, we have been unable to achieve even coverage of the full mode even with extended training.
This behavior is consistent with the topological constraints discussed in Sec.~\ref{sec:topo-matching}, which imply it is difficult for the flow to learn to puncture the spherical Gaussian prior and move density away from $\overline{\phi} \approx 0$. 
Instead, it learns the suboptimal solution of stretching the spherical Gaussian out into an approximation of a ring. Tunneling to the (topologically distinct) global minimum would require a substantial reconfiguration of the model.

\subsection{Architectural approaches}

In this section, we test architectural remedies to subspace collapse in complex scalar field theory.
Other than the architectural augmentations described below, the models are structured as the baseline.
In all cases, the baseline reverse KL self-training is employed.

\subsubsection{Equivariant flows}

\emph{Symmetrized models:}
Section~\ref{sec:arch_methods} presents a method to symmetrize an arbitrary transformation over a finite group. 
Here, we apply it to symmetrize the baseline flow architecture over a global $Z_N$ as
\begin{equation}
    \phi_A\rightarrow \frac{1}{N}\sum_{n=1}^N (z^*)^n h(z^n \phi_A | z^n \phi_F)
\end{equation}
where $z = \exp[2 \pi i / N]$ is an $N$th root of unity, and $h$ is the base transformation kernel defined in Eq.~\eqref{eq:complex-affine}.
Equivariance under global U(1) is recovered as $N \rightarrow \infty$. 
However, as we show below, enforcing even a small subgroup of the global U(1) may prevent subspace collapse.

\emph{Canonicalized models:}
Unlike with symmetrization, the canonicalization approach described in Sec.~\ref{sec:canonical} can encode exact equivariance under continuous symmetries.
For this application, in the notation of Sec.~\ref{sec:canonical}, we choose $c(\phi_F) = (\overline{\phi}_F)^* / | \overline{\phi}_F |$, i.e.~the conjugate of the phase part of the magnetization of the frozen sites, and $\overline{c}(\phi_F) = c(\phi_F)^*$. Using this to canonicalize an arbitrary transformation $\phi'_A = h(\phi_A|\phi_F)$ yields
\begin{equation}
    \phi'_A = \frac{\overline{\phi}_F}{|\overline{\phi}_F|} h \left( \frac{(\overline{\phi}_F)^*}{|\overline{\phi}_F|} \phi_A ~ \middle| ~ \frac{(\overline{\phi}_F)^*}{|\overline{\phi}_F|} \phi_F \right) ~ .
\end{equation}
Note that, as with the real version presented in Sec.~\ref{sec:real-arch-results}, this modified flow is not strictly a diffeomorphism due to discontinuities at $\phi_F = 0$. As in that case, topological restrictions do not apply, but the discontinuities present separate difficulties as noted below.

\subsubsection{Topology matching}
\label{sec:cx-topo-matching}

To construct a prior distribution which matches the ring-shaped topology of the target distribution, we employ essentially the same approach as for the real case.
For this application, we construct a mixture of width-$
\sigma$ complex Gaussians smeared over a ring of radius $R$.
Each sample $z(x)$ from this distribution is obtained as $z(x) = z_0(x) + R e^{i \theta}$ where $\theta$ is drawn uniformly from $[0, 2\pi)$, and $z_0(x)$ is drawn from $r_\sigma(z)$, in this case an independent width-$\sigma$ complex Gaussian on each site $x$.
The marginalization over $\theta$ necessary to compute the mixture density $r_\text{ring}$ can be performed analytically, obtaining
\begin{align}
\begin{split}
    r_\text{ring}(z) &= \frac{1}{2\pi}\int d\theta\prod_x 
    r_\sigma\left( z(x) - R e^{i \theta} \right)\\
    &=I_0\left(\frac{VR}{\sigma^2}|\overline{z}| \right) \prod_x \frac{1}{2\pi\sigma^2}\exp\left( -\frac{|z(x)|^2+R^2}{2\sigma^2} \right),
\end{split}
\label{eq:ring_prior}
\end{align}
where $I_0$ is a modified Bessel function, $V$ is the number of lattice sites, and $\overline{z} = \sum_x z(x)/V$.
Fig.~\ref{fig:priors} shows an example of the magnetization density of this distribution.
Setting $R=0$ recovers $r_\sigma(z)$.

As with the bimodal prior in the real case, the prior ring thickness $\sigma/R$ cannot be trivially rescaled by affine architectures. We treat it as a hyperparameter.
Similar to the real case, collapse may still occur if training fails to match the prior's topology onto the target, and instead compresses the full prior ring onto a segment of the target ring.
In this application, we observe that using thin rings makes subspace collapse less likely, but can result in undersampling near $\overline{\phi} \approx 0$. We treat this by increasing the ring width slowly over the course of training. 
Specifically, we begin training with with $R=\sigma=1$, then
every 200 epochs, increase $\sigma$ in steps of 0.5 until $\sigma = 3.5$, at which point the prior is held fixed.
Simultaneously, we decrease the learning rate by a factor 0.9 every 200 epochs; this smoother and faster schedule versus the baseline helps prevent collapse in this case.

\begin{figure}
    \centering
    \includegraphics[width=\linewidth]{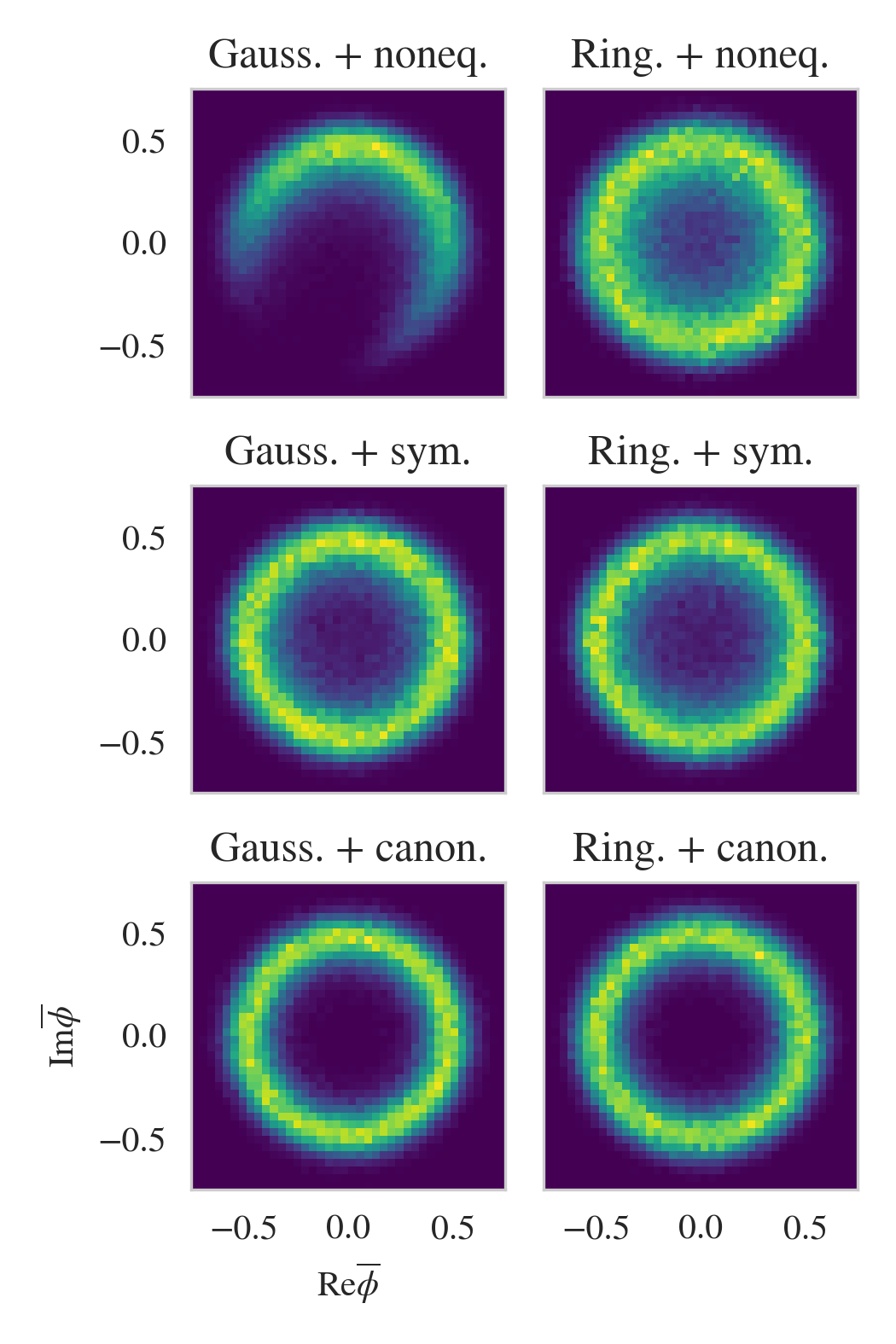}
    \caption{Complex magnetization distributions for models trained targeting complex scalar field theory with $m^2=-4$ and $\lambda=3$ on an $8 \times 8$ lattice, for flows constructed with nonequivariant (top), symmetrized (middle), and canonically equivariant (bottom) architectures and either ring prior (left) or Gaussian prior (right). Symmetrization is done over $Z_4$.}
    \label{fig:arch_maps}
\end{figure}

Figure~\ref{fig:arch_maps} illustrates the qualitative success of using equivariant architectures and/or topology-matched priors in avoiding subspace collapse.
Versus the baseline, all of the architectural modifications produce model distributions with the same ring-shaped structure as the target.
More quantitatively, the (exact, in some cases) phase independence of the phase-conditioned forwards KL divergences in Fig.~\ref{fig:arch_dkl} demonstrates that the theory's U(1) global symmetry is well-respected in all cases but the baseline.
Also shown in Fig.~\ref{fig:arch_dkl} are radius-conditioned KL divergences. Overall, the modified architectures demonstrate significantly improved performance over the baseline model in the high-density region near $|\overline{\phi}| \approx 0.5$. 
The topology-matched nonequivariant model is an improvement globally.
However, performance for the equivariant models at $\overline{\phi} \approx 0$ is worse than the baseline.
This may be due to insensitivity to undersampled regions in reverse KL self-training, discontinuities for the canonicalized flows, and/or other effects.

\subsubsection{Mixture models}

\begin{figure}
    \centering
    \includegraphics[width=\linewidth]{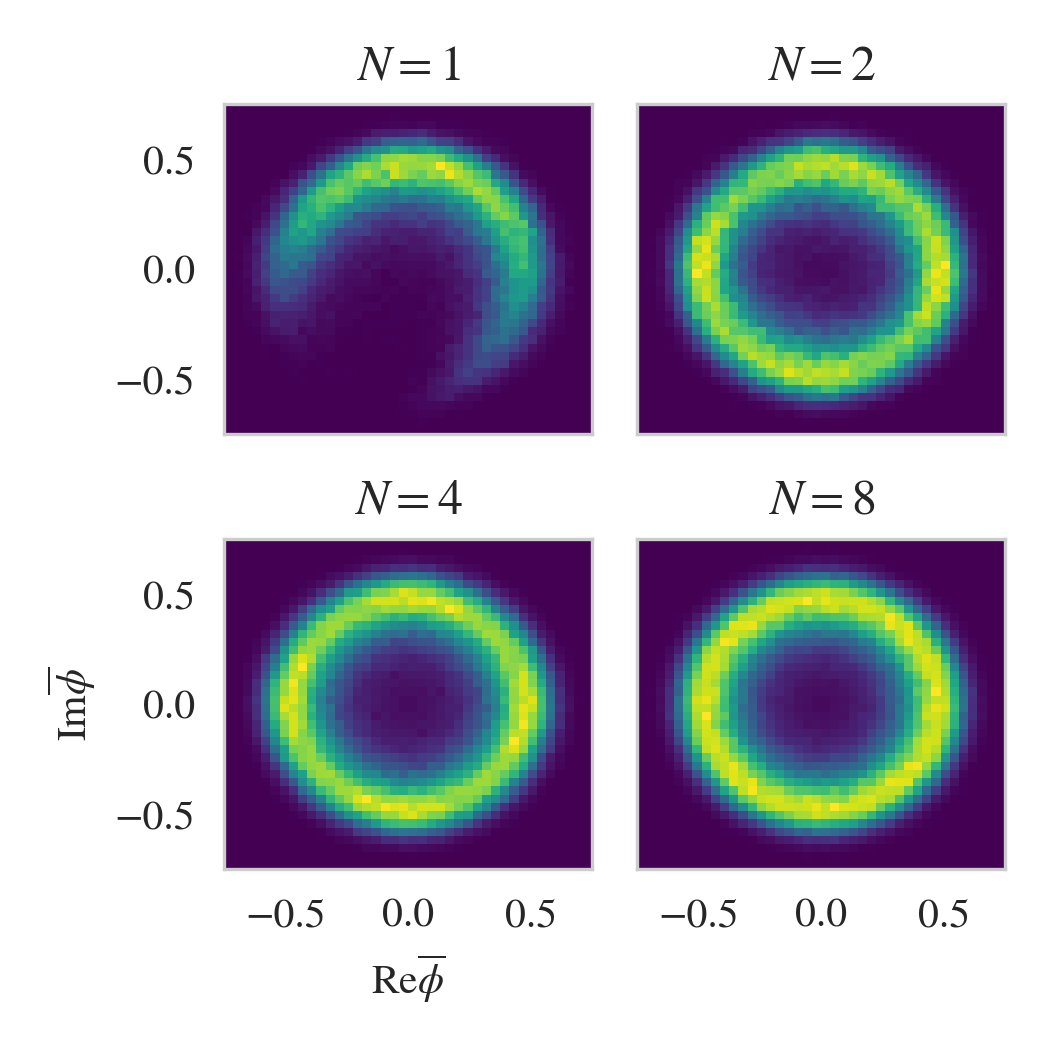}
    \caption{For the target complex scalar field theory, magnetization distributions under mixture models constructed by symmetrizing a subspace-collapsed nonequivariant model over $Z_N$, demonstrating restoration of U(1) global symmetry with increasing $N$.}
    \label{fig:epf_heatmaps}
\end{figure}

\begin{figure}
    \centering
    \includegraphics[width=\linewidth]{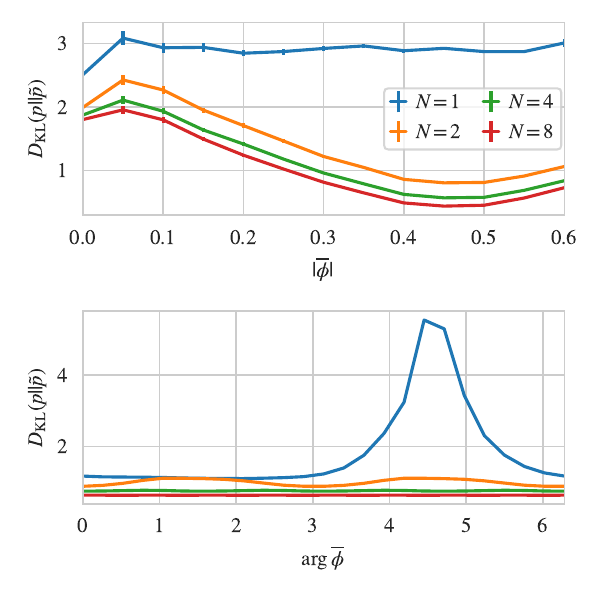}
    \caption{For the target complex scalar field theory, radius- (top) and phase-conditioned (bottom) forwards KL divergences for mixture models constructed by symmetrizing a subspace-collapsed nonequivariant model over $Z_N$, computed using 128000 target-distributed samples generated with HMC.
    }
    \label{fig:expostfacto}
\end{figure}

Lastly, we apply the symmetrized single-model mixture construction of Sec.~\ref{sec:symm-mixtures} to re-introduce the U(1) symmetry broken by subspace collapse in the baseline model. 
Specifically, we symmetrize over a $Z_N$ subgroup using Eq.~\eqref{eqn:symm-mix-model} in the form
\begin{equation}
\tilde{p}_{\text{mix}}(\phi) = \frac{1}{N} \sum_{n=0}^{N-1} \tilde{p}(z^n \phi),
\end{equation}
where $z = \exp[2 \pi i / N]$.
Figs.~\ref{fig:epf_heatmaps} and \ref{fig:expostfacto} show the resulting densities and phase- and radius-conditioned KL divergences for a range of $N$.
In this case, a subgroup as small as $Z_2$ is sufficient to alleviate the worst effects of mode collapse, though the KL divergence remains weakly phase-dependent for small $N$.  
However, mismodeling of the inner tails is not completely alleviated, as expected given that the symmetrization does not modify the (marginal) radial magnetization density.
Considering also the symmetrized equivariant architectures above, as a general tool, symmetrization with respect to only a small subgroup of U(1) can be sufficient to restore the mode structure topology.

\subsection{Training approaches}
\label{sec:extended-training-results}

Here, we present results of applying similar training methods as tested for real scalar field theory in Sec.~\ref{sec:multimodal-comparison}, namely forwards KL training, adiabatic retraining, and flow distance regularization, suitably adapted to the complex case.

\subsubsection{Forwards KL training}
\label{sec:extended-fkl}

As discussed in Sec.~\ref{sec:forwards-kl-training}, sampling training data from a fully symmetrized mixture is intractable for a continuous symmetry.
However, naive data augmentation may still be applied with the full symmetry group.
In this case, forwards KL self-training is performed via random phase augmentation to model samples using Eq.~\eqref{eq:fwd_dkl} with
\begin{equation}\begin{aligned}
    D_{\mathrm{KL}}^{\text{fwd,rw}}(p||\tilde{p})\approx \frac{1}{N} \sum_i \frac{p(\phi_i)}{\tilde p(\phi_i)} \log \frac{p(\phi_i)}{\tilde{p}(e^{i\alpha_i}\phi_i)}, \quad (\phi_i \sim \tilde p)
    \label{eq:random_phase}
\end{aligned}\end{equation}
where $\alpha_i$ is sampled uniformly from $[0,2\pi)$ and the reweighting factors $p/\tilde{p}$ taken as independent of the model parameters in gradient computations.
We do not test training with data drawn from a mixture partially symmetrized over a $Z_N$ subgroup, but mixtures of this type are considered as architectures in their own right above.

As in Sec.~\ref{sec:multimodal-comparison}, we also consider forwards KL training with HMC-generated data to set a best-case baseline for the forwards KL self-training.
Rather than generating samples on-the-fly, we draw batches of 1024 configurations at random (with replacement) from an ensemble of $1.28 \times 10^5$.

\subsubsection{Adiabatic retraining \& flow-distance regularization}

\begin{figure*}
    \centering
            \subfloat[Forwards KL, $\phi\sim p$]{{
        \includegraphics[width=0.48\linewidth]{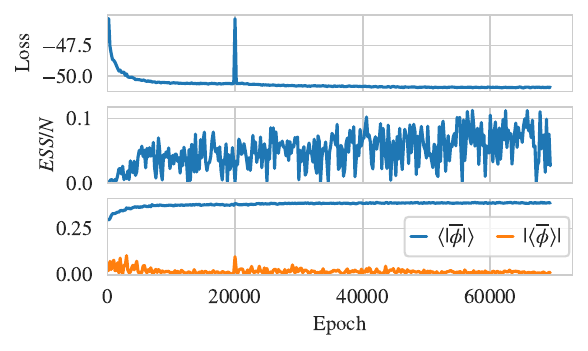}
    }}
    \!
        \subfloat[Forwards KL, $\phi\sim \tilde p_{\mathrm{mix}}$]{{
        \includegraphics[width=0.48\linewidth]{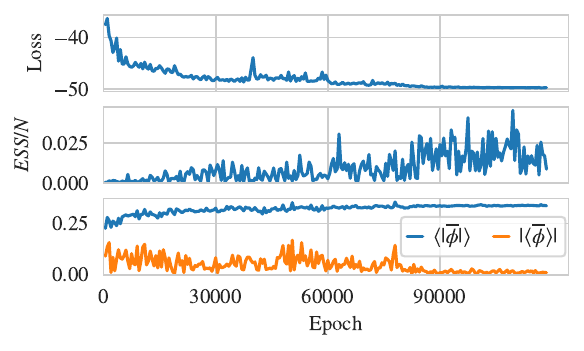}
    }}
    \!
    \subfloat[Flow distance]{{
        \includegraphics[width=0.48\linewidth]{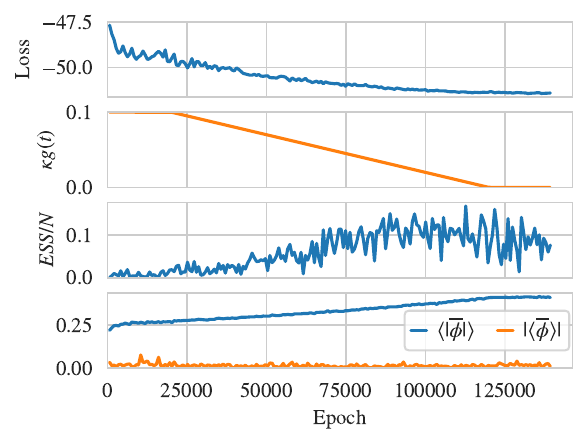}
    }}
    \!
    \subfloat[Adiabatic]{{
        \includegraphics[width=0.48\linewidth]{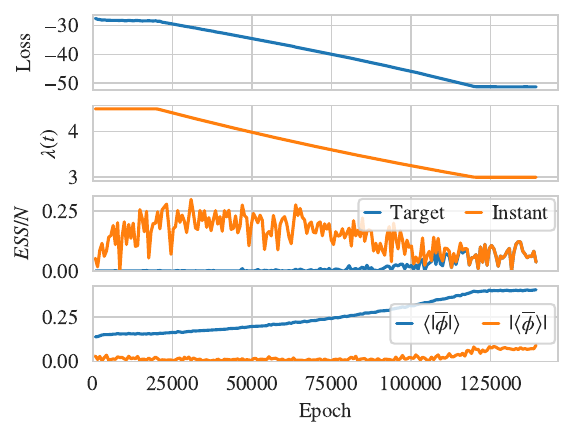}
        \label{fig:adia}
    }}
    \caption{
        For the target complex scalar field theory with $m^2=-4$ and $\lambda=3$ on an $8 \times 8$ lattice, training dynamics for the baseline architecture trained with (a) forwards KL training using HMC samples ($\phi\sim p$), (b) forwards KL self-training with naive data augmentation ($\phi\sim\tilde{p}_{\mathrm{mix}}$), (c) flow distance regularization, and (d) adiabatic retraining. In all cases, ``Loss'' is the unnormalized reverse KL divergence. For adiabatic retraining, the ESS is computed for $\lambda(t)$ (Instant) and $\lambda=3$ (Target).
    }
    \label{fig:training2}
\end{figure*}

For distance regularization, we add a regularizer that penalizes the absolute mean squared error versus the complex Gaussian prior. 
The loss is  the complex generalization of Eq.~\eqref{eqn:flow-distance-loss}.
The strength of this regularizer is decreased linearly over training from $\kappa=0.1$ at $t_1=20000$ epochs to $\kappa=0$ at $t_2=120000$, with segments of training at constant $\kappa$ before and after this linear decrease. This schedule is depicted in Fig.~\ref{fig:training2}.

For adiabatic retraining (Fig.~\ref{fig:training2}), we vary $\lambda=4.5\rightarrow 3.0$ during training using an exponential interpolation:
\begin{equation}
\lambda(t) = \lambda_ie^{-(t-t_1)/\tau}, \quad t_1 < t < t_2
\label{eqn:adiabatic-retraining-inst-action-complex}
\end{equation}
with $\tau=(t_2-t_1)/\ln(\lambda_i/\lambda_f)$, $\lambda_i=4.5, \lambda_f=3$, $t_1=20000$, and $t_2=120000$.
$\lambda$ is held constant for epochs before $t_1$ and after $t_2$.
This schedule is depicted in Fig.~\ref{fig:training2}.

\begin{figure}
    \centering
    \includegraphics[width=\linewidth]{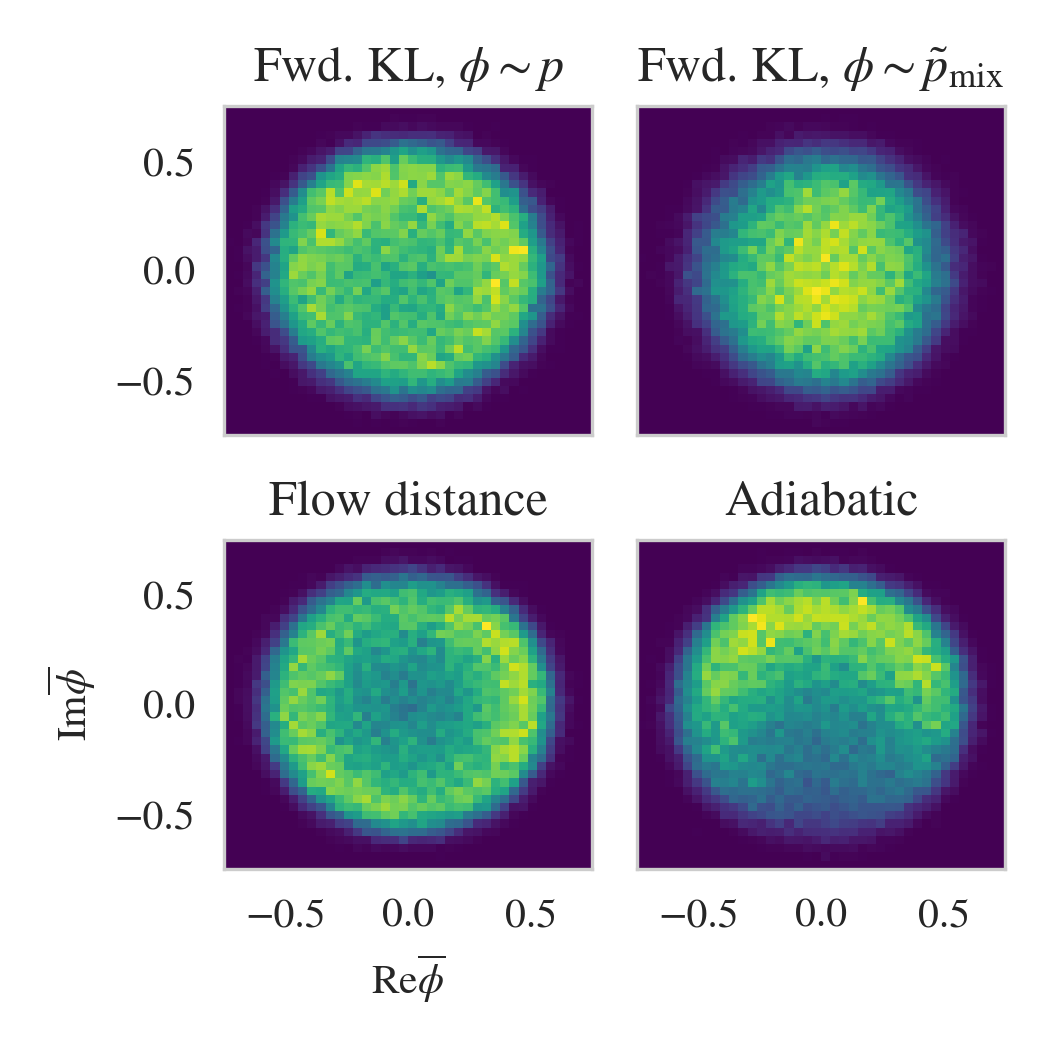}
    \caption{Complex magnetization distributions for models trained targeting the same complex scalar field theory, with $m^2=-4$ and $\lambda=3$ on an $8 \times 8$ lattice, using the baseline architecture trained with forwards KL with HMC samples ($\phi \sim p$), forwards KL self-training with naive data augmentation ($\phi \sim \tilde{p}_\mathrm{mix}$), flow distance regularization, and adiabatic retraining.}
    \label{fig:training1}
\end{figure}

\begin{figure*}
    \centering
    \includegraphics[width=\linewidth]{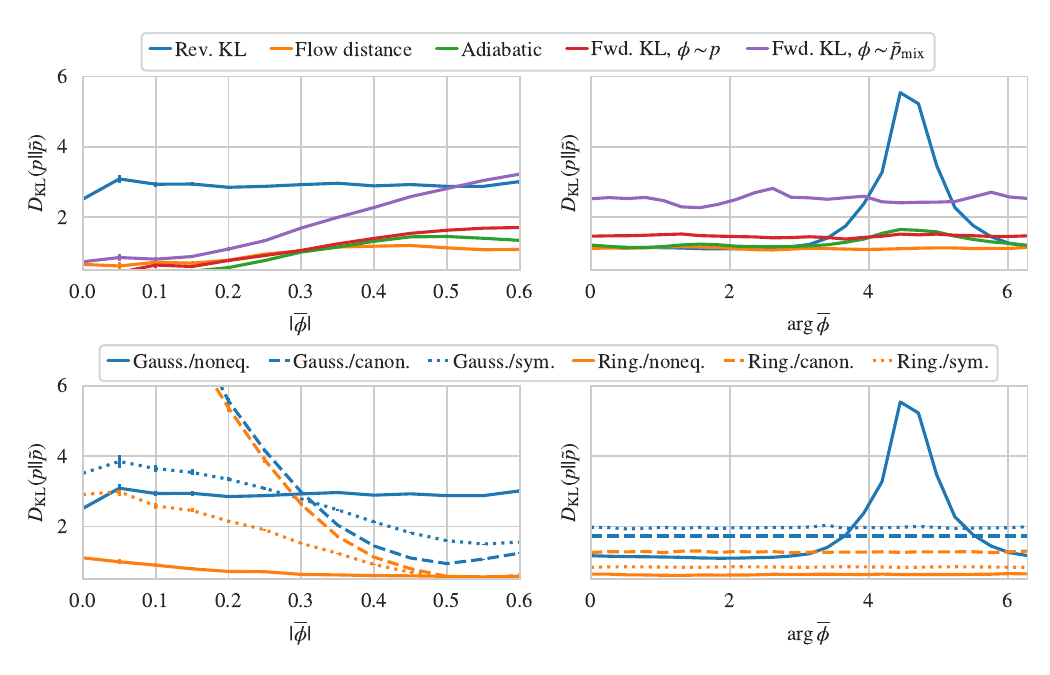}
    \caption{
        Performance metrics for models trained targeting complex scalar field theory with $m^2=-4$ and $\lambda=3$ on an $8 \times 8$ lattice. Top: radius-conditioned (left) and phase-conditioned (right) forwards KL divergences for the training-based methods.
        The baseline scheme is Rev.~KL.
        Bottom: analogous plots for the architecture-based methods.
        Here, each plot represents a different combination of prior (Gauss.: spherical Gaussian, Ring.: Eq.~\eqref{eq:ring_prior}) and flow architecture (noneq.: nonequivariant, canon.: canonicalized equivariant, sym.: $Z_4$-symmetrized equivariant). The baseline scheme is Gauss./noneq.. All plots were computed using 128,000 target-distributed samples (generated with HMC) with bootstrapped errors.
    }
    \label{fig:arch_dkl}
\end{figure*}

Figure~\ref{fig:training2} illustrates training dynamics of example applications of these four methods to the baseline architecture, and Fig.~\ref{fig:training1} shows the magnetization distributions of the resulting models.
As clear from comparing with Fig.~\ref{fig:gne_heatmap}, these methods prevent the collapse that occurs for the baseline: $\langle |\overline\phi| \rangle$ increases during training for all methods, indicating density being pushed outwards to match onto the extended target mode. Meanwhile, $|\langle \overline\phi \rangle|$ remains near zero, suggesting subspace collapse is not as severe as in baseline training.
This manifests as a better approximation of the global U(1) symmetry, as quantified in the phase-conditioned metrics of Fig.~\ref{fig:arch_dkl}.

However, in all cases, there is an obvious excess in density near $\overline{\phi} \approx 0$ versus the target (cf.~Figs.~\ref{fig:training1} and \ref{fig:priors}).
In the case of forwards KL self-training, the magnetization distribution lacks even a suggestion of the central ``hole''.
This failure to capture the target's structure occurs even though the training dataset has good coverage across the full symmetry orbit by construction.
Meanwhile, flow-distance regularization and adiabatic retraining produce ringlike but noticeably asymmetric models.
In the adiabatic retraining case, the evolution of $\vev{|\phi|}$ and $|\vev{\phi}|$ indicates that this asymmetry arises as the target becomes increasingly annular: the initially symmetric model begins to ``roll off'' to one side.
We have been unable to avoid this behavior with slower-moving schedules.

Taken together, these observations suggest that the topological limitations of the architecture, as discussed in Sec.~\ref{sec:topo-matching}, are a greater obstacle than in the bimodal real example.
Architectural improvements over the baseline (as outlined in the previous section) are practically required.
Nevertheless, this is still an improvement upon the reverse KL baseline.
As quantified by the radius-conditioned metrics of Fig.~\ref{fig:arch_dkl}, near $\overline{\phi} \approx 0$, the severe undersampling by the baseline model results in worse model quality than the oversampling by these models with modified training.

\subsection{Comparison}

We have demonstrated training and architecture-based methods for avoiding subspace collapse in a theory with an extended mode. Both classes of methods produce models which better respect the global U(1) symmetry of the target theory than the naive baseline approach.
However, the architecture-based methods 
yield qualitatively more accurate models of the mode structure.
Our results suggest this can be attributed to topological constraints: the baseline flow architecture is ill-equipped to change the topology of a spherical Gaussian prior. Using a topology-matched prior, or introducing equivariance, serves to alleviate this constraint.

We observe that the different architecture methods are not necessarily synergistic when combined. 
For example, among all of the models considered here, only the topology-matched nonequivariant model has a nontrivial ESS ($\approx$ 20\%, computed on target samples per the discussion of Sec.~\ref{sec:sampling}).
This means that, in this example, combining topology matching with equivariance produces worse-quality models than topology matching alone.
Of course, caution is required in comparing the resulting architectures with each other and the baseline, since the neural networks play different roles in each kind of transformation.
This precludes a truly fair comparison, even when the neural network structures are matched.
Tuning of architecture and training hyperparameters may provide better modeling and present a different hierarchy.
Finally, we note that, while not explored here, combination with training-based methods like distance regularization may help with problems like the inner tail undersampling seen for equivariant models.

\section{Comparison between models for real and complex scalar field theories}

In both the real and complex case, all the methods demonstrated are able to alleviate mode collapse, a qualitative improvement over the training baseline---with varying degrees of success otherwise.
Considering together the experiments of this section and Sec.~\ref{sec:multimodal-comparison}, it is clear that modeling the low-density inner tail region near $\overline{\phi} = 0$ is a challenge in general.
For the complex theory, topological constraints on the baseline architecture result in severe oversampling of inner tail regions, which is only partially corrected using specialized training schemes.
Meanwhile, the models undersample the inner tail in most examples for the real theory and for most of the architecture-based approaches in the complex case.
The common factor in both cases is the scarcity of training data: because these inner regions make up a small fraction of the total probability mass, their effect on the loss is poorly resolved.
While this represents a mismodeling of only a small fraction of the target distribution, the local severity of this mismodeling can lead to very poor performance when these models are employed for sampling.
This effect is explored in the next section.

\section{Sampling using multimodal models in real scalar field theory}
\label{sec:sampling}

In this section, we examine the utility of the models constructed in Sec.~\ref{sec:multimodal-comparison} in generating target-distributed samples for real scalar field theory using different sampling schemes.
We find that, when used with flow-based MCMC, these models generically produce pathologically inefficient samplers due to undersampling in the inner tails.
The way this issue arises is subtle, so we note its effects in various direct checks of sampler quality and discuss how it can be better diagnosed using target-distributed validation data.
To demonstrate its severity, we perform a case study of the asymptotic performance of flow-based MCMC using an example model.
Finally, we demonstrate that the problems of flow-based MCMC can be regulated by augmenting with HMC to produce an efficient composite MCMC sampler.

\subsection{Flow-based MCMC}

\begin{figure}
    \centering
    \subfloat[\centering  ]{{
        \includegraphics[width=\linewidth]{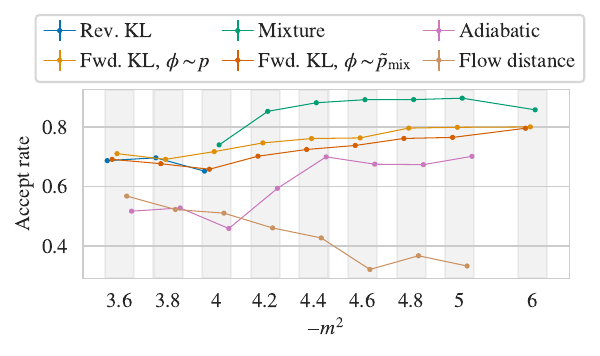}
        \label{fig:cf_ar_M2}
    }}
    \!
    \subfloat[\centering  ]{{
        \includegraphics[width=\linewidth]{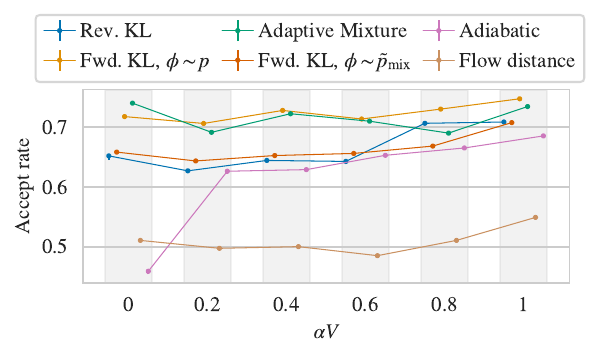}
        \label{fig:cf_ar_alpha}
    }}
    \caption{
        MCMC acceptance rate for models constructed using the different approaches described in the text, for real scalar field theory with \protect\subref{fig:cf_ar_M2} fixed $\alpha=0$ and varying $m^2$ and \protect\subref{fig:cf_ar_alpha} fixed $m^2=-4$ and varying $\alpha$.
        All estimates are made on $10^6$ model-distributed samples, with errors computed by taking bootstrap draws from fixed datasets and applying independence Metropolis within each.
        Error bars are smaller than the markers for all points.
        Additional variance due to seed dependence is not quantified.
        These data should not be interpreted as ranking the methods; see the discussion in Sec.~\ref{sec:results_comparison}.
    }
    \label{fig:cf_ar}
\end{figure}

Figure~\ref{fig:cf_ar} shows the flow-based MCMC acceptance rate for models constructed using the different approaches presented in this work, modeling the variance in acceptance rate for different model-generated datasets by applying Independence Metropolis (as described in Sec.~\ref{sec:flow-review}) to different bootstrap draws from a fixed sample of $10^6$ configurations.
The resulting estimates have little error, and all models yield samplers with measured acceptance rates $\sim 30\% - 90\%$.
This metric gives a similiar picture as the forwards KL divergence (cf.~Fig.~\ref{fig:cf_fkl}).

\begin{figure}
    \centering
    \includegraphics[width=\linewidth]{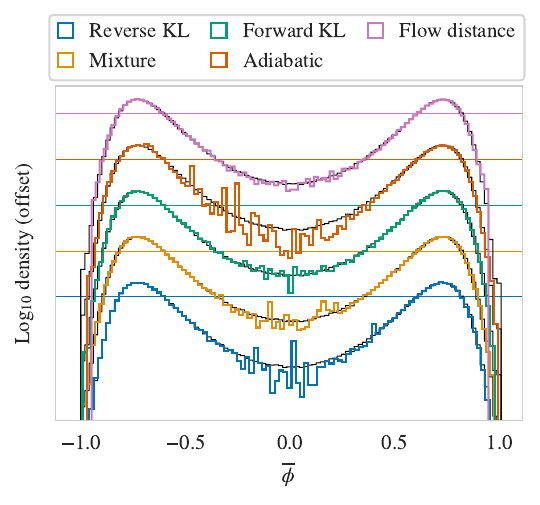}
    \caption{
        Distribution of $\overline{\phi}$ for ensembles generated using flow-based MCMC with models constructed using the different approaches described in the text, all for real scalar field theory with $m^2=-4$ and $\alpha=0$.
        The log distributions are offset for clarity; the horizontal line of the same color indicates density 1 for the corresponding distribution, and lines are separated by factors of 10.
        The true distribution of $\overline{\phi}$, estimated using AHMC, is shown in black behind each curve.
    }
    \label{fig:cf_mcmc_hist_a0}
\end{figure}

\begin{figure}
    \centering
    \includegraphics[width=\linewidth]{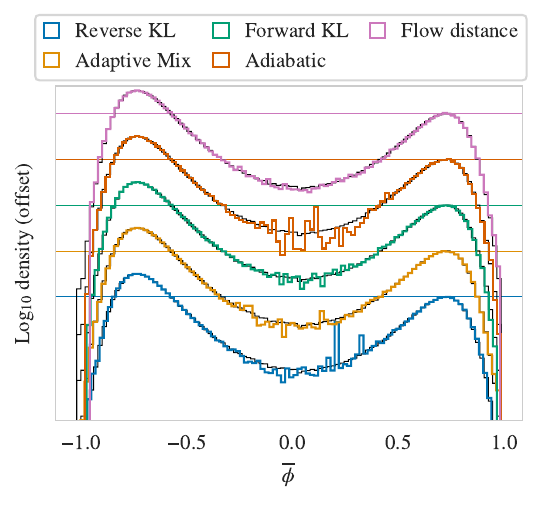}
    \caption{
        As in Fig.~\ref{fig:cf_mcmc_hist_a0}, but for $m^2=-4$ and $\alpha = 0.8/V$.
    }
    \label{fig:cf_mcmc_hist_a8}
\end{figure}

Figures~\ref{fig:cf_mcmc_hist_a0} and \ref{fig:cf_mcmc_hist_a8} show histograms of $\overline{\phi}$ for $m^2=-4$ and $\alpha = 0$ and $0.8/V$, respectively, with each constructed from $10^6$ target-distributed samples generated using flow-based MCMC with each model.
All samplers correctly reproduce the distribution under the target, especially around the peaks.
However, we generically observe large fluctuations around the target density in the inner tails; these fluctuations are not due to statistical limitations in estimating histogram bin counts (compare with the smooth outer tails at lower densities), but rather due to large numbers of replicated samples near $\overline{\phi} \approx 0$ arising from long rejection runs triggered by high-weight configurations.
This reflects a large uncertainty in the estimated density near $\overline{\phi} \approx 0$ and is a symptom of the problems with these samplers.

\begin{figure}
    \centering
    \subfloat[\centering  ]{{
        \includegraphics[width=\linewidth]{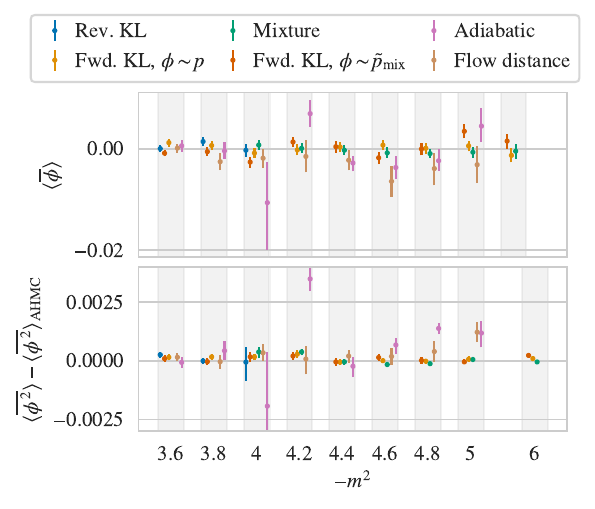}
        \label{fig:cf_obs_M2}
    }}
    \!
    \subfloat[\centering  ]{{
        \includegraphics[width=\linewidth]{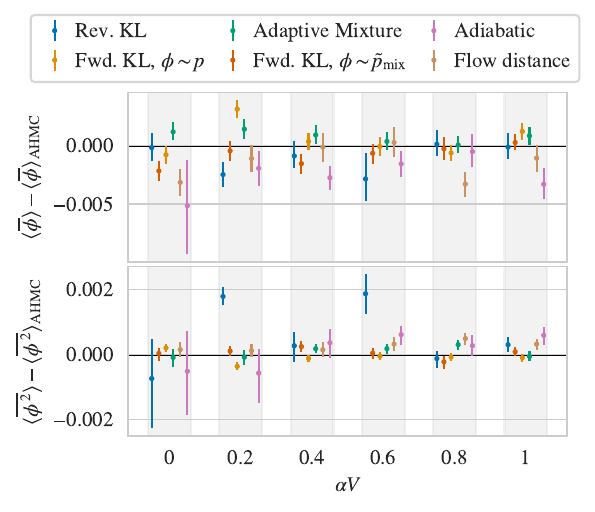}
        \label{fig:cf_obs_alpha}
    }}
    \caption{
        Estimates of $\langle \overline{\phi} \rangle$ and $\langle \overline{\phi^2} \rangle$,  computed with flow-based MCMC using models obtained using the different approaches described in the text, for real scalar field theory with \protect\subref{fig:cf_obs_M2} fixed $\alpha=0$ varying $m^2$ and \protect\subref{fig:cf_obs_alpha} fixed $m^2=-4$ varying $\alpha$.
        Observables are computed by drawing $10^6$ configs from each model, applying independence Metropolis to construct a Markov chain, binning by $10^4$, and bootstrapping to estimate the error.
        Markers denote the median over bootstraps, and error bars extend to the 15.9th ($-1\sigma$) and 89.1th ($+1\sigma$) percentiles.
        Except for $\langle \overline{\phi} \rangle$ when $\alpha = 0$, which is zero by symmetry, ground-truth values computed on ensembles of $10^6$ samples generated with AHMC have been subtracted from each observable, without accounting for the error in the AHMC estimate.
        Estimates that are consistent with zero within error indicate agreement with ground truth.
        Additional variance due to seed dependence in training and sampling is not quantified.
    }
    \label{fig:cf_obs}
\end{figure}

Figure~\ref{fig:cf_obs} shows the results of using the various different models with flow-based MCMC to compute the observables $\vev{\overline{\phi}}_p$ and $\vev{\overline{\phi^2}}_p$, compared against ``ground truth'' computed using AHMC.
We apply Independence Metropolis to a sample of $10^6$ configurations drawn from each model to generate an ensemble of target-distributed configurations. For both HMC samples and flow samples, we first bin the data to remove autocorrelations\footnote{For flow-based samplers we use bins of size $10^4$. For HMC we use bins of size $10^3$ to avoid any irregularities that might arise because our AHMC dataset is composed of 100 ensembles of $10^4$ configurations. Means are stable and error estimates approach an asymptotic value under additional binning until finite-sample effects arise.} then bootstrap to estimate uncertainties.
Most observables computed using flow-based MCMC agree within errors with the AHMC results (or the analytic value $\vev{\overline{\phi}}=0$ when $\alpha=0$) and no systematic bias is apparent.

However, a few points in Fig.~\ref{fig:cf_obs} lie many standard deviations away from the true values, indicating their errors are severely underestimated.
Additionally, we observe a larger-than-expected range of estimated errors given the apparently comparable quality of these models as measured by the forwards KL divergence and acceptance rate. 
We also obtain significantly different error estimates when analyzing different samples of $10^6$ configurations drawn from the same model, indicating non-convergence of variance estimates even for these large sample sizes.
We trace these effects to the presence or absence in each sample set of high-weight configurations which trigger long rejection streaks.
This implies that the Markov chains have not entered the smoothly converging regime even at length $10^6$, and thus that flow-based MCMC converges pathologically slowly when used with these models.

Further cause for concern is the large variances of model-sample ESS estimates even when computed from samples of $10^6$ configurations. Drawing different samples from the same models, we observe significantly different estimates, again indicating non-convergence.
We trace this effect to the same high-weight configurations that break convergence of error estimations; we observe individual configurations that change the ESS by $O(1)$ factors, as discussed further in the next subsection.

\begin{figure}
    \centering
    \subfloat[\centering  ]{{
        \includegraphics[width=\linewidth]{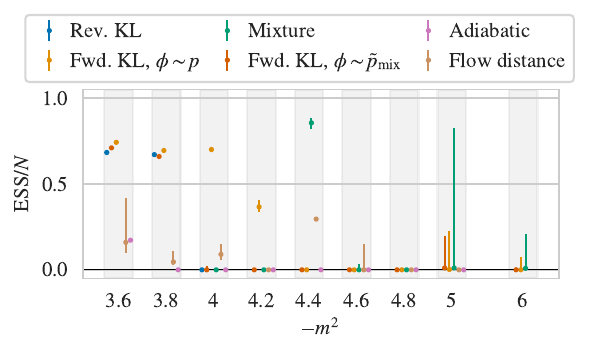}
        \label{fig:cf_ess_M2}
    }}
    \!
    \subfloat[\centering  ]{{
        \includegraphics[width=\linewidth]{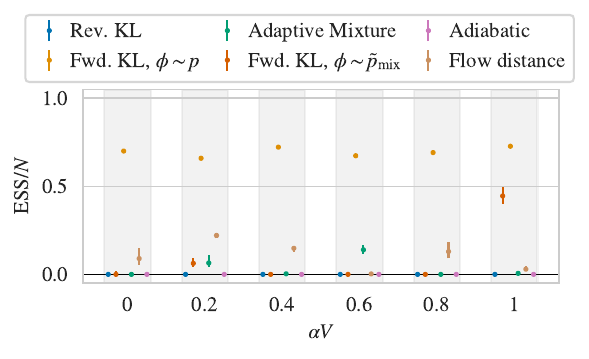}
        \label{fig:cf_ess_alpha}
    }}
    \caption{
        ESS per configuration estimated using target-distributed samples for models constructed using the different approaches described in the text, for real scalar field theory with \protect\subref{fig:cf_ess_M2} fixed $\alpha=0$ and varying $m^2$ and \protect\subref{fig:cf_ess_alpha} fixed $m^2=-4$ and varying $\alpha$.
        All estimates are made on shared ensembles of $10^6$ samples, and errors are computed by bootstrapping.
        Markers denote the median over bootstraps, and error bars extend to the 15.9th ($-1\sigma$) and 89.1th ($+1\sigma$) percentiles.
        Additional variance due to seed dependence in training and sampling is not quantified.
    }
    \label{fig:cf_ess}
\end{figure}

To circumvent finite-sample limitations, we use a target-sample estimator of the asymptotic ESS per configuration,
\begin{equation}\begin{aligned}
    \text{(ESS)}/N 
    & = \frac{1}{ \int d\phi ~ p(\phi) \frac{\tilde{p}(\phi)}{p(\phi)} w(\phi)^2 } 
    \\ &= \frac{1}{ \int d\phi ~ p(\phi) w(\phi) } 
    \approx \frac{ N^2 }{ \left( \sum_i w_i \right) \left( \sum_i \frac{1}{w_i} \right) } \; ,
\label{eqn:asymp-target-ess}
\end{aligned}\end{equation}
where $\phi_i \sim p$ and the estimator incorporates the stochastic estimator of Eq.~\eqref{eqn:stochastic-Z} to allow use with unnormalized reweighting factors.
The resulting ESS estimates for the different models are shown in Fig.~\ref{fig:cf_ess}, measured using the same shared ensembles of $10^6$ configurations of target-distributed samples used for the forwards KL divergence estimates in the previous section.
We observe that $\text{ESS}/N \approx 0$ for most of the models.
This should not be read as a deficiency of the ESS as a metric, but rather as a direct indication that these models will give poor-quality results when used with reweighting or direct resampling.
As motivated further below, this also diagnoses the unsuitability of these models for use with flow-based MCMC.

The high variance in some of these target-sample ESS estimates, as well as the analysis in the next subsection, suggests that some or all of these estimates may still be poorly converged.
However, some features of Fig.~\ref{fig:cf_ess} merit discussion.
We see $\text{ESS}/N \approx 0$ for all models with $m^2 < -4.4$; the large errors on some models reflect cases where only a few high-weight samples cause the ESS to be small, similar to what is observed using the model-sample estimator.
In many cases, forwards KL training with AHMC data is able to produce models with nonzero ESS where the other schemes cannot.
A few models trained using other procedures have nontrivial ESS, but there is no pattern to this apparent success, and this could be due to the particular target samples used for these estimates.
In a few spot checks, we observe strong seed dependence; for example, in Fig.~\ref{fig:cf_ess_M2} the model trained with AHMC data for $m^2 = -4.2$ has nontrivial ESS but, as discussed further below, the model studied in the next section (trained similarly, for the same parameters) has $\text{ESS} \approx 0$.

\subsection{Case study: flow-based MCMC}
\label{sec:case_study}

To better understand the apparent convergence problems observed in the previous section, we perform a careful study of one particular model.
We find that it exhibits exactly the issue discussed in Sec.~\ref{sec:challenges}, where a rarely sampled-from region of configuration space is also highly underweighted in the model relative to the target, leading to pathologically slow convergence.

The model in question was trained targeting ${m^2 = -4.2}$ and $\alpha = 0$ with forwards KL training using AHMC samples. 
The model discussed here is different than the one presented in Fig.~\ref{fig:cf_ess_M2}, but trained identically up to the random seed. Unlike for that model, we measure $\text{ESS} \approx 0$ using the target-sample estimator; meanwhile, the forwards KL divergence and MCMC acceptance rate are worse outside error for the present model, but closely comparable in magnitude.
This indicates that  different initial weights and different sets of training data can yield models with drastically reduced ESS without qualitatively affecting the forwards KL divergence or MCMC acceptance rate.
As explored below, this low observed value of the ESS signals severe problems with flow-based MCMC using this model which are not diagnosed by the other two metrics.

For all numerics in this section, we use a single dataset of $10^6$ model-distributed samples drawn from the flow model and another set of $1.28 \times 10^6$ target-distributed samples generated with AHMC.
We normalize all reweighting factors $w(\phi) = p(\phi)/\tilde{p}(\phi)$ such that $\vev{w}_{\tilde{p}} = 1$ using the stochastic $Z$ estimator of Eq.~\eqref{eqn:stochastic-Z} evaluated on the target samples.
Most of the ensuing discussion is phrased in terms of 
the distribution of reweighting factors under the model,
\begin{equation}\begin{aligned}
    \tilde{\rho}(w) &\equiv \left\langle \delta\left( w - \frac{p(\phi)}{\tilde{p}(\phi)} \right) \right\rangle_{\tilde{p}} = \int d\phi ~ \tilde{p}(\phi) \delta\left( w - \frac{p(\phi)}{\tilde{p}(\phi)} \right) \, .
\end{aligned}\end{equation}
which contains all observable-independent information about model quality.
Although correlations between reweighting factors and observables can further affect the quality of specific measurements, we are interested in model quality generically.

\begin{figure}
    \centering
    \includegraphics[width=\linewidth]{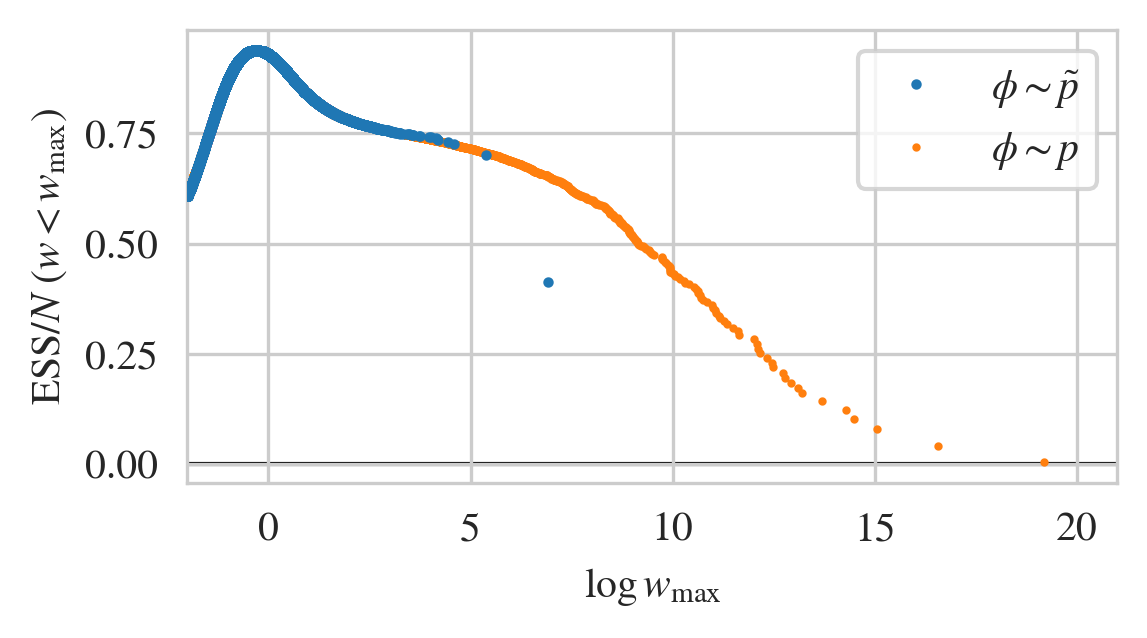}
    \caption{
        Running value of the ESS per configuration (for reweighting from the model to target) computed including the subset of configurations with $w < w_{\mathrm{max}}$, using either model (blue) or target (orange) samples and the corresponding estimators as described in the text.
    }
    \label{fig:case_study_ess_run}
\end{figure}

Figure~\ref{fig:case_study_ess_run} demonstrates that the large variation in the ESS estimates derived from different model sample datasets, as well as the low values of the ESS estimated using target samples, are due to configurations from the high-weight tail of $\tilde{\rho}(w)$.
The two sets of points in the figure correspond to model- and target-sample estimates of the ESS including all samples with $w \le w_{\text{max}}$, varying $w_{\text{max}}$.
Considering the model-sample estimator in isolation, as the cut $w_{\text{max}}$ is increased, the ESS appears to be converging to a value $\sim 0.7$ until two high-weight ``outlier'' configurations significantly reduce it; the highest-weight configuration reduces the ESS by a factor $\sim 2$.
Comparison with the target-sample estimator indicates that these apparent outlier configurations correspond to an unsampled high-weight tail of configurations.
The discrepancy between the highest-weight model-estimator point and the target-sample curve can be attributed simply to coarse sampling in the tail.
The target-sample estimator smoothly approaches a near-zero value, although the increasingly coarse sampling suggests the exact value estimated is still unreliable, with $\sim 5$ samples accounting for the difference between $0.15$ and $\approx 0$.

\begin{figure}
    \centering
    \includegraphics[width=\linewidth]{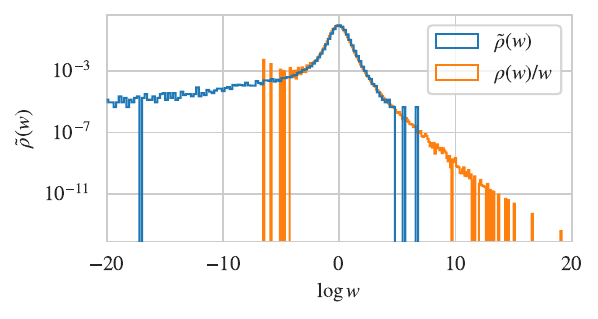}
    \caption{
        The distribution of reweighting factors under the model, $\tilde{\rho} = \vev{\delta(w - p/\tilde{p})}_{\tilde{p}}$, with the blue histogram made using $10^6$ model-distributed samples and orange histogram made using $1.28 \times 10^6$ target-distributed samples generated with AHMC then reweighted.
        The two isolated ``outlier'' points from Fig.~\ref{fig:case_study_ess_run} are visible at large $w$ in the blue histogram.
    }
    \label{fig:case_study_w_dists}
\end{figure}

Figure~\ref{fig:case_study_w_dists} confirms directly the existence of an undersampled high-weight tail in $\tilde{\rho}(w)$.
The histogram in blue corresponds to $\tilde{\rho}(w)$ estimated directly using model-distributed samples.
This distribution can also be obtained by reweighting the distribution of reweighting factors under the target ${\rho(w) = \vev{\delta(w-p/\tilde{p})}_p}$,
\begin{equation}\begin{aligned}
    \tilde{\rho}(w) &= \int d\phi ~ p(\phi) \frac{\tilde{p}(\phi)}{p(\phi)} \delta\left( w - \frac{p(\phi)}{\tilde{p}(\phi)} \right)
    = \rho(w) / w
\end{aligned}\end{equation}
which we apply to a histogram estimate of $\rho(w)$ made with target-distributed samples to produce the histogram in orange.
In Fig.~\ref{fig:case_study_w_dists}, the reweighted histogram satisfies $\int dw ~ \rho(w) \approx 1$, and matches well with the direct estimate where the two both have support; however, it also reveals a long tail of high-weight configurations which are not sampled in $10^6$ draws from the model.

\begin{figure*}
    \centering
    \includegraphics{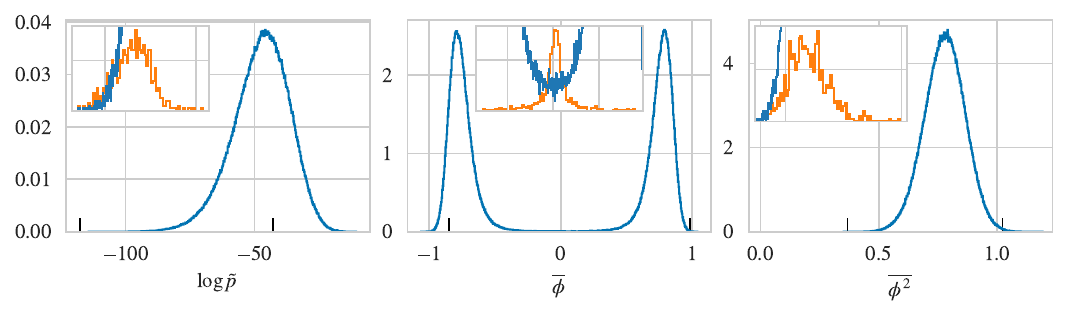}
    \caption{
        Distributions of $\log \tilde{p}$, $\overline{\phi}$, and $\overline{\phi^2}$ under the target $p$.
        The blue lines include all data, while the orange lines show the undersampled tail and include only configurations with $w > 5$, corresponding roughly to where the model distribution (with $10^6$ samples) cuts off in Fig.~\ref{fig:case_study_w_dists} and where the two ``outlier'' points sit in Fig.~\ref{fig:case_study_ess_run}.
        The additional spines on the x axes indicate the extent of the insets.
    }
    \label{fig:case_study_obs_densities}
\end{figure*}

Identifying the tail as all samples with $\log w > 5$ (corresponding roughly to the first outlier in Fig.~\ref{fig:case_study_ess_run} and Fig.~\ref{fig:case_study_w_dists}), Fig.~\ref{fig:case_study_obs_densities} examines where these high-weight configurations lie with respect to the distributions of other observables under the target.
The tail configurations come from regions of very low density in the model, accounting for their infrequent sampling.
As suggested by the rejection-induced finite-sample fluctuations near $\overline{\phi} \approx 0$ in Figs.~\ref{fig:cf_mcmc_hist_a0} and \ref{fig:cf_mcmc_hist_a8}, the high-weight configurations come from the inner tails of the bimodal distribution.
This corresponds to samples from the lower-valued tail of the distribution of $\overline{\phi^2}$, explaining the more severe effect on the variance of this observable in Fig.~\ref{fig:cf_obs}, and between different model-generated samples as discussed above, compared with $\overline{\phi}$ where duplications due to rejections enter near the midpoint of a symmetric distribution.

Due to its undersampling, we cannot tractably study the effect of the high-weight tail by directly sampling using flow-based MCMC.
Instead, we can use target-distributed samples to estimate the asymptotic properties of flow-based MCMC by considering the asymptotic behavior of the Markov process that defines it.
After equilibration, the current configuration of any Markov chain generated using flow-based MCMC is target-distributed, and proposed updates are independent model-distributed configurations.
Proposals are accepted with probability
\begin{equation}
    A(\phi \rightarrow \phi') \equiv \min \left( 1, \frac{w(\phi')}{w(\phi)} \right) \equiv A(w \rightarrow w')
\end{equation}
so the asymptotic acceptance rate is
\begin{equation}\begin{aligned}
    A \equiv A(p \rightarrow \tilde{p})
    &= \int d\phi d\phi' ~ p(\phi) A(\phi \rightarrow \phi') \tilde{p}(\phi')
    \\&= \int dw dw' ~ \rho(w) A(w \rightarrow w') \tilde{\rho}(w')
\end{aligned}\end{equation}
where in the second line we use that the acceptance rate is a function of $w$ and $w'$ only to rewrite the expression in terms of the distributions of reweighting factors.
Evaluating this expression using target samples for the expectation over $p$, and model samples for the expectation over $\tilde{p}$, we find estimates consistent with the acceptance rates obtained by directly sampling using flow-based MCMC, as expected given that the high-weight tails comprise a small fraction of the mass of $p$.\footnote{We would expect a large correction for a mode-collapsed model where the high-weight tail contains $\sim 1/2$ of the mass of $p$.}

We can extend this reasoning to estimate the asymptotic distribution of rejection runs and thus the autocorrelation time.
We relegate the derivation to Appendix~\ref{sec:asymptotics}, and summarize the results here:
in terms of the ``acceptance rate out of a configuration'' with weight $w$,
\begin{equation}
    A(w \rightarrow \tilde{p}) = \int dw' A(w \rightarrow w') \tilde{\rho}(w')
\end{equation}
the asymptotic acceptance rate is
\begin{equation}
    A(p \rightarrow \tilde{p}) = \int dw \rho(w) A(w \rightarrow \tilde{p})
\end{equation}
and the length $r$ of the rejection run following the initial acceptance of a configuration with weight $w$ is (geometrically) distributed as
\begin{equation}
    p(r | w) = A(w \rightarrow \tilde{p}) \left[ 1 - A(w \rightarrow \tilde{p}) \right]^r
\end{equation}
from which we can derive the expected rejection length under the target
\begin{equation}
    \vev{R}_p = \int dw ~ \rho(w) \frac{1 - A(w \rightarrow \tilde{p})}{A(w \rightarrow \tilde{p})}
\end{equation}
which is directly and trivially related to the rejection-run estimator~\cite{Albergo:2019eim} for the integrated autocorrelation time as
\begin{equation}
    \tau_{\mathrm{int}} = \frac{1}{2} + \vev{R}_p.
\end{equation}

\begin{figure}
    \centering
    \includegraphics[width=\linewidth]{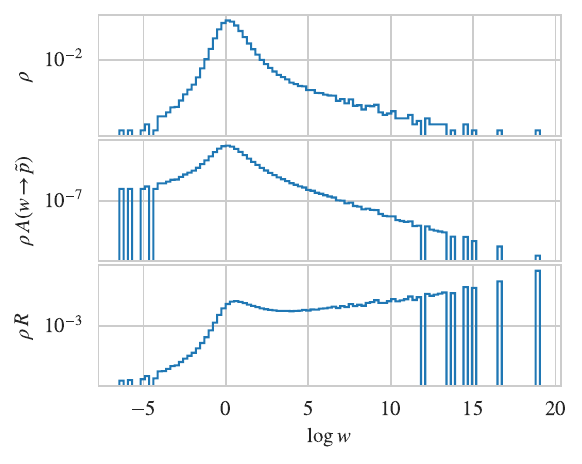}
    \caption{
        Top panel: for the case-study model, distribution of reweighting factors under the target, $\rho(w) = \vev{\delta(w - p/q)}_p$, estimated using $1.28 \times 10^6$ samples generated using AHMC.
        Bottom panels: integrands constructed from the distribution of reweighting factors as described in the text.
        Integrating over the second panel will yield the asymptotic acceptance rate, while integrating over the third panel gives the integrated autocorrelation time up to an offset of $+1/2$ as described in the text.
    }
    \label{fig:case_study_integrand_stack}
\end{figure}

Applying these expressions, as illustrated in Fig.~\ref{fig:case_study_integrand_stack}, we find immediately that the autocorrelation time either diverges or is too long to estimate even using target samples.
Using model samples to estimate $A(w \rightarrow \tilde{p})$ as a function of $w$ and applying them to a histogram estimate of $\rho(w)$, we see that the integrand of the asymptotic acceptance rate has little mass present in the tail beyond $\log w \sim 5$, as expected given the consistent direct and asymptotic estimates of the acceptance rate.
However, the integrand of the integrated autocorrelation time grows as $w$ increases, and because the variables of the theory are not compact, there is in principle no upper bound on the maximum $w$.
Due to this lack of an upper bound we cannot conclude that the autocorrelation time diverges, as it is possible that the integrand turns over at even higher $w$; however, even if this is the case, the sampler is clearly pathological.
By comparison, the autocorrelation time measured on ensembles generated using flow-based MCMC with this model is only $O(1)$, even on datasets of $O(10^6)$ samples.

It should be noted that the high-weight tails which exist in these models are not necessarily a generic feature of flow models. 
Real scalar field theory has non-compact variables, and so the reweighting factors are unbounded; by contrast, theories with compact variables are guaranteed to have a maximum $w$, so high-weight tails are naturally regulated, and the possible divergence of the flow-based MCMC integrated autocorrelation time observed in Sec.~\ref{sec:case_study} cannot occur.

\subsection{Discussion: performance metrics for flow-based MCMC}

The failure of the forwards KL divergence and MCMC acceptance rate to diagnose the pathology explored in the previous subsection indicates that they are unsuitable metrics for flow-based MCMC.
They fail because the problems with flow-based MCMC are caused by poor modeling in low-density regions.
The forwards KL divergence, $\int d\phi ~ p(\phi) \log w(\phi)$, is insensitive to these regions: the weight $p$ is small therein, and the logarithm damps the effect of large $w$.
The acceptance rate provides only a lower bound on the autocorrelation time~\cite{Albergo:2019eim}, and this bound is particularly weak when the sampler may undergo long runs of rejections, as observed here.
To illustrate: a Markov chain in which every other proposal is accepted has a much shorter autocorrelation time than one where half of the samples are replicates due to a single long rejection run, even though both chains have a 50\% acceptance rate.
The acceptance rate also fails to qualitatively penalize mode-collapsed models: for a symmetric bimodal system, the finite-sample acceptance rate will tend to be overestimated only by a factor of 2 versus the asymptotic one, as the rejections that populate the missed mode can only make up half of the total samples.
A low KL divergence and high acceptance rate are necessary, but insufficient.

The ESS is more useful for this purpose, but must be interpreted carefully when evaluated using model-distributed samples in direct studies. 
Unlike the other metrics, the ESS penalizes models with broad distributions of reweighting factors that, as discussed above, lead to long MCMC autocorrelation times.
Similar arguments apply for Eq.~\eqref{eqn:asymp-target-ess}.
The ESS also distinguishes mode-collapsed models from non-pathological ones, as the denominator clearly becomes large when $w(\phi) \gg 1$ over some mode of the target.
In practice, the ESS is a more useful metric, especially when measured with target-distributed validation data. We found that high variance in the ESS estimated with model samples indicated low asymptotic ESS as estimated with target samples, and that this indicated problems with the models for use with flow-based MCMC. Low observed values of the ESS should be taken seriously, even if the MCMC acceptance rate is high.

\subsection{Case study: mixing flow-based MCMC and HMC}
\label{sec:hybrid-sampling}

As motivated in Sec.~\ref{sec:augmented-MCMC}, composing different MCMC updates can produce samplers with better properties than either type of update alone.
Specifically, we consider a Markov chain constructed by alternating $n_F$ flow-based MCMC updates with $n_H$ trajectories of HMC.
Each type of update is applied completely independently and ends in its own Metropolis test, so neither algorithm needs any modification to use the two in combination.

In this scheme, the limit $n_H \gg n_F$ corresponds to augmented HMC as discussed in Sec.~\ref{sec:aug-hmc-results}, except the augmentation step is a learned flow-based MCMC proposal rather than some mode-hopping transformation constructed by hand.
As long as the flow model samples from all modes and has a nontrivial acceptance rate, this augmentation solves the problem of freezing in HMC.
While this is not useful when simpler mode-hopping transformations are known, it may be possible to train multimodal flow models in the more general case (using e.g.~flow-distance regularization) to provide machine-learned augmentation steps.
However, in this case mixing of the Markov chain is primarily driven by HMC, and any critical slowing down beyond freezing will not be alleviated.

\begin{figure}
    \centering
    \includegraphics[width=\linewidth]{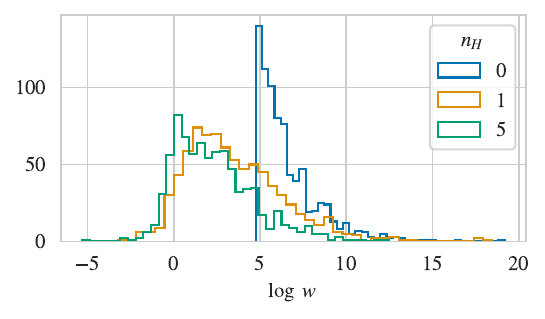}
    \caption{
        Effect of different numbers of HMC steps $n_H$ on a distribution of 800 samples from the underweighted tail with $\log w \gtrsim 5$.
        Histograms for $n_H \ne 0$ include a marginalization over 100 different HMC chains out of each configuration.
    }
    \label{fig:mixin_densities}
\end{figure}

More interestingly, we can consider the regime where ${n_F > n_H = 1}$, i.e.~flow-based MCMC augmented with occasional applications of HMC.
Above, we saw that the asymptotic performance of purely flow-based updating with these models is compromised by long rejection runs triggered when high-weight configurations are accepted.
Mixing in occasional HMC updates must solve this problem, as long rejection streaks will be terminated after enough HMC steps occur to move the chain to a better-modeled part of configuration space.
The practical question to answer is thus how many HMC updates must be applied with what frequency to regulate the long rejection runs.
For our case-study model and the HMC parameters we consider in this paper, Fig.~\ref{fig:mixin_densities} demonstrates that even a single HMC step moves the bulk of the tail samples with $\log w \gtrsim 5$ to smaller reweighting factors.

Asymptotic analysis of the behavior of the composite algorithm is complicated by the need to marginalize over all possible momentum draws as well as HMC's dependence on the previous configuration, which means that the distribution of reweighting factors no longer captures all information about overall sampling performance.
However, we can make estimates by considering the asymptotics of flow proposals made on top of fixed HMC evolutions.
Between HMC steps, proposals are made completely independently to update a fixed configuration.
We can model the distribution of rejection run lengths under the composite algorithm, starting from some initial configuration $\phi_0$, as a generalization of the geometric distribution
\begin{equation}
    p(r|\phi_0, \bm{\Pi}) = \frac{
        \prod_{t \le r} (1 - A(t))
    }{
        \sum_{r'} \prod_{t \le r'} (1 - A(t))
    }
    \label{eqn:asymptotic-mixin-rej-dist}
\end{equation}
where $A(t) = A(\phi_t \rightarrow \tilde{p})$ is a time-dependent acceptance rate out of the configuration evolved by $t$ steps of HMC, $\phi_t$.
We have made explicit the dependence on the (infinite) set of momentum fields $\bm{\Pi}$ drawn for use in all ensuing HMC steps, which must be marginalized over to quantify asymptotic sampling performance.
We approximate this marginalization by averaging over 100 HMC chains evolved from $\phi_0$ using different random seeds.
After 100 trajectories the unnormalized geometric distributions are zero within numerical precision, even for $n_F=1$, indicating that expectations computed using the truncated distributions are converged and accurate.

\begin{figure*}
    \centering
    \includegraphics{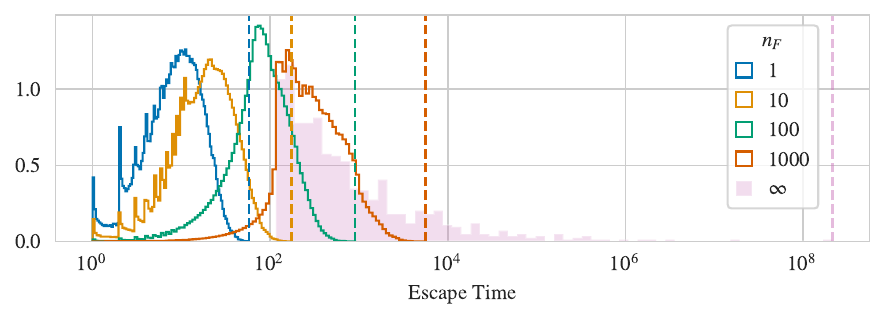}
    \caption{
        Distribution of expected number of rejections following configurations from the undersampled tail with $w \gtrsim 5$, for a Markov chain evolved by alternating $n_F$ flow-based MCMC proposals with single HMC trajectories.
        The escape time metric counts either a flow proposal or an HMC trajectory as one step.
        The distribution corresponds to $\sum_r r p(r|\phi)$, where $p(r|\phi)$ is as in Eq.~\eqref{eqn:asymptotic-mixin-rej-dist}, computed as described in the text for 800 different tail samples $\phi$, and marginalized over 100 different HMC chains run out of each sample and over all different delays before the first HMC step occurs (i.e.~after $0 \ldots n_F$ flow proposals, with even weight).
        Each dashed vertical line indicates the largest value for the distribution shown in the same color.
    }
    \label{fig:mixin_escape_times}
\end{figure*}

The results, shown in Fig.~\ref{fig:mixin_escape_times}, demonstrate that even infrequent HMC steps will truncate long rejection runs.
In the purely flow-based MCMC limit $n_F = \infty$, the typical run length following the tail samples is $\sim 10^3$, but the longest is $\sim 10^8$.
A single HMC step per thousand flow steps, $n_F = 1000$, does not change the typical run length appreciably, but does truncate the maximum run length to $\sim 10^4$. This indicates that the neighborhoods of high-weight configurations are not extensive (as would be the case with a mode-collapsed model).
As the frequency of HMC steps increases, the maximum escape time out of the tail configurations decreases progressively, and the shape of the distribution changes more substantially.

\begin{figure}
    \centering
    \includegraphics[width=\linewidth]{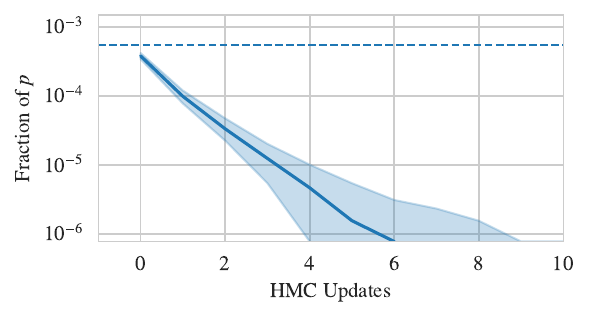}
    \caption{
        Time for HMC updates to move out of the high-weight tail (where $w > 5$) after first entering it, computed as described in the text.
        The dashed line indicates the fraction of $p$ in the high-weight tail, $\sim 0.05\%$.
        At zero HMC steps, the value indicates the fraction of $p$ which one HMC update newly moves into the high weight tail, $\sim .03\%$, which decays as HMC moves these configurations back out of the tail.
        The solid line denotes the median fraction, while the band spans the 1st to 99th percentiles.
    }
    \label{fig:mixin_tail_decay}
\end{figure}

The HMC updates must move configurations into the undersampled regions of configuration space to compensate for the truncation of the rejection runs that would have otherwise populated them.
This could reintroduce the problem.
However, as shown in Fig.~\ref{fig:mixin_tail_decay}, while the HMC augmentation introduces high-weight configurations more frequently in a relative sense, this happens only infrequently in an absolute sense.
Using our target-distributed ensemble of $1.28 \times 10^6$ configurations, we estimate that only $\sim 0.05\%$ of the mass of $p$ is in the high-weight tail where $\log w > 5$.
We apply 1000 HMC updates to each configuration in the full ensemble, generating an ensemble of Markov chains.
Updating one target-distributed ensemble with HMC yields another, but will move some fraction of the original ensemble into the tail and some fraction out.
We find that each HMC update moves $\sim .03\%$ of $p$ into the tail, which corresponds roughly to the probability that an HMC augmentation step will introduce a problematic high-weight configuration.
Subsequent updates rapidly move these configurations back out of the tail, as shown in Fig.~\ref{fig:mixin_tail_decay} and similarly demonstrated in Fig.~\ref{fig:mixin_escape_times}.
Heuristically, the mismodeled regions are only a few HMC trajectories wide.

These results demonstrate that augmenting flow-based MCMC with occasional HMC updates regulates the issues arising in flow-based MCMC, providing a sampler with all the advantages of independence Metropolis without the risk of slow convergence.
A more quantitative comparison of the relative performance of HMC and the composite algorithm is highly application-dependent~\cite{Abbott:2022zsh} and beyond the scope of this exploratory study. 
The performance of the composite algorithm depends not only on the model, but also on the tuning of the HMC augmentation step, whose optimal parameters depend on the structure of the lattice theory as well as the model being augmented.
We also note that this algorithm is unlikely to outperform cluster updates or other specialized sampling algorithms, when they are available for the theory of interest~\cite{Hoshen:1976zz,Wolff:1987jk,Swendsen:1987ce,Edwards:1988ba,Wolff:1988uh,BrowerTamayo89,Hasenbusch:1989kx,Sinclair:1992vm,Bietenholz:1995zk,Evertz:2000rk,Prokofev:2001ddj,Kawashima:2004,Azcoiti:2009md,Gattringer:2015baa,Bernard:2009,Michel:2015,Nishikawa:2015,Hasenbusch:2018ztj,Lei:2018vis}.

\section{Conclusion}
\label{sec:conclusion}

This work investigates how flow-based methods can be applied to distributions with nontrivial mode structure, where the most straightforward training methods are prone to mode (or subspace) collapse. 
We present a set of general approaches, both architecture- and training-based, to treat this issue. 
Each method has strengths and weaknesses, and each is applicable in different circumstances, depending on the prior information available about the target mode structure.
We numerically demonstrate their application to two different LQFT examples: one multimodal, and one with an extended mode. All of the methods investigated are able to alleviate collapse in both cases (with varying degrees of success).
We conclude that mode collapse is not an obstacle to flow modeling, especially in the frequent case where the mode structure arises from spontaneous symmetry breaking.

In our numerical investigations, we observe that all methods considered struggle to model the (inner) tails of the target distribution.
Accurate modeling of nontrivial structure across a hierarchy of density scales is, of course, an inherently difficult problem.
The engineering of architecture and training scheme necessary to treat this issue will likely require heavy specialization for each target theory~\cite{Abbott:2022zsh}.
This will be important to explore in applications to targets of physical interest like QCD.

Mismodeling of the inner tail regions presents an obstacle to using otherwise high-quality models with reweighting or flow-based MCMC.
These issues manifest in subtle ways in direct sampling studies, much like with topological freezing in gauge theories~\cite{Albergo:2022qfi}.
Different metrics are better-suited to diagnose these problems than others.
The ESS computed with target-distributed validation data is particularly powerful in this regard.

Tail mismodeling is not, however, an obstacle to useful flow-based sampling in general. As demonstrated in Sec.~\ref{sec:sampling}, intermixing steps of HMC with flow-based MCMC regulates the pathologies exhibited by flow-based MCMC alone. The resulting composite sampling algorithm combines the merits of both its components. For theories of physical interest, hybrid approaches like this one may serve as a bridge until higher-quality models are available~\cite{Abbott:2022hkm}.
Further study and characterization of their performance properties will be important.

Besides what has been tested here, there is plenty of opportunity for further improvement. This includes both new methods for constructing models, as well as new ways to use them for sampling or inference.
We present some interesting possibilities suggested by this study in Appendix~\ref{sec:other-approaches}.

In summary, the results presented in this work show that nontrivial mode structures are not an obstacle to flow-based methods.
Simultaneously, this study demonstrates that not all flow-based approaches to inference are equivalent, and problems with one approach do not mean all others must fail.
There is a broad class of methods to be explored, each with their own advantages, disadvantages, and scaling properties; it remains an open question which will be most useful for lattice field theory.

\begin{acknowledgements}
We thank Danilo J. Rezende, S\'{e}bastien~Racani\`{e}re, and Julian Urban for useful comments on a draft of this manuscript and Yoshihiro Saito for helpful discussions. GK, DB, DCH, and PES are supported in part by the U.S. Department of Energy, Office of Science, Office of Nuclear Physics, under grant Contract Number DE-SC0011090. PES is additionally supported by the National Science Foundation under CAREER Award 1841699 and under EAGER grant 2035015, by the U.S. DOE Early Career Award DE-SC0021006, by a NEC
research award, and by the Carl G and Shirley Sontheimer Research Fund. KC was supported by the National Science Foundation under the awards OAC1836650. MSA thanks the Flatiron Institute and is supported by the Carl Feinberg Fellowship in Theoretical Physics. 
CCH and JWC are partly supported by 
the National Science and Technology Council, Taiwan, under Grants 112-2112-M-002-027 and 113-2112-M-002-012 and the Kenda Foundation. KFC is supported by the Ministry of Science and Technology, Taiwan, under Grant No. 109-2628-M-002-002.
This work is supported by the National Science Foundation under Cooperative Agreement PHY-2019786 (The NSF AI Institute for Artificial Intelligence and Fundamental Interactions, http://iaifi.org/). 
This manuscript has been authored by Fermi Forward Discovery Group, LLC under Contract No. 89243024CSC000002 with the U.S. Department of Energy, Office of Science, Office of High Energy Physics.
This work is also associated with an ALCF Aurora Early Science Program project, and used resources of the Argonne Leadership Computing Facility, which is a DOE Office of Science User Facility supported under Contract DEAC02-06CH11357. SP acknowledges the financial support of the MIT Undergraduate Research Opportunities Program (UROP).
The authors acknowledge the MIT SuperCloud and Lincoln Laboratory Supercomputing Center~\cite{reuther2018interactive} for providing HPC resources that have contributed to the research results reported within this paper.
Numerical experiments and data analysis used PyTorch~\cite{NEURIPS2019_9015}, Horovod~\cite{sergeev2018horovod}, NumPy~\cite{harris2020array}, SciPy~\cite{2020SciPy-NMeth}, and pandas~\cite{jeff_reback_2020_3715232,mckinney-proc-scipy-2010}.
Figures were produced using matplotlib~\cite{Hunter:2007} and seaborn~\cite{Waskom2021}.
\end{acknowledgements}

\bibliography{multimodal}

\appendix

\section{Asymptotic analysis of flow-based MCMC autocorrelations}
\label{sec:asymptotics}

Because each proposal in a flow-based MCMC chain is completely independent of any previous proposal and of the current state of the chain, the acceptance rate given that the current configuration is $\phi_i$ (the ``acceptance rate out of $\phi_i$'') is always
\begin{equation}\begin{alignedat}{2}
    a_i &\equiv a(\phi_i) 
    = A(\phi_i \rightarrow \tilde{p}) 
    &&= \int d\phi' ~ A(\phi_i \rightarrow \phi') \tilde{p}(\phi') 
    \\
    & = a(w_i)
    = A(w_i \rightarrow \tilde{p})
    &&= \int dw' ~ A(w_i \rightarrow w') \tilde{\rho}(w').
\end{alignedat}\end{equation}
This value is independent of how many updates have been proposed and can be thought of as a property of a configuration or, as emphasized in the second line, as a function of the weight of the current configuration $w_i$.
In terms of this quantity, the global asymptotic acceptance rate is
\begin{equation}
    A \equiv \int d\phi_i ~ p(\phi_i) a(\phi_i)
    = \int d w_i ~ \rho(w_i) a(w_i).
\end{equation}

Because each proposal is independent, the number of rejections following the initial acceptance of $\phi_i$ is geometrically distributed, i.e.
\begin{equation}
    p_i(r) \equiv p(r | \phi_i) = p(r|w_i) = a_i \left[ 1 - a_i \right]^r
\end{equation}
where $r$ is the number of rejections before a new configuration is accepted (such that $r=0$ means the next proposal is accepted).
The expected number of rejections following the initial acceptance of $\phi_i$ is thus
\begin{equation}\begin{gathered}
    R_i \equiv R(\phi_i) = R(w_i) = \sum_r r ~ p_i(r) = \frac{ 1 - a_i }{ a_i }
    \\ \Leftrightarrow \frac{1}{1+R_i} = a_i
    \label{eqn:rej_run_ev}
\end{gathered}\end{equation}
such that the expected rejection run length under the target is
\begin{equation}\begin{aligned}
    \vev{R}_p 
    &= \int d\phi ~ p(\phi) R(\phi)
    = \int dw ~ \rho(w) R(w)
    \\&\approx \frac{1}{N} \sum_i R_i, \quad (\phi_i \sim p) \, .
\end{aligned}\end{equation}

Using the distributions of rejection runs following individual configurations $p(r|\phi)$, we can write an expression for the overall distribution of rejection runs $p(r)$ produced by flow-based MCMC:
\begin{equation}\begin{aligned}
    p(r) &= 
        \int d\phi ~ \frac{p(\phi)}{1 + R(\phi)} p(r | \phi)
    ~ \bigg/
        \int d\phi ~ \frac{p(\phi)}{1 + R(\phi)}
    \\&= \frac{1}{A} \int d\phi ~ p(\phi)  a(\phi)  p(r | \phi)
    \\&= \frac{1}{A} \int dw ~ \rho(w)  a(w)  p(r | w)
    \\&\approx \frac{1}{A} \frac{1}{N} \sum_i a_i ~ p_i(r) \; , \quad (\phi_i \sim p) \, .
    \label{eqn:rej-run-dist}
\end{aligned}\end{equation}
In the first line, the factor $1/(1+R(\phi))$ compensates for the difference in counting between $p(\phi)$, which encodes every appearance of $\phi$ including duplications due to rejections, versus the counting of rejection runs following each initial acceptance of $\phi$.
The integral in the denominator of the first line normalizes $p(r)$ such that $\sum_r p(r) = 1$, and in the second line we recognize using Eq.~\eqref{eqn:rej_run_ev} that this integral is simply the asymptotic acceptance rate $A$.
Note that $\vev{R}_p$ is different from the average rejection run length, which can be computed straightforwardly from Eq.~\eqref{eqn:rej-run-dist} as $\sum_r r p(r) = (1-A)/A$.

We can write an asymptotic estimator for the rejection-run autocorrelation function $p_{t \text{rej}} = \rho(t)/\rho(0) = \mathrm{AC}(t)$ (defined in Ref.~\cite{Albergo:2019eim}) in terms of the above expression for $p(r)$,
\begin{equation}\begin{aligned}
    \mathrm{AC}(t) 
    &= A \sum_{r \ge t} \sum_{r' \ge r} p(r')
    = A \sum_{r \ge t} (r-t+1) p(r) 
    \\&= \frac{1}{N} \sum_i a_i \frac{(1-a_i)^t}{a_i}
    = \frac{1}{N} \sum_i p_i(t) (1 + R_i) 
\end{aligned}\end{equation}
where the factor of the acceptance rate $A$ fixes overcounting of rejection runs (for a sample of $N$ there are $AN$ newly accepted configurations, each of which is followed by $r \ge 0$ rejections), and we see in the second line that we recover the intuitive expression ${\mathrm{AC}(t) = \vev{\left[ 1 - A(\phi \rightarrow \tilde{p}) \right]^t}_{\phi \sim p}}$.
From this we can compute the integrated autocorrelation time for the rejection-run estimator,
\begin{equation}\begin{aligned}
    \tau_{\mathrm{int}} 
    &= \frac{1}{2} + \sum_{t=1} \mathrm{AC}(t) = -\frac{1}{2} + \sum_{t=0}  \mathrm{AC}(t)
    \\&= -\frac{1}{2} + \frac{1}{N} \sum_i (1+R_i) = \frac{1}{2} + \vev{R}_p
    \label{eqn:rej_run_ac_time}
\end{aligned}\end{equation}
where we have used that $\sum_t p(t|\phi_i) = 1$.
We see from Eq.~\eqref{eqn:rej_run_ac_time} that the integrated autocorrelation time is directly and trivially related to $\vev{R}_p$.

\section{Extensions \& other approaches}
\label{sec:other-approaches}

\subsection{Model construction}

In this work we demonstrate basic applications of the methods described in Secs.~\ref{sec:arch_methods} and \ref{sec:multimodal-approaches} to construct models which capture the nontrivial mode structures of real and complex scalar $\phi^4$ theory in two dimensions.
The results of this study suggest some obvious extensions, listed below.

Adiabatic retraining and flow-distance regularization may each be used in combination with forwards KL self-training, rather than with reverse KL self-training.
More generally, each of these procedures may be considered as a deformation of the target, and so may be used with any training procedure.
Similarly, combining the improvements in architecture and training methods may yield better models than obtained here. 

Flow-distance regularization and (to a lesser extent) adiabatic retraining require careful design of a schedule to avoid mode collapse.
Schedulers designed to detect and prevent mode-collapse can make these methods more practical to apply, but require an investigation of metrics to anticipate mode collapse during training.

Self-training with the reweighted forwards KL divergence is generically more expensive than reverse KL self-training, and especially so when training samples are drawn from a mixture. However, we observe that already-bimodal models are typically stable against mode collapse under reverse KL self-training. This suggests using forwards KL self-training only early in training, switching to reverse KL after circumventing the regime where mode collapse may occur.

The forwards KL self-training scheme presented in this paper allows for data augmentation in combination with self-training generically, which may have useful applications other than putting in mode structures by hand.

This study considered only the reverse KL and forwards KL divergences (and simple generalizations thereof) as losses.
As noted in Ref.~\cite{Wu:2021tfb}, the reverse KL divergence is an integral over $\tilde{p} \log \tilde{p}/p$, so the loss is only sensitive to large discrepancies between $\tilde{p}$ and $p$ when $\tilde{p}$ is not small.
A similar argument with $\tilde{p} \rightarrow p$ applies for using the forwards KL divergence as a loss.
At finite sample sizes, this problem manifests as poorly resolved contributions to the gradients from rarely sampled regions, making it difficult for any training scheme using KL divergences as losses to achieve accurate modeling of those regions.
As discussed in Sec.~\ref{sec:case_study} (see Fig.~\ref{fig:case_study_obs_densities}), problematic high-weight configurations arise from mismodeling in exactly these low-density regions.
Further, as noted by Ref.~\cite{DelDebbio:2021qwf}, the mode-seeking effect of reverse KL self-training tends to produce models which undersample from the tails of distributions.
Training with different losses may be able to produce models with better-behaved distributions of reweighting factors.
Similarly, it may be possible to use self-training schemes which produce additional samples from underweighted regions (as data augmentation with the forwards KL self-training procedure does for the different modes in this work).

Mixtures are a common prescription for smoothing out pathological reweighting factor distributions~\cite{artbowen_mc_textbook}.
In this work we constructed symmetrized mixtures over the global $Z_2$ symmetry of the target, which is not a symmetry of our model architecture.
We considered only mixtures constructed with mode-collapsed unimodal flow models, but the resulting models compared well with truly bimodal flow models, and our results suggested some of these mixtures had nontrivial ESS where bimodal flow models failed.
Applying the symmetrized mixture construction to bimodal flow models trained using the other methods may help regulate the distribution of reweighting factors and provide models that can be used with flow-based MCMC.
There are additional symmetries that could be exploited in this manner, like the $Z_2$ subgroup of translational symmetries broken by the checkerboard masking used in our coupling layers.

Since we observe undersampled high-weight tails nearly ubiquitously, for all the methods considered, it may also be the case that the (fixed) affine architecture we consider here may be insufficiently expressive to capture the bimodal distribution.
The problem could be a consequence of the non-universality of individual affine coupling layers, which cannot produce multimodal conditional distributions in isolation \cite{wehenkel2020}. 

For complex scalar field theory, one could consider representing the complex variables in a polar basis, rather than the Cartesian basis employed in this work. 
However, the expressivity of allowed transformations is severely limited by the restricted domain of the radial variables.
Preliminary investigations found indications that the coordinate singularity at $\phi=0$ lead to inner-region undersampling similar to that observed with canonicalized equivariant models in this work.

\subsection{Flow-based sampling \& inference}

In this work we considered the performance of flow-based MCMC, reweighting, and the composite flow/HMC sampling algorithm.
However, these are only a few means of inferring properties of the target distribution using flow models, and the broader class of flow-based approaches to sampling and inference remains yet to be explored.
We list some interesting options below.

Ref.~\cite{Wu:2021tfb} encountered similar problems with high-weight configurations as seen in this study. Rather than augmenting flow-based MCMC steps with HMC, that work regulated the effect of high-weight configurations using augmented MCMC (as described in Sec.~\ref{sec:augmented-MCMC}) to propose updates by transformations which are symmetries of the target, but not the model, with the reasoning that all configurations in the symmetry orbit are unlikely to be high-weight. This method can be used in combination with or instead of the HMC augmentation used here, although it is unclear whether this approach would be as helpful for the translationally symmetric architectures considered in this work. 
This provides a less expensive alternative to constructing mixtures over the same symmetry groups, which require evaluation of the model for all configurations in the symmetry orbit, rather than possibly only a randomly chosen subset. 

Defensive mixtures~\cite{hesterberg1995weighted} provide another means of incorporating target samples, wherein samples are drawn stochastically from either the model distribution or from the target (necessitating a separate reservoir of target-distributed samples generated with e.g.~HMC).
This approach directly clips the distribution of reweighting factors, imposing a maximum $w$ even for non-compact variables.
Unfortunately, this construction requires the normalized target density ${p = \exp[-S]/Z}$, which is not generally available in the lattice field theory context.
However, it may be possible to use flow-based stochastic estimators of $Z$~\cite{Nicoli:2020njz} to estimate $w$ for use with reweighting; further work is required to determine the conditions for exactness of this approach.
It is unclear whether an approach based on stochastic $Z$ estimation can be reconciled with MCMC sampling algorithms like the pseudo-marginal method~\cite{andrieu2009pseudo}.

Stochastic normalizing flows~\cite{wu2020stochastic} generalize fully deterministic coupling layers to stochastically sampled conditional distributions, while still admitting exact and tractable computation of importance weights.
This framework allows incorporation of updates with (or inspired by) traditional MCMC algorithms directly into the models, rather than using them to augment deterministic flow proposals as done in this work.
This approach is explored in a lattice context its relation to non-equilibrium thermodynamics in Refs.~\cite{Caselle:2022acb,Caselle:2022esc}.

Validated variational inference~\cite{pmlr-v108-huggins20a} may provide an alternative to MCMC or reweighting, where error estimates from sampling statistics are augmented with systematic errors provided by formal bounds on discrepancies between expectations under different distributions.
In this framework, these explicit error bounds may be directly related to distances between distributions and can be formally incorporated into the variational inference workflow. This can be achieved by using a distance measure that incorporates an underlying metric in the space of probability measures, e.g.~integral probability metrics \cite{muller_1997}, or by bounding such a distance with a linear combination of scale invariant {$\alpha$-divergences}. The latter is achievable without considerable computational overhead beyond what is normally necessary in variational inference contexts. 

It may be possible to improve upon reweighting using recent developments in stabilizing importance ratios, e.g.~by fitting tails with generalized Pareto distributions~\cite{vehtari2021pareto}. Such approaches may also provide new diagnostics of model quality based on the reliability of the importance weight.

\clearpage

\end{document}